\documentclass[aps,prd,superscriptaddress,nofootinbib,amsmath,amsfonts,preprintnumbers,groupedaddress,showpacs,9pt,english]{revtex4-2}
\usepackage{amsmath}
\usepackage{makecell}
\usepackage{amsmath}
\usepackage{amssymb}
\usepackage{babel}
\usepackage{wrapfig}
\usepackage{cancel}
\usepackage{amsmath}

\usepackage{relsize,exscale}
\makeatletter
\usepackage{amsmath}
\usepackage{array}
\newlength{\Wone}
\newlength{\Wtwo}
\newlength{\Wthr}

\usepackage{makecell}
\newcommand{\beq}{\begin{equation}}
\newcommand{\eeq}{\end{equation}}
\newcommand{\bea}{\begin{eqnarray}}
\newcommand{\eea}{\end{eqnarray}}

\newcommand{\del}{\delta}
\newcommand{\bet}{\beta}

\newcommand{\dd}{\text{d}}
\newcommand{\comm}[1]{}

\newcommand{\Ri}{\mathcal{R}}
\usepackage{array,multirow,graphicx}
\usepackage{dcolumn}
\usepackage{newlfont}
\usepackage{bm}
\usepackage[colorlinks,citecolor=blue,urlcolor=blue,linkcolor=blue]{hyperref}
\usepackage[figtopcap]{subfigure}
\usepackage{color}

\usepackage{scalerel}
\usepackage{tikz}
\usetikzlibrary{svg.path}
\definecolor{orcidlogocol}{HTML}{A6CE39}
\tikzset{
  orcidlogo/.pic={
    \fill[orcidlogocol] svg{M256,128c0,70.7-57.3,128-128,128C57.3,256,0,198.7,0,128C0,57.3,57.3,0,128,0C198.7,0,256,57.3,256,128z};
    \fill[white] svg{M86.3,186.2H70.9V79.1h15.4v48.4V186.2z}
                 svg{M108.9,79.1h41.6c39.6,0,57,28.3,57,53.6c0,27.5-21.5,53.6-56.8,53.6h-41.8V79.1z M124.3,172.4h24.5c34.9,0,42.9-26.5,42.9-39.7c0-21.5-13.7-39.7-43.7-39.7h-23.7V172.4z}
                 svg{M88.7,56.8c0,5.5-4.5,10.1-10.1,10.1c-5.6,0-10.1-4.6-10.1-10.1c0-5.6,4.5-10.1,10.1-10.1C84.2,46.7,88.7,51.3,88.7,56.8z};}}
\newcommand\orcid[1]{\href{https://orcid.org/#1}{\mbox{\scalerel*{
\begin{tikzpicture}[yscale=-1,transform shape]
\pic{orcidlogo};
\end{tikzpicture}
}{|}}}}
\begin{document}

\date{\today}

\title{Holographic Dark Energy as a Source for Wormholes in Modified Gravity}

\author{G.G.L. Nashed$^{1,2}$}\email{nashed@bue.edu.eg}
\author{A.~Eid$^{3}$}\email{amaid@imamu.edu.sa}
\affiliation{$^{1}$Centre for Theoretical Physics, The British University, P.O. Box 43, El Sherouk City, Cairo 11837, Egypt\\$^{2}$Center for Space Research, North-West University, Potchefstroom 2520, South Africa\\ $^{3}$Department of Physics, College of Science, Imam Mohammad Ibn Saud Islamic University (IMSIU), Riyadh, Kingdom of Saudi Arabia}
\date{}

\begin{abstract}
Traversable wormhole solutions are explored in $f(\mathcal{R},\mathbb{T})$ gravity, a curvature--matter extension in which $\mathcal{R}$ is the Ricci scalar and $\mathbb{T}$ denotes the trace of the energy--momentum tensor. To generate explicit wormhole models, we prescribe holographic dark-energy densities based on entropy formalism proposed by R\'enyi, Moradpour, and Bekenstein--Hawking, namely
\[
\rho_{\textit{R}} = \frac{\alpha}{4\alpha_1 r^4 c^2 \kappa}\ln\!\left(1+\pi \alpha_1 r^2\right), \qquad
\rho_{\textit{M}} = \frac{\alpha}{4\pi r^2 c^2 \kappa \left(\pi \alpha_1 r^2 + 1\right)}, \qquad
\rho_{\textit{BH}} = \frac{\alpha}{4 c^2 \kappa r^2},
\]
with $\alpha$ and $\beta$ carrying dimensions of $L^{-2}$. The corresponding shape functions obtained from the field equations satisfy the standard throat and flare-out requirements for traversability. We then study how varying $\alpha$ and $\beta$ affects (i) the balance of forces associated with equilibrium and (ii) the status of the energy conditions. In particular, the null energy condition is found to be violated, indicating that exotic matter (or an effective exotic sector) is required to support the wormhole geometry. The spatial structure of the solutions is further visualized through embedding surfaces.
\end{abstract}

\maketitle

\section{Introduction}
In theoretical physics, wormholes are regarded as unusual spacetime geometries that may arise as solutions, classical or quantum, to the field equations of gravity. They have long been considered potential mechanisms for advanced concepts such as faster-than-light propulsion, shortcuts for interplanetary travel, and even time machines. The earliest formulation appeared in the mid of 1930's when Einstein and Rosen described what is now called the Einstein-Rosen bridge (ERB) \cite{Einstein:1935tc}, a connection between two identical universes. Although initially presented as a particle model, this interpretation proved inadequate. Wheeler later reintroduced the concept, emphasizing its role as a possible link between distant regions of spacetime \cite{Misner:1957mt}. A major advance came from Morris and Thorne \cite{Morris:1988cz}, who argued that wormholes could be traversable under certain conditions, sparking wide interest in exploring geometries that allow stable throats.

The simplest example of such a solution is embedded in the Schwarzschild metric, giving rise to the so-called Schwarzschild wormhole. This solution represents an eternal black hole, yet unlike black holes, which are well-established objects defined as regions of spacetime from which light cannot escape, wormholes remain hypothetical \cite{LIGOScientific:2016aoc,Khodadi:2022dff,LIGOScientific:2016lio,Errehymy:2023xpc}. Within general relativity (GR), sustaining a wormhole requires exotic matter, a serious obstacle that has driven researchers toward modified gravity theories. These frameworks extend GR by altering the structure of spacetime geometry and can naturally address both cosmic acceleration and the exotic matter issue.

One of the earliest modifications, $f(R)$ gravity, extends the Einstein-Hilbert action by replacing the Ricci scalar with a function of curvature \cite{DeFelice:2010aj,Nashed:2021gkp,starobinsky1979relict}. Since then, a wide range of theories, such as $f(R, T)$ gravity \cite{Harko:2011kv,Nashed:2023pxd,Errehymy:2024spg}, noncommutative models \cite{Rahaman:2013ywa,Rahaman:2012pg}, Ricci-inverse gravity \cite{Mustafa:2024ark}, $f(R, G)$ \cite{Ashraf:2023bfg}, $f(Q)$  \cite{Mustafa:2023kqt,Kiroriwal:2024ymu,Kiroriwal:2023nul}, unimodular gravity \cite{Agrawal:2022atn}, Rastall-type models \cite{Errehymy:2024lhl,Battista:2024gud,Mehdizadeh:2015jra,Nashed:2021pkc,Boehmer:2012uyw,
Nashed:2021pah,Mustafa:2020kng,Shirafuji:1996im,Chaudhary:2023mfx,DeFalco:2021ksd,Malik:2023rov}
, Gauss-Bonnet extensions \cite{Errehymy:2024yey,Mishra:2021ato,Hussain:2022lxb}, Born-Infeld \cite{Eiroa:2012nv,Harada:2008rx}, Einstein-Cartan \cite{Bronnikov:2015pha,Nashed:2009hn,Bronnikov:2016xvj,ElHanafy:2014efn,Nashed:2005kn,Mehdizadeh:2017tcf} braneworld scenarios \cite{LaCamera:2003zd,Parsaei:2020hke,Kar:2015lma}, Chaplygin gas \cite{Javed:2023jqk,Eiroa:2007qz,Sharif:2014xwa,Nashed:2015pga,Lobo:2005vc,Esmakhanova:2011ar}, and thin-shell approaches  \cite{Javed:2023coi,Mustafa:2022obk,Sadiq:2024srd,Fatima:2024vvt,Fatima:2024dqh} have been employed to construct traversable wormhole solutions \cite{Furey:2004rq,Godani:2018blx}.

Among these, the $f(R, T)$ framework is particularly novel because its Lagrangian depends simultaneously  $R$ and  $T$. This opens new avenues for analyzing compact objects like black holes and wormholes. Initial investigations \cite{Harko:2011kv,Xu:2016rdf,Nashed:2023uvk,Jeakel:2023hss} examined its cosmological implications, including accelerated expansion, inflation, and perturbation dynamics. Within this setting, Khatri et al. \cite{Khatri:2024fef} studied how generalized uncertainty principle (GUP) corrections influence Casimir wormholes, while  Rajabi and Nozari \cite{Rajabi:2017alf} analyzed wormholes in unimodular gravity.

Though still confined to theory, wormholes are typically described as tunnel-like structures with a throat connecting two distant, asymptotically flat regions. They may be static, if the throat size is fixed, or dynamic, if the radius evolves in time. Ongoing research continues to search for observational strategies to distinguish such phenomena from black holes and other astrophysical objects.


Today, {gravitational lensing} serves not only as a fundamental test of General Relativity (GR) but also as a powerful probe of the universe's unseen components, such as dark matter (DM), dark energy (DE), and even extrasolar planets. Unlike ordinary luminous matter, DM reveals itself primarily through its gravitational effects. As early as the 1930s, Zwicky applied the virial theorem to galaxy clusters and inferred the existence of hidden mass~\cite{Zwicky:1937zza}. Subsequent cosmological surveys have confirmed that baryonic matter---comprising stars, gas, and planets---accounts for only about $5\%$ of the total cosmic content. The remaining $95\%$ is dominated by DM ($\approx27\%$) and DE, which together shape the evolution of the universe. On galactic scales, DM plays a crucial role in governing the formation and distribution of galaxies~\cite{Trujillo-Gomez:2010jbn}.

Another avenue of research emerges from the interplay between \textit{wormhole physics and holographic dark energy (HDE)}. Grounded in the Bekenstein--Hawking entropy--area relation~\cite{Bekenstein:1973ur,Hawking:1975vcx,Nashed:2001im,Cohen:1998zx,Chen:2023azy}, HDE models have been extended to incorporate quantum corrections, particularly through R\'enyi entropy, leading to modified energy densities with improved thermodynamic consistency~\cite{Moradpour:2017shy,Tsallis:1987eu}. Such extensions not only provide mechanisms for cosmic acceleration but also establish conditions that allow wormholes to remain stable. For example, Chaudhary \emph{et al.} investigated R\'enyi HDE in the context of $f(Q)$ gravity~\cite{Chaudhary:2025wsd}, while Paul and collaborators examined multiple HDE density profiles within the same framework, confirming the possibility of stable wormhole solutions~\cite{Paul:2025vem}.

Work on traversable wormholes has increasingly turned to modified gravity theories, where the effective energy-momentum tensor (EMT) can sometimes play the role that exotic matter traditionally occupied. Instead of relying exclusively on hypothetical fluids, researchers have demonstrated that frameworks such as $f(R)$, $f(T)$, $f(R, T)$, and Gauss-Bonnet gravity allow for wormhole configurations that are compatible with various dark energy candidates, including holographic models, phantom energy, quintessence, and Chaplygin-type gases. Collectively, these approaches suggest that the long-standing problem of exotic matter may be less restrictive than once believed.


The present research builds on this body of work by exploring wormholes in $f(R, T)$ gravity with three distinct holographic dark energy density profiles associated with the R\'enyi, Moradpour, and Bekenstein--Hawking entropy formalisms. A particular focus is the extent of null energy condition (NEC) violation, since this signals the exotic contributions required to preserve a wormhole throat. By comparing these density functions, we aim to demonstrate that traversable wormholes can arise within this framework without relying solely on exotic matter.

The article proceeds as follows. The introduction is presented in Section I, while Section II develops the modified gravitational equations corresponding to wormhole solutions in $f(R,T)$ theory. Section III analyzes the chosen holographic energy profiles and their associated shape functions and also show the embedding diagrams. Section IV examines the energy density and condition plots. Section V study the gravitational lensing produced by the wormholes.   Section VI evaluates equilibrium criteria for stability.   Section VII  concludes with the main findings.


\section{Outline of  $f(\mathcal{R}, \mathbb{T})$}\label{Sec:Overview_of_f(R,T)}

This section is devoted to outlining the essentials of $f(\mathcal{R}, \mathbb{T})$ theory. The gravitational framework is developed in the metric formulation by adopting the Levi--Civita connection, as discussed in~\cite{Harko:2011kv}. By contrast, the Palatini approach regards the affine connection as an entity independent of the metric, its form being determined through the variation of the Lagrangian~\citep[cf.][]{Wu:2018idg}. Within this framework, the action of the $f(\mathcal{R}, \mathbb{T})$ model can be written as:
\beq
S=\int{\dd^4x\sqrt{-g}}\left[ \frac{f(\Ri, { \mathbb{T}})}{\kappa}+\mathcal{L}_m \right]\,.
\label{f(R,T) action}
\eeq
In this context, ${\mathrm \kappa}$ is introduced as the Einstein coupling parameter, expressed as
${\mathrm \kappa = \tfrac{8\pi G}{c^4}}$, with ${\mathrm G}$ standing for Newton's gravitational constant and ${\mathrm c}$ for the speed of light. When the matter Lagrangian $\mathcal{L}_m$ is taken to be a function of the metric, the matter energy-momentum tensor is obtained through the definition:
\beq
 { \mathbb{T}_{\mu\nu}}=\frac{-2}{\sqrt{-g}}\frac{\del \left(\sqrt{-g} \mathcal{L}_m \right)}{\del g^{\mu\nu}}\,.
\eeq
For an anisotropic fluid, the corresponding energy-momentum tensor yields:
\begin{equation}\label{Tmn-anisotropy}
    { \mathbb{T}{^\mu}{_\nu}=  (p_{t}+\rho c^2)w{^\mu} w{_\nu}+p_{t} \delta ^\mu _\nu + (p_{r}-p_{t}) v{^\mu} v{_\nu}}\,.
\end{equation}
We describe the fluid by an energy density $\rho(r)$ and two distinct pressure components: a radial pressure $p_r(r)$ and a transverse pressure $p_t(r)$, specified with respect to the time--like four--velocity $w^\alpha$. The vector $v^\alpha$ is chosen as a unit radial space--like vector, leading to the diagonal energy--momentum tensor
\[
T^\alpha{}_\beta=\mathrm{diag}(-\rho c^2,\,p_r,\,p_t,\,p_t).
\]  The modified Einstein equations follow from the variation of the action~\eqref{f(R,T) action} with respect to the metric~\cite{Harko:2011kv}:
\beq
f_\Ri \Ri_{\mu\nu}-\frac{1}{2}f\, g_{\mu\nu}+S_{\mu\nu}f_\Ri=\kappa T_{\mu\nu}-f_T\left(T_{\mu\nu}+ \Theta_{\mu\nu} \right)\equiv { \mathbb{T}}_{\mu\nu}\,.
\label{f(R,T) EOM}
\eeq
Here we introduce the shorthand notations
$f_{\mathcal{R}} \equiv \tfrac{\partial f(\mathcal{R},\mathbb{T})}{\partial \mathcal{R}}$
and
$f_{\mathbb{T}} \equiv \tfrac{\partial f(\mathcal{R},\mathbb{T})}{\partial \mathbb{T}}$.
In addition, let
$S_{\mu\nu}$ denote the differential operator
$S_{\mu\nu} \equiv \left(g_{\mu\nu}\Box - \nabla_{\mu}\nabla_{\nu}\right)$.
For the quantity $\Theta_{\mu\nu}$ we define
\begin{equation}
\Theta_{\mu\nu} \equiv g^{\alpha\beta}
\frac{\partial \mathbb{T}_{\alpha\beta}}{\partial g^{\mu\nu}}
= -2\,\mathbb{T}_{\mu\nu} + g_{\mu\nu}\,\mathcal{L}_m
- 2 g^{\alpha\beta}
\frac{\partial^2 \mathcal{L}_m}{\partial g^{\mu\nu}\, \partial g^{\alpha\beta}} \, .
\end{equation}
Adopting the prescription of Harko \emph{et al.}~\cite{Harko:2011kv}, we set $\mathcal{L}_m = p$
(contrast with the alternative choice $\mathcal{L}_m = -p$ discussed in the same reference).
With this assumption, one obtains
$\Theta_{\mu\nu} = -2 \mathbb{T}_{\mu\nu} + p\, g_{\mu\nu}$.
Contracting Eq.~\eqref{f(R,T) EOM} results in
\beq
f_\Ri \Ri-2 f+3 \Box f_\Ri=\kappa  { \mathbb{T}}\,.
\label{eq:trace_eom}
\eeq
{
In $f(R,T)$ gravity the matter Lagrangian $L_m$ is not unique for anisotropic fluids, and common choices in the literature include $L_m = p$ and $L_m = -\rho$. Different choices modify the tensor $\Theta_{\mu\nu}$ and hence alter the explicit form of the effective source terms in the field equations. In this work we adopt $L_m = p$, which is used consistently in deriving both the field equations and the generalized Tolman--Oppenheimer--Volkoff (TOV) relation.

Choosing $L_m = -\rho$ would change the numerical coefficients entering the energy--condition combinations and may shift the corresponding quantitative parameter bounds. However, the qualitative behavior, in particular the violation of the radial null energy condition (NEC) near the throat, remains a generic feature of the solutions in linear $f(R,T)$ gravity. Moreover, the generalized TOV equilibrium follows from the modified conservation equation together with the field equations; therefore, for any fixed and consistent choice of $L_m$, the force balance is preserved and the total residual vanishes identically.
 }
In  $f(\mathcal{R},\mathbb{T})$ gravity, $\mathbb{T}$ given by  Eq.~(\ref{eq:trace_eom}) no longer remains linear in $\mathcal{R}$; instead, it turns into a second-order differential relation, unlike the algebraic form encountered in standard GR. Hence, nonlinear choices of the function $f(\mathcal{R},\mathbb{T})$ naturally produce a nonzero scalar curvature $\mathcal{R}$. This feature arises from the fact that ${ \mathbb{T}}$ does not vanish in the corresponding action. Accordingly, the covariant divergence of ${ \mathbb{T}}_{\mu\nu}$ takes the form~\cite{BarrientosO:2014mys}:
\beq
\nabla^{\mu}{ \mathbb{T}}_{\mu\nu}=\frac{f_{ \mathbb{T}}}{ \kappa-f_{ \mathbb{T}}} \left[ \left({ \mathbb{T}}_{\mu\nu}+ \Theta_{\mu\nu} \right) \nabla^{\mu} \ln{f_{ \mathbb{T}}} + \nabla^{\mu}\Theta_{\mu\nu}-\frac{1}{2}\nabla_{\nu}{ \mathbb{T}} \right]\,.
\label{Tmn divergence}
\eeq
In $f(\mathcal{R},\mathbb{T})$ gravity, the fact that the matter energy-momentum tensor is not conserved gives rise to an additional force. Such a force introduces modifications to the gravitational interaction beyond the predictions of General Relativity and may become relevant both at galactic scales and within the solar system~\cite{Harko:2011kv}.

Energy-momentum conservation holds only in the special case where the function reduces to $f(\mathcal{R},\mathbb{T})=\mathcal{R}$. This condition allows one to design corresponding $f(\mathcal{R},\mathbb{T})$ models~\citep[see also][]{Pretel:2021kgl}, although their field equations typically require numerical treatment. Since the aim of this work is to explore analytic configurations, such models will not be addressed here.

Equation~\eqref{f(R,T) EOM} can be reformulated in a form reminiscent of GR, with the Einstein tensor $G_{\mu\nu}$ appearing on the left-hand side, while the right-hand side incorporates an effective energy-momentum tensor that includes contributions from both the matter sector and curvature terms, namely:
\beq
G_{\mu\nu}=\frac{1}{f_\Ri}\left[{ \mathbb{T}}_{\mu\nu} +\frac{f-\Ri\,f_\Ri}{2}g_{\mu\nu}-S_{\mu\nu}f_\Ri \right] \equiv { \mathbb{T}}^{(\text{eff})}_{\mu\nu}\,.
\label{Gmn_Teff}
\eeq
For a generic choice of $f(\mathcal{R},\mathbb{T})$, the structure of the effective tensor ${\mathbb{T}}^{(\text{eff})}_{\mu\nu}$ becomes highly intricate, which makes the extraction of closed-form analytical solutions extremely difficult. Nevertheless, in certain specific models one can obtain simplifications. For instance, in the special case
\beq
f(\Ri,T)=\Ri+h(\mathbb{T})\,.
\label{f(R,T) separable}
\eeq
In the absence of an explicit $\mathcal{R}$--$\mathbb{T}$ interaction and assuming linearity in the Ricci scalar, the curvature sector of the effective energy--momentum tensor reduces to a much simpler form. Consequently, the field equations admit a more tractable representation, which can be written as:
\beq
G_{\mu\nu} = \kappa { \mathbb{T}}_{\mu\nu}+\frac{h}{2}g_{\mu\nu}+h_{ \mathbb{T}} \left({ \mathbb{T}}_{\mu\nu}-p\,g_{\mu\nu} \right)\,.
\label{GR+T effects EOM}
\eeq
For the sake of analytical simplicity and to capture the minimal modification of Einstein's theory, we consider a linear realization of the $f(\mathcal{R},\mathbb{T})$ framework, characterized by the functional choice:
\beq
f(\Ri,{ \mathbb{T}})=\Ri+\beta { \mathbb{T}}\,.
\label{eq:linear_f(R,T)}
\eeq
For the linear realization of the theory, the field equations simplify considerably, where $\beta$ appears as a dimensional constant parameter. Under this assumption, analytic solutions to the field equations can be obtained without difficulty~\citep[see, e.g.,][]{Hansraj:2018jzb, Bhar:2021uqr, Pretel:2020oae, Pretel:2021kgl}.

It should be emphasized that in the case of nonlinear extensions of $f(\mathcal{R},\mathbb{T})$ gravity, the resulting field equations acquire a far more intricate structure. A useful perspective on this complexity can be gained by examining the dynamical equivalence with scalar-tensor formulations of the theory and by employing the Palatini approach~\citep[cf.][]{Rosa:2021teg, Rosa:2022cen}.

 \section{Wormhole Configurations and Field Equations in Linear $F(R,\mathbb{T})$ Gravity}\label{sec2}

We begin our analysis with a static, spherically symmetric background takes the form:
\begin{equation} \label{1}
ds^{2} = e^{\nu(r)} dt^{2} - e^{\lambda(r)} dr^{2} - r^{2} d\theta^{2} - r^{2} \sin^{2}\theta\, d\phi^{2}.
\end{equation}
In this framework, the metric functions are given by $\nu(r)=2\zeta(r)$ and $e^{\lambda(r)}=\left(1-\frac{s(r)}{r}\right)^{-1}$.
Here, $s(r)$ denotes the shape function, which governs the spatial structure of the wormhole, while $\zeta(r)$ represents the redshift function, encoding the effects of gravitational time dilation. Accordingly, the metric components can be written as
$\nu(r)=2\zeta(r)$ and
$e^{\lambda(r)}=\left(1-\tfrac{s(r)}{r}\right)^{-1}$.
The function $s(r)$, known as the shape function, determines the spatial structure of the wormhole, while $\zeta(r)$, termed the redshift function, encapsulates the effects of gravitational time dilation.
For a wormhole to be traversable, the shape function $s(r)$ must satisfy the flaring-out condition~\cite{M_S_Morris}.
At the throat, located at $r=r_{0}$, this imposes the constraints $s(r_{0})=r_{0}$ and $s'(r_{0})<1$, where $s'(r)$ denotes the derivative of $s(r)$ with respect to the radial coordinate.

In addition, asymptotic flatness requires that $s(r)/r \to 0$ as $r \to \infty$, ensuring that the spacetime approaches Minkowski geometry at large distances. To prevent the occurrence of event horizons, the redshift function $\zeta(r)$ is assumed to remain finite throughout the manifold.

Within GR, satisfying these conditions requires the existence of exotic matter near the throat, namely matter that violates the standard energy conditions. Using Eq.~ we get the Ricci scalar as:
\begin{align}
\mathcal{R}=\frac {2\,rs  \zeta''  +2\,rs   \zeta'^{2}+3\,\zeta' s  -2\,{r}^{2}\zeta''  -2\,{r}^{2}\zeta'^{2}-4\,\zeta'  r+r \zeta's'  +2\,s' }{{r}^{2}}.
\end{align}

Substituting this into the field equations yields~\cite{Tayde_1}:
\begin{eqnarray}\nonumber
\nonumber\rho&=&\frac{1}{e^{\lambda}}\left[\bigg(\frac{\nu'}{r}-\frac{\nu'\lambda'}{4}+\frac{\nu''}{2}+\frac{\nu'^2}{4}\bigg)
f_{\mathcal{R}}(\mathcal{R},\mathbb{T})+
\bigg(\frac{\lambda'}{2}-\frac{2}{r}\bigg)f'_{\mathcal{R}}(\mathcal{R},\mathbb{T})-f''_{\mathcal{R}}(\mathcal{R},\mathbb{T})\right.\\\label{4}
&-&\left.\frac{f(\mathcal{R},\mathbb{T})}{2}e^{\lambda}\right], \nonumber\\
p_r &=&
\frac{1}{e^{\lambda}(1+f_{\mathbb{T}}(\mathcal{R},\mathbb{T}))}\left[\bigg(\frac{\lambda'}{r}-
\frac{\nu'\lambda'}{4}-\frac{\nu''}{2}-\frac{\nu'^2}{4}\bigg)
f_{\mathcal{R}}(\mathcal{R},\mathbb{T})+\bigg(\frac{\nu'}{2}+\frac{2}{r}\bigg)f'_{R}(\mathcal{R},\mathbb{T})\right.\\
&&\left.+\frac{(\mathcal{R},\mathbb{T})}{2}e^{\lambda}\right]\label{5}
-\frac{{\rho}f_{\mathbb{T}}(\mathcal{R},\mathbb{T})}{(1+f_{\mathbb{T}}(\mathcal{R},\mathbb{T}))},\\\nonumber
\nonumber p_t &=& \frac{1}{e^{\lambda}(1+f_{\mathbb{T}}(\mathcal{R},\mathbb{T}))}\left[\bigg(\frac{(\lambda'-\nu')r}{2}-e^{\lambda}+1\bigg)\frac{f_{R}(\mathcal{R},\mathbb{T})}{r^2}+
\bigg(\frac{\nu'-\lambda'}{2}+\frac{1}{r}\bigg)f'_{R}(\mathcal{R},\mathbb{T})\right.\\
&&\left.+f''_{\mathcal{R}}(\mathcal{R},\mathbb{T})+\frac{(\mathcal{R},\mathbb{T})}{2}e^{\lambda(r)}\right]
-\frac{{\rho}f_{\mathbb{T}}(\mathcal{R},\mathbb{T})}{(1+f_{\mathbb{T}}(\mathcal{R},\mathbb{T}))}\,, \label{6}
\end{eqnarray}
where $\nu$ and $\lambda$ are defined after Eq.~\eqref{1}.
It can be observed that the above equations appeared as much
complicated to find the explicit expressions of $\rho$, $p_r$ and
$p_t$, since $(\mathcal{R},\mathbb{T})$ has direct dependence on trace of
stress-energy tensor. Within this framework, a convenient choice takes the form
\[
f(\mathcal{R},\mathbb{T}) = \mathcal{R} + \beta\,\mathbb{T},
\]
where $\beta$ is a constant parameter with appropriate physical dimensions. Here, we set this choice for $f(\mathcal{R},\mathbb{T})$ and simplify the above
equations (\ref{4})-(\ref{6}) as follows\footnote{{ Throughout this work we use SI units, with $\kappa=8\pi G/c^4$, and $\rho$ denotes the mass density, so that $\rho c^2$, $p_r$, and $p_t$ all have the dimensions of energy density (pressure) and appear consistently in the field equations and the generalized TOV relation.
}}:
\begin{equation}
\label{swhfrt1}\frac{s^{\prime}}{r^{2}}={c^2 \kappa}\rho+2\beta\left(  2\rho
-\frac{p_{r}+2p_{t}}{3}\right)  ,
\end{equation}
\begin{equation}
\label{swhfrt2}\frac{1}{r}\left[  \frac{s}{r^{2}}+2\zeta^{\prime}\left(
\frac{s}{r}-1\right)  \right]  =-{ \kappa} p_{r}+2\beta\left[  \rho-\frac{2}%
{3}(2p_{r}+p_{t})\right]  ,
\end{equation}
\begin{equation}
\label{swhfrt3}\frac{1}{2r}\left[  \frac{1}{r}\left(  \zeta^{\prime
}s+s^{\prime}-\frac{s}{r}\right)  +2(\zeta^{\prime\prime}+\zeta^{\prime
2})s-\zeta^{\prime}(2-s^{\prime})\right]  -(\zeta^{\prime\prime}%
+\zeta^{\prime2})=-{ \kappa} p_{t}+2\beta\left(  \rho-\frac{p_{r}+5p_{t}}%
{3}\right)  .
\end{equation}

To close, we collect the energyinequality criteria commonly used to test the matter sector in wormhole spacetimes:
\begin{itemize}
\item \textbf{Weak energy condition (WEC):} $\rho \ge 0,\; \rho + p_r \ge 0,\; \rho + p_t \ge 0$.
\item \textbf{Null energy condition (NEC):} $\rho + p_r \ge 0,\; \rho + p_t \ge 0$.
\item \textbf{Dominant energy condition (DEC):} $\rho \ge 0,\; \rho \pm p_r \ge 0,\; \rho \pm p_t \ge 0$.
\item \textbf{Strong energy condition (SEC):} $\rho + p_r \ge 0,\; \rho + p_t \ge 0,\; \rho + p_r + 2p_t \ge 0$.
\end{itemize}

\section{Wormhole in \texorpdfstring{$f(\mathcal{R},\mathbb{T})=\mathcal{R}+\beta\,\mathbb{T}$}{f(\mathcal{R},\mathbb{T})=\mathcal{R}+\beta\,\mathbb{T}} theory}
\label{sec3}
Here, we focus on a widely studied $f(\mathcal{R},\mathbb{T})$ framework defined by the linear relation
\begin{equation}
\label{17}
f(\mathcal{R},\mathbb{T})= \mathcal{R}+\beta\,\mathbb{T},
\end{equation}
where, $\beta$ is a constant parameter with appropriate physical dimensions.
This linear form was originally introduced by Xu et al.~\cite{Y_Xu}, who demonstrated its capability to account for an exponentially expanding universe.
More recently, Loo et al.~\cite{Avik2023CQG} applied the same framework to Bianchi type-I cosmology, making use of observational constraints such as Hubble parameter measurements and Type Ia supernova data.

When Eq.~\eqref{17} is employed together with a constant redshift function, the general field equations \eqref{swhfrt1}--\eqref{swhfrt3} reduce to a simplified form\footnote{In this study, to adjust the units we put $\beta=\kappa\beta_1$.},

\begin{align}\label{sfe}
\rho &= \frac{s'(8\beta_1 + +3)}{3c^2\kappa(4\beta_1 + +1)(2\beta_1+1)\, r^2}, \quad
p_r= -\frac{12s\beta_1 - 4r\beta_1s' +3s}{3\kappa(4\beta_1 + +1)(2\beta_1+1)\, r^3},   \quad
p_t  = -\frac{12s\beta_1 - 4r\beta_1s'-3rs' +3s}{6\kappa(4\beta_1 + +1)(2\beta_1+1)\, r^3}.
\end{align}

The following analysis considers wormhole geometries under three different holographic dark energy density profiles.
\subsection{Constructing Wormhole Geometries}

We restrict our study to holographic dark energy densities, the resulting conditions on the shape function, and their graphical representation. 
In cosmology, the Ricci Holographic Dark Energy (RHDE) density profile, constructed using  the holographic principle~\cite{Hu:2006ar,Myung:2004ch}, is often employed to model dark energy.
This formulation is fundamental for investigating both the expansion history of the universe and its dynamical behavior.
The particular choice of density distribution affects the amount and localization of exotic matter near the wormhole throat, as well as its variation with radial distance.
In this study, we adopt the standard RHDE profile~\cite{Manoharan:2022qll} to construct the corresponding wormhole shape function, which is expressed as follows:
\begin{align}\label{Re}
\rho_{Re}=\frac{\alpha}{\alpha_1 c^2\kappa r^4}\ln\left(1+\pi \alpha_1 r^2\right), \quad \mbox{where $\alpha$ and $\alpha_1$ have dimension of inverse $length^2$}.
\end{align}
where $\alpha_1$ denotes the R\'enyi parameter.
The origin of this model lies in the R\'enyi entropy, an alternative entropy framework beyond Boltzmann--Gibbs statistics, capable of describing systems with nonlocal interactions and fractal characteristics~\cite{Elizalde:2020mfs}.

This profile plays a crucial role in smoothing singular behaviors and regulating the distribution of matter in wormhole geometries, thereby reducing, and in some cases even eliminating, the violation of the null energy condition (NEC), especially when considered alongside corrections from modified gravity.

A comparison of Eqs. \eqref{sfe} and \eqref{Re} yields the following solution of the shape  function $s(r)$

\begin{align}\label{21}
&s(r) = {\frac {48}{\sqrt {\pi\alpha_1^3}\left( 8\beta_1+3\right){r}{r_{th}}}}  \left[ \frac{1}2r \left( \frac{1}2+\beta_1\right) \sqrt {\pi\alpha_1} \left( \beta_1+\frac{1}4 \right) \alpha\ln  \left( 1+\pi\alpha_1{r_{th}}^{2} \right) +r_{th} \left( -\frac{1}2 \left( \frac{1}2+\beta_1\right) \sqrt {\pi\alpha_1} \left( \beta_1+\frac{1}4 \right) \right.\right.\nonumber\\
 &\left.\left.\alpha\ln  \left( 1+\pi\alpha_1{r}^{2} \right)+\alpha_1 \left\{  \left( \frac{1}2+\beta_1\right)  \left( \beta_1+\frac{1}4 \right) \pi\alpha \left\{ \arctan\Theta- \arctan \Theta_0\right\}  +\frac{r_{th}}6 \left( \frac{3}8+\beta_1\right) \sqrt {\pi\alpha_1} \right\}r \right) \right] ,
\end{align}
where  $\theta= \left( {\frac {\pi\alpha_1r}{\sqrt {\pi\alpha_1}}} \right)$ and $\Theta_0=\left( {\frac {\pi\alpha_1r_{th}}{\sqrt {\pi\alpha_1}}} \right)$.

Substituting Eq.~\eqref{21} into  Eqs.~\eqref{sfe} yields the energy density and pressures:
\begin{align}
&p_r(r)  = \frac {1}{\alpha_1 \left( 8\beta_1+3 \right) r_{th} \left( 1+2 \beta_1 \right) \kappa{r}^{4}}\left[\alpha\ln \left( 1+\pi \alpha_1{r}^{2} \right)\left\{22  \beta_1r_{th}+ 32  {\beta_1}^{2}r_{th}-24 {\beta_1}^{2}  r-18 \beta_1 r-3 r+3 r_{th}\right\}-3r\alpha_1{r_{th}}^{2}\right.\nonumber\\
&\left.-8r\alpha_1{r_{th} }^{2}\beta_1+\sqrt {\alpha_1}\arctan \left( \sqrt {\pi\alpha_1}r_{th}\right)\left\{6rr_{th}\sqrt {\pi }\alpha -6rr_{th}\sqrt {\pi }\alpha +36rr_{th}\sqrt {\pi }\alpha  \beta_1-48rr_{th}\sqrt {\pi }\alpha {\beta_1}^{2}-36rr_{th}\sqrt {\pi }\alpha\beta_1\right\}\right.\nonumber\\
&\left.+48rr_{th}\sqrt { \pi }\sqrt {\alpha_1}\alpha\arctan  { \beta_1}^{2}\right], \nonumber\\
&p_t(r) = -\frac {1}{2\alpha_1 \left( 8\beta_1+3 \right) r_{th} \left( 1+2\beta_1 \right) \kappa{r}^{4}}\left[\alpha\ln\left( 1+\pi \alpha_1{r}^{2} \right) \left\{28  \beta_1r_{th}+32  {\beta_1}^{2}r_{th}-24{\beta_1}^{2}   r- 18\beta_1 r-3 r+6 r_{th}\right\}-3r\alpha_1{r_{th}}^{2}\right.\nonumber\\
&\left.-8r\alpha_1{r_{th}}^{2}\beta_1+\sqrt {\alpha_1}\arctan \left( \sqrt {\pi\alpha_1}r_{th}\right)\left\{6rr_{th}\sqrt {\pi }\alpha-6rr_{th}\sqrt {\pi }\alpha+36rr_{th}\sqrt {\pi } \alpha\beta_1-48rr_{th}\sqrt {\pi }\alpha{\beta_1}^{2}-36rr_{th}\sqrt {\pi } \alpha\beta_1\right\}\right.\nonumber\\
&\left.+48r r_{th}\sqrt {\pi \alpha_1}\alpha\arctan \left( \sqrt {\pi\alpha_1}r_{th} \right) {\beta_1}^{2}\right].
\end{align}
where the density of this model, $\rho(r)$, is given by Eq.\eqref{Re}.
\begin{figure}
\centering
\subfigure[Asymptotic flatness]{\includegraphics[width=.25\textwidth]{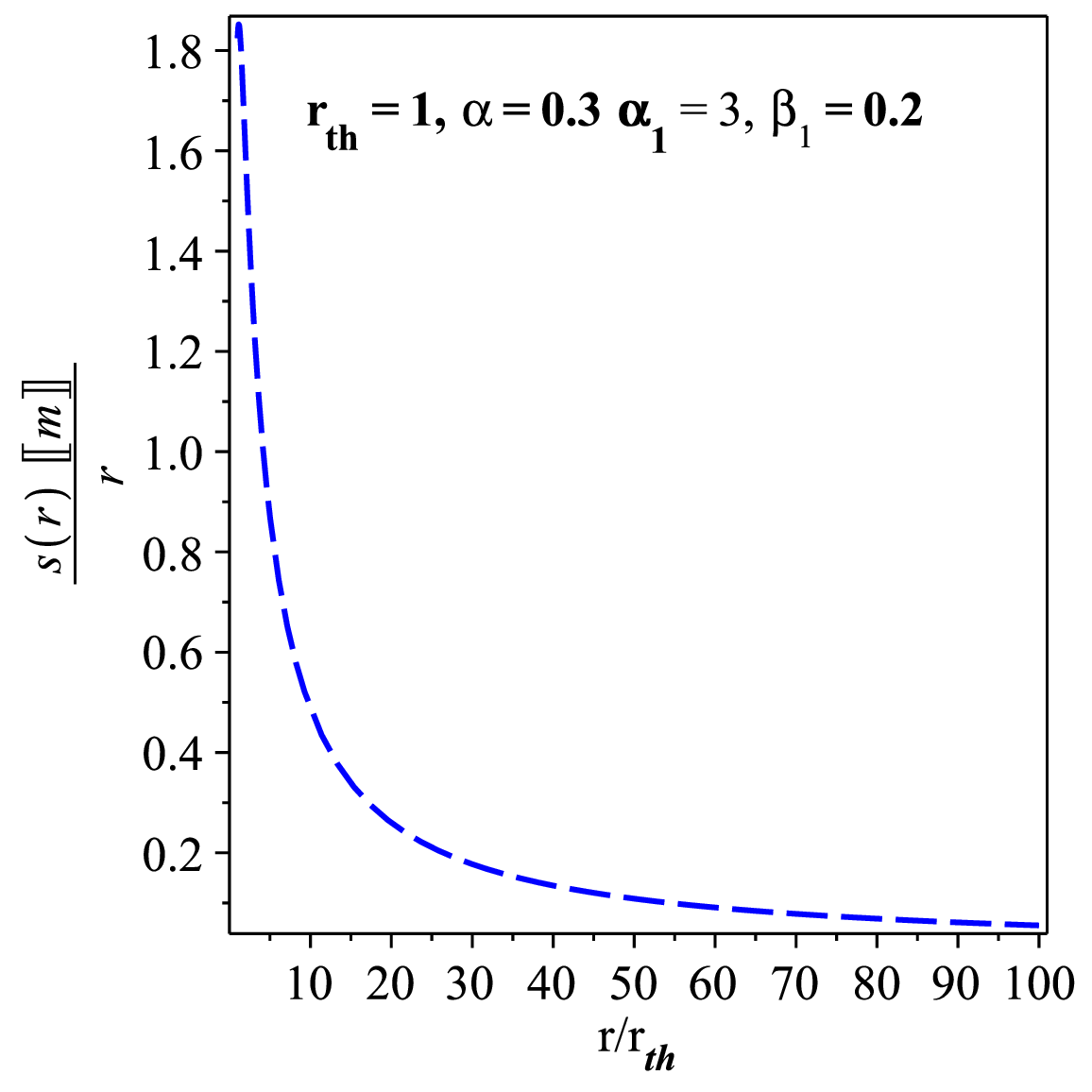}}\hspace{1cm}
\subfigure[Flaring-out at the throat]{\includegraphics[width=.25\textwidth]{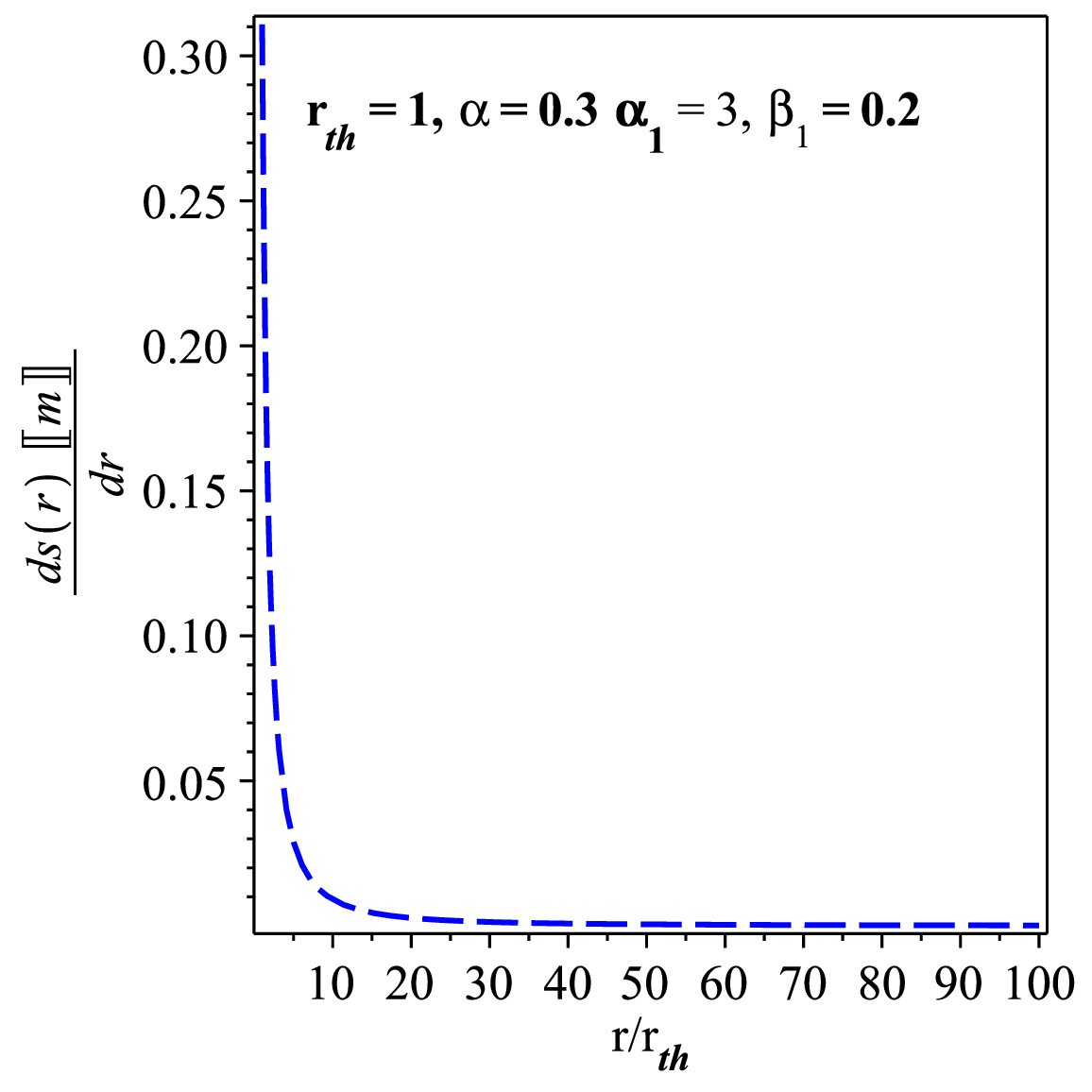}}
\subfigure[Energy density]{\includegraphics[width=.25\textwidth]{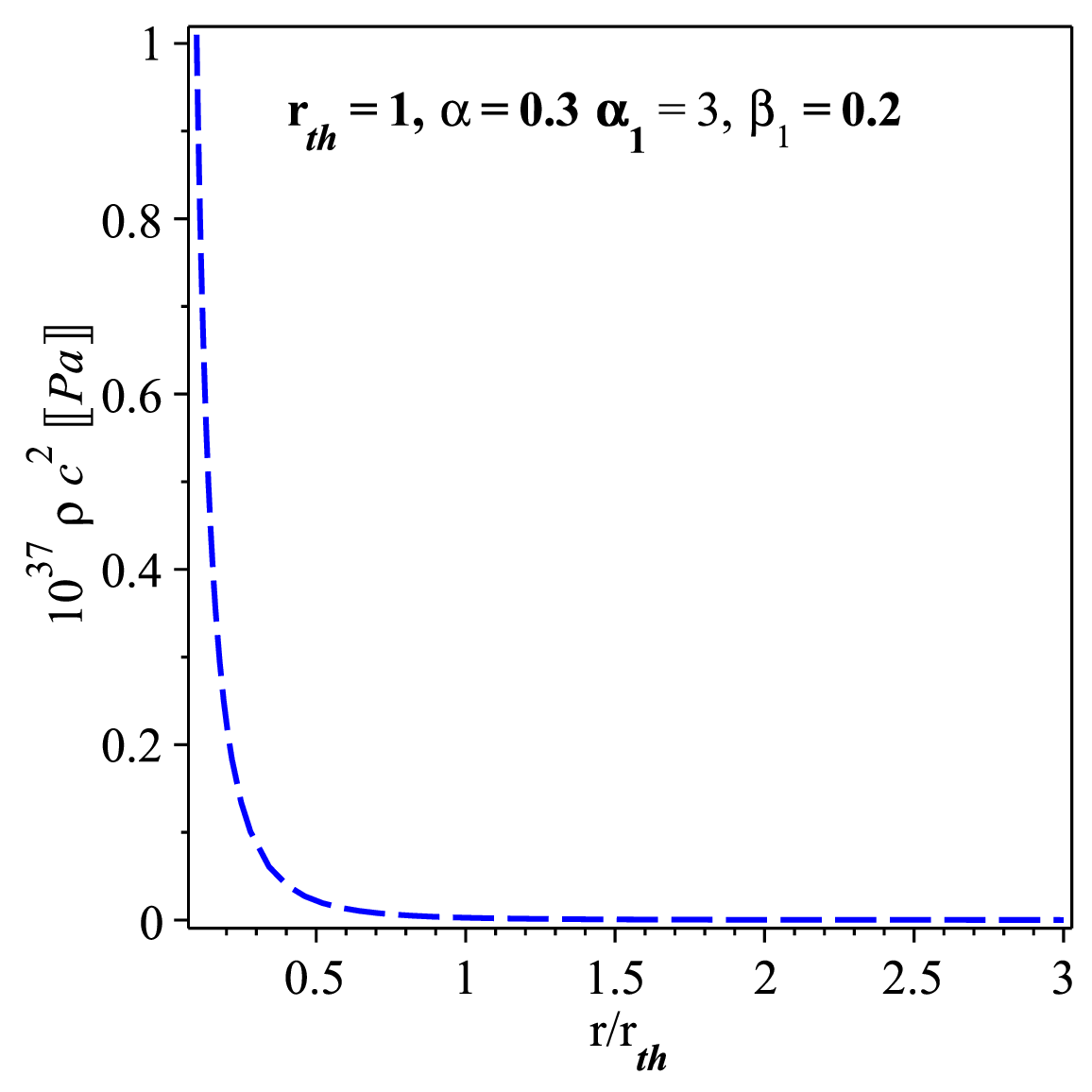}}\\
\subfigure[NEC for radial pressure]{\includegraphics[width=.25\textwidth]{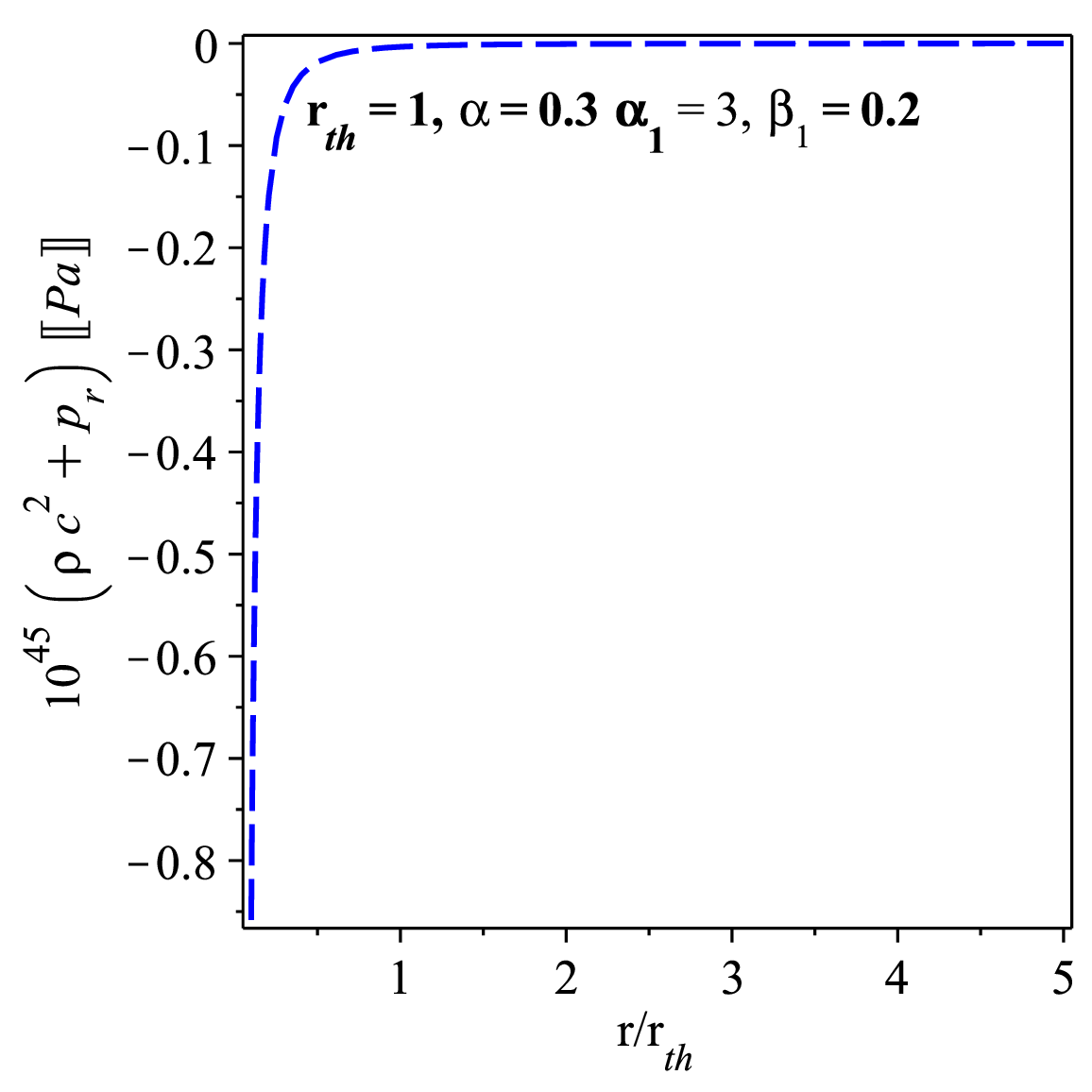}}\hspace{1cm}
\subfigure[NEC for tangential pressure]{\includegraphics[width=.25\textwidth]{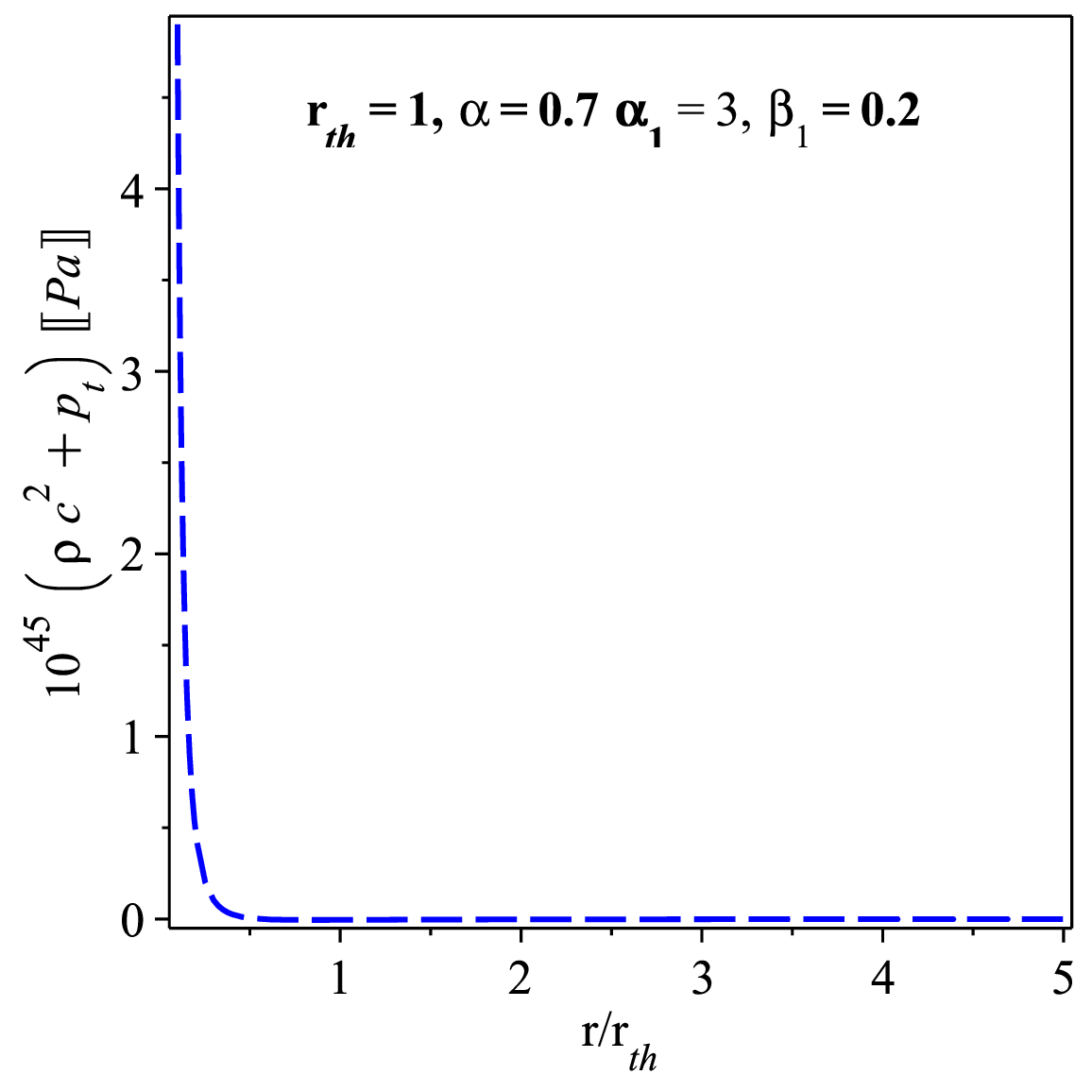}}
\caption{The density given by Eq.\eqref{Re}:The figure depicts how the shape function and associated energy conditions vary with the radial coordinate
 $r$, using representative parameter values listed in the plot.}
\label{Fig:Model1_beta}
\end{figure}

Figures~\ref{Fig:Model1_beta}(a)--\ref{Fig:Model1_beta}(e) display the behavior of the shape function and the associated energy conditions for representative values of the parameters.

To visualize the wormhole geometry, we consider a spatial slice at $t = \mathrm{const.}$ and $\theta = \pi/2$, reducing the metric to
\[
ds^2 = \frac{dr^2}{1 - s(r)/r} + r^2\,d\phi^2.
\]
Embedding this surface in three-dimensional Euclidean space with cylindrical coordinates $(r,\phi,z)$ yields
\[
ds^2 = \left[ 1 + \left( \frac{dz}{dr} \right)^2 \right] dr^2 + r^2\,d\phi^2,
\]
from which the embedding profile is given by
\begin{equation}\label{eq:surface_int}
z(r) = \pm \int_{r_{th}}^r \frac{d\zeta_1}{\sqrt{\zeta_1/s(\zeta_1) - 1}}.
\end{equation}

\begin{figure}
\centering
\subfigure[~2 D embedding]{\includegraphics[width=0.25\textwidth]{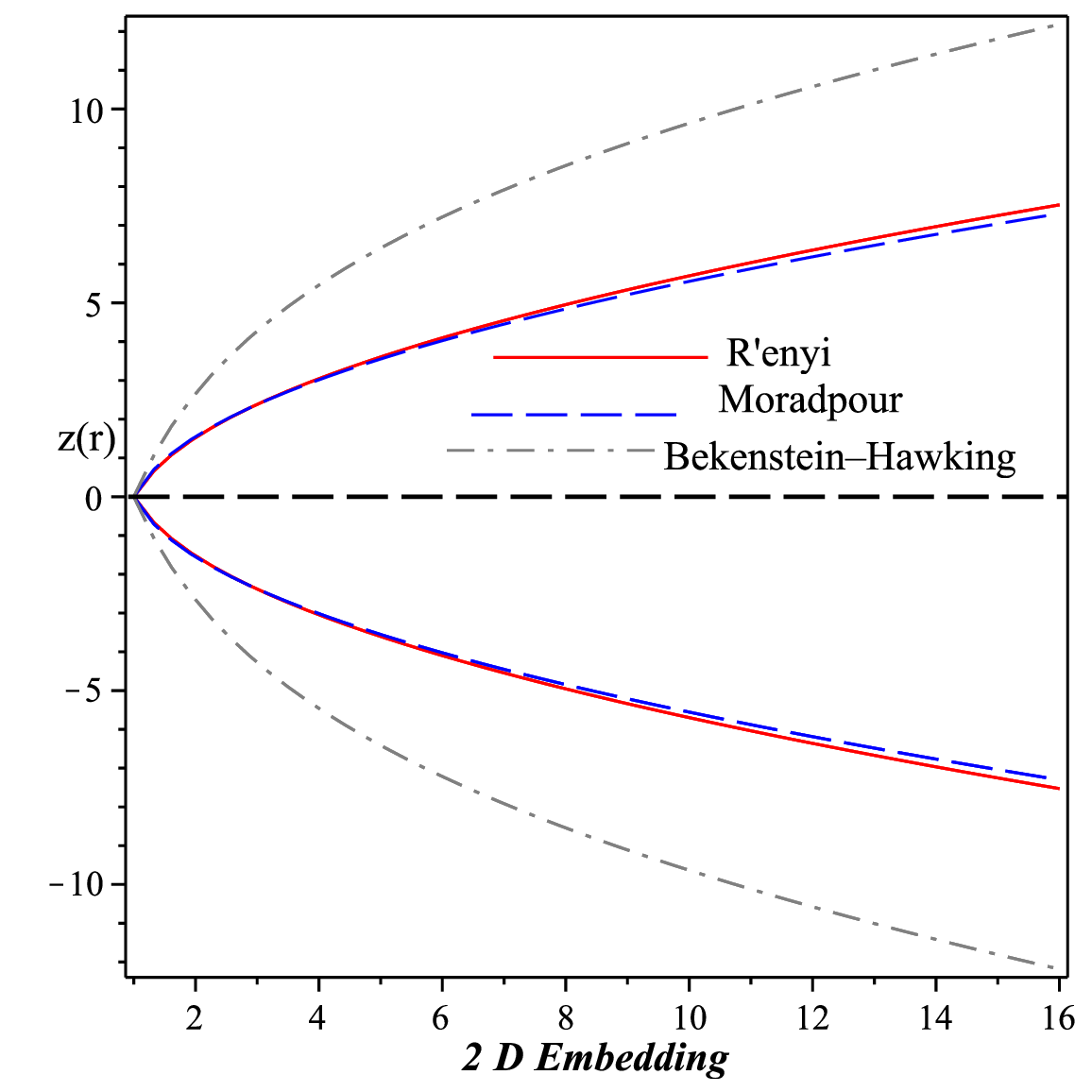}}\hspace{1cm}
\subfigure[~3 D embedding]{\includegraphics[width=0.25\textwidth]{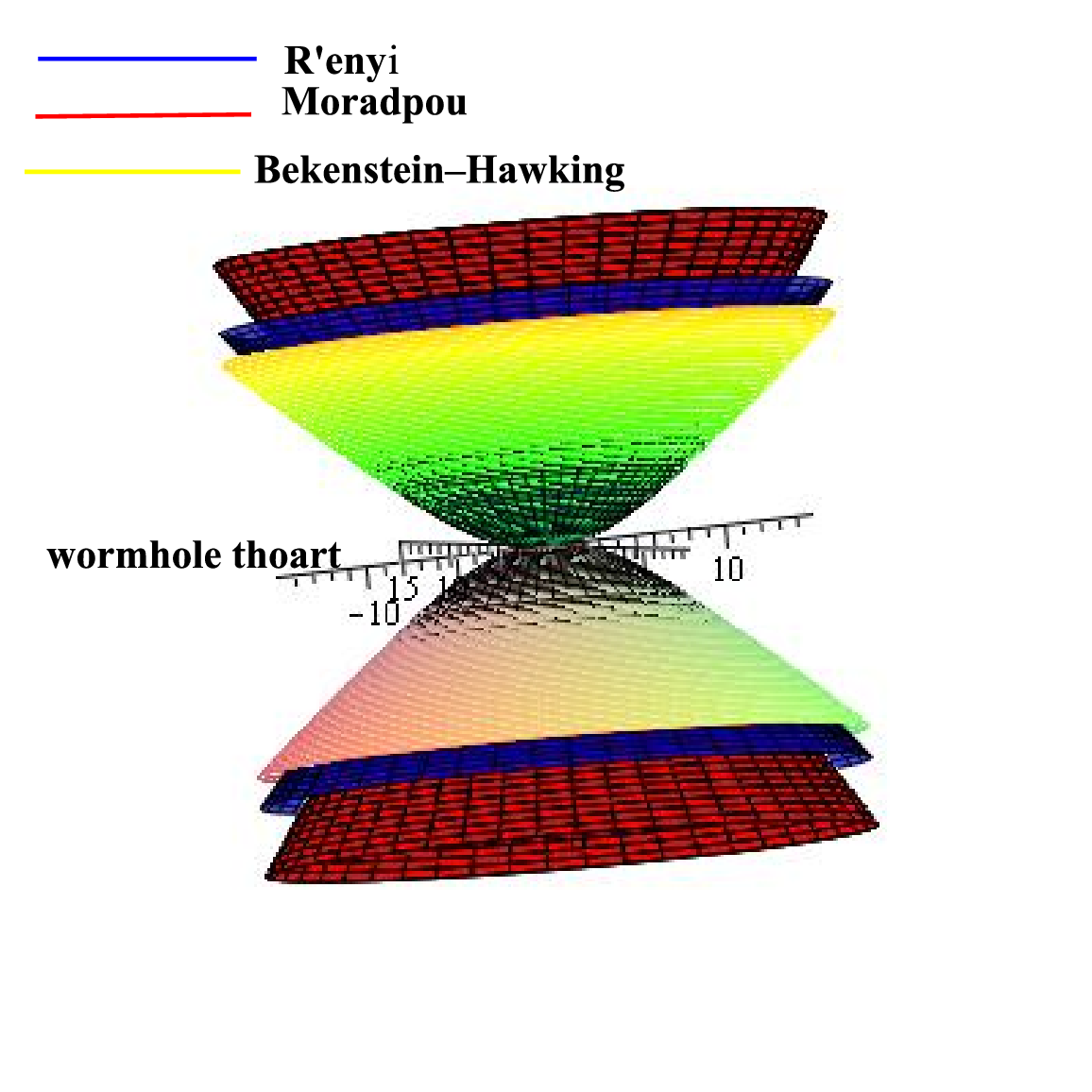}}
\caption{Two-dimensional embedding diagram $z(r)$ for the wormhole geometry.
The solid red curve corresponds to the R\'enyi holographic model
($\alpha=0.3$, $\alpha_1=3$, $\beta_1=0.2$),
the blue dashed curve to the Moradpour model with the same parameters,
and the gray dashdotted curve to the Bekenstein--Hawking case.
All curves are plotted for $r_{\textrm th}=1$.
}
\label{Fig:emedding1}
\end{figure}
\subsection{Second model: the Moradpour entropy--based holographic dark energy density}

In the Moradpour holographic dark energy model, the usual Bekenstein--Hawking entropy is replaced by the R\'enyi entropy, leading to a modified holographic framework. This generalized framework alters the effective energy conditions and, for suitable parameter regimes, permits the existence of wormhole configurations with enhanced stability. Within this approach, deviations from general relativity affect the wormhole dynamics and can give rise to solutions with enhanced stability. Within this framework, exotic matter--like behavior emerges dynamically and is sufficient to sustain traversable wormhole solutions. As a consequence, the null energy condition whose violation is required to support a traversable wormhole'is not satisfied in this model. Moreover, holographic corrections modify the dark energy equation of state, enabling the Moradpour HDE framework to support a broader range of wormhole configurations and stability regimes. The Moradpour energy density is given by
\begin{align}\label{Mad}
\rho_{Md}=\frac{\alpha}{4 c^2\kappa r^2\left(1+\pi \alpha_1 r^2\right)}, \quad \mbox{where $\alpha$ and $\alpha_1$ as usual  have dimension of inverse $length^2$}.
\end{align}

\newpage
\begin{table}[t]
\centering
\caption{Comparison of R\'enyi and Moradpour holographic dark energy models.}
\label{tab:HDE_final}
\footnotesize
\renewcommand{\arraystretch}{1.15}
\setlength{\tabcolsep}{3pt}

\begin{tabular}{|l|l|l|}
\hline
\textbf{Item} & \textbf{R\'enyi HDE} & \textbf{Moradpour HDE} \\
\hline
Energy density &
$\rho_{Re}(r)=\dfrac{\alpha}{4\alpha_1 c^2\kappa r^4}\ln\!\left(1+\pi\alpha_1 r^2\right)$ &
$\rho_{Md}(r)=\dfrac{\alpha}{4c^2\kappa r^2(1+\pi\alpha_1 r^2)}$ \\
\hline
Regular field equations &
\multicolumn{2}{c|}{$(4\beta_1+1)(2\beta_1+1)(8\beta_1+3)\neq 0$} \\
\hline
Throat regularity &
\multicolumn{2}{c|}{$s(r_{th})=r_{th}$ imposed by construction} \\
\hline
Flaring--out condition &
$0<\alpha<\alpha^{\max}_{Re}$, $\alpha^{\max}_{Re}=
\dfrac{4\alpha_1 r_{th}^2(8\beta_1+3)}
{3(4\beta_1+1)(2\beta_1+1)\ln(1+\pi\alpha_1 r_{th}^2)}$ &
$0<\alpha<\alpha^{\max}_{Md}$, $\alpha^{\max}_{Md}=
\dfrac{4(1+\pi\alpha_1 r_{th}^2)(8\beta_1+3)}
{3(4\beta_1+1)(2\beta_1+1)}$ \\
\hline
Positivity of $\rho$ &
Sufficient: $\alpha>0$, $\alpha_1>0$ &
Sufficient: $\alpha>0$, $1+\pi\alpha_1 r^2>0$ \\
\hline
Asymptotic flatness &
$\rho_{Re}=\mathcal{O}(\ln r/r^4)\Rightarrow s(r)/r\to0$ &
$\rho_{Md}=\mathcal{O}(r^{-4})\Rightarrow s(r)/r\to0$ \\
\hline
\end{tabular}
\end{table}

Moradpour et al. \cite{Moradpour:2017fmq}, inspired by Tsallis statistics, extended entropy to a non-additive form, yielding a modified HDE density expression that introduces corrections to the conventional area law. In highly curved spacetimes, such as wormholes, this framework provides greater flexibility in energy distribution by incorporating deviations from extensively. With thermodynamically motivated modifications to matter content, the Moradpour HDE density can mitigate energy condition violations in wormhole geometries and offer more physically consistent models.   By equating Eqs. \eqref{sfe}  and \eqref{Mad}, we get the shape function to have the form:
\begin{equation}\label{212}
s(r) = \frac { 1}{\sqrt {\pi\alpha_1}\left( 32\beta_1+12 \right)} \left[32 \left\{\frac{3\alpha}4 \left( \frac{1}2+\beta_1 \right)  \left( \beta_1+\frac{1}4 \right)\left\{ \arctan\Theta -\arctan \Theta_0\right\} + \sqrt {\pi\alpha_1}r_{th} \left( \frac{3}8+\beta_1 \right)  \right\}\right].
\end{equation}
{
For a constant redshift function, throat regularity requires $s(r_{th})=r_{th}$, which is imposed in the integration of the shape functions for both the R\'enyi and Moradpour profiles. The flaring--out condition $s'(r_{th})<1$ follows from the field equations,
\[
s'(r)=\frac{3c^2\kappa(4\beta_1+1)(2\beta_1+1)}{8\beta_1+3}\,r^2\rho(r),
\]
leading to explicit bounds on the model parameters once $\rho(r_{th})$ is substituted. Positivity of the density is ensured for $\alpha>0$ with $\alpha_1>0$ (R\'enyi) and $1+\pi\alpha_1 r^2>0$ for $r\ge r_{th}$ (Moradpour). In both cases the density decays faster than $r^{-2}$, implying $s(r)/r\to 0$ as $r\to\infty$ and hence asymptotic flatness. The allowed parameter ranges used in the figures satisfy all these conditions.
}

Substituting Eq.~\eqref{212} into  Eqs.~\eqref{sfe} yields the energy density and pressures:
\begin{align}
&p_r(r)  = -{\frac {6}{\sqrt {\pi\,\alpha_1} \left( 8 \beta_1+3 \right) \left( 1+\pi\alpha_1{r}^{2} \right)  \left( 2 \beta_1+1 \right) {r}{\kappa}}}\left[  \left( 1+\pi\alpha_1{r}^{2} \right) \alpha \left( \beta_1+\frac{1}4 \right)  \left( \frac{1}2+\beta_1 \right)\left\{ \arctan\Theta-  \arctan \Theta_0 \right\} \right.\nonumber\\
&\left.-\frac{1}3\,\sqrt {\pi\,\alpha_1} \left\{ \alpha r{\beta_1}^{2}+ \left(\frac{1}2\,r\alpha -4\,r_{th}\,\pi\,\alpha_1{r}^{2}-4 \,r_{th}\right) \beta_1-\frac{3}2\,r_{th}\, \left( 1+\pi\,\alpha_1{r}^{2} \right)  \right\}  \right]\,,\nonumber\\
&p_t(r) =  {\frac {3}{\sqrt {\pi\,\alpha_1} \left( 8 \beta_1+3 \right) \left( 1+\pi\alpha_1{r}^{2} \right)  \left( 2 \beta_1+1 \right) {r}{\kappa}}}  \left[  \left( 1+\pi\alpha_1{r}^{2} \right) \alpha \left( \beta_1+\frac{1}4 \right)  \left( \frac{1}2+\beta_1 \right)\left\{ \arctan\Theta-  \arctan \Theta_0\right\}\right. \nonumber\\
&\left.-\frac{1}3\sqrt {\pi\alpha_1} \left\{  \left( \frac{1}2+\beta_1 \right) \left( \beta_1+\frac{3}4 \right) r\alpha-4 \left( \beta_1+\frac{3}8 \right) r_{th} \left( 1+\pi\alpha_1{r}^{2} \right)  \right\}  \right].
\end{align}
where the density of this model, $\rho(r)$, is given by Eq.\eqref{Mad}.

\begin{figure}
\centering
\subfigure[Asymptotic flatness]{\includegraphics[width=.25\textwidth]{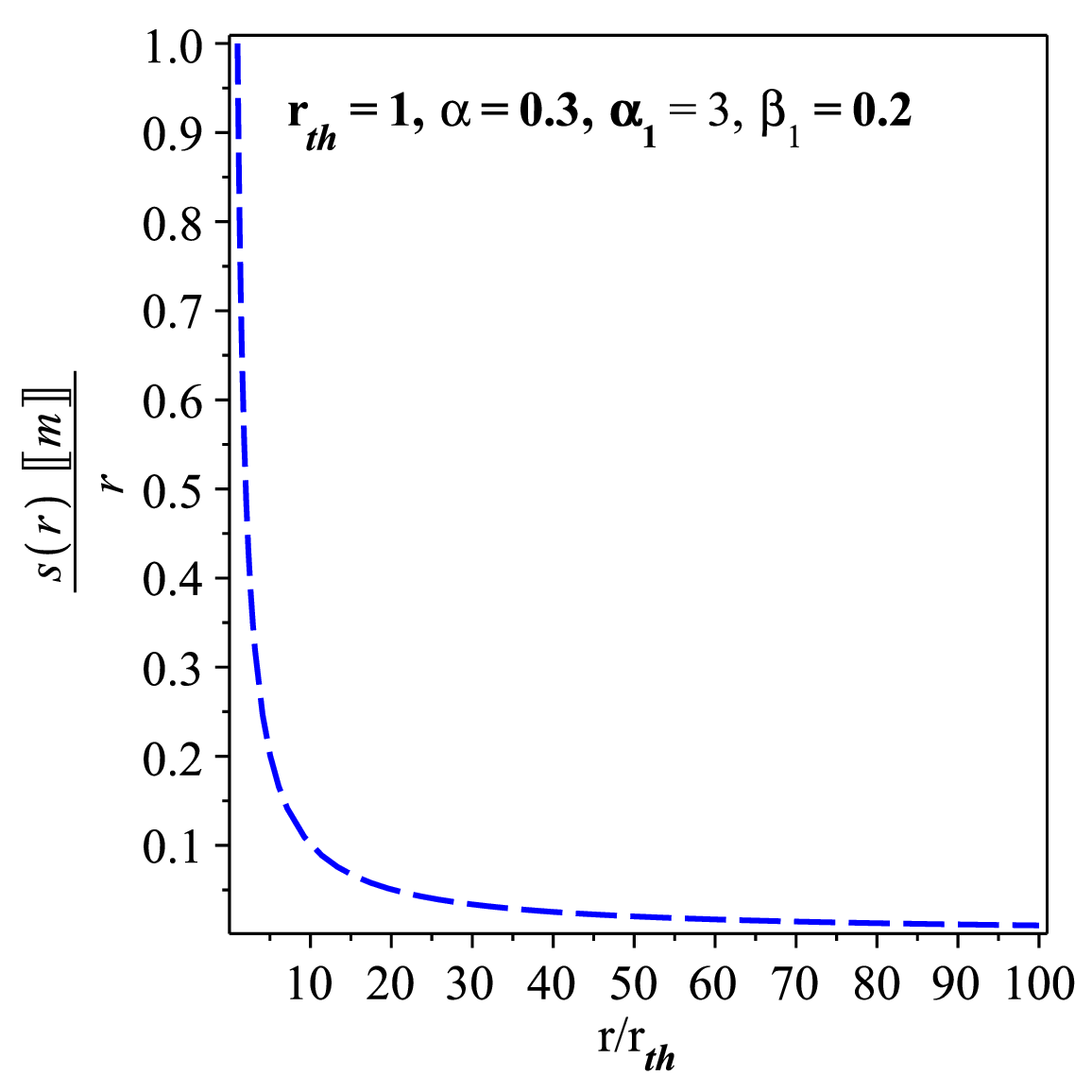}}\hspace{1cm}
\subfigure[Flaring-out at the throat]{\includegraphics[width=.25\textwidth]{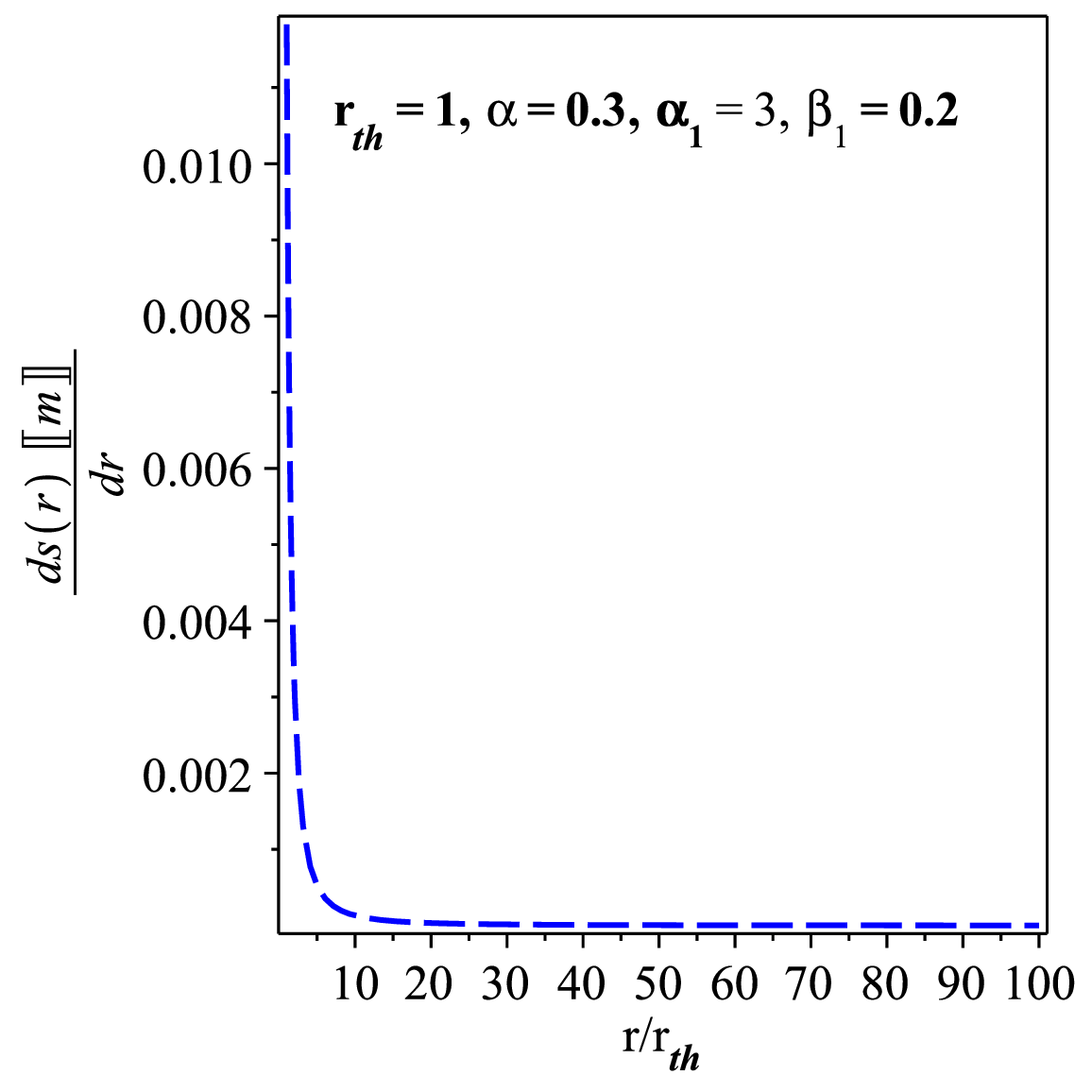}}
\subfigure[Energy density]{\includegraphics[width=.25\textwidth]{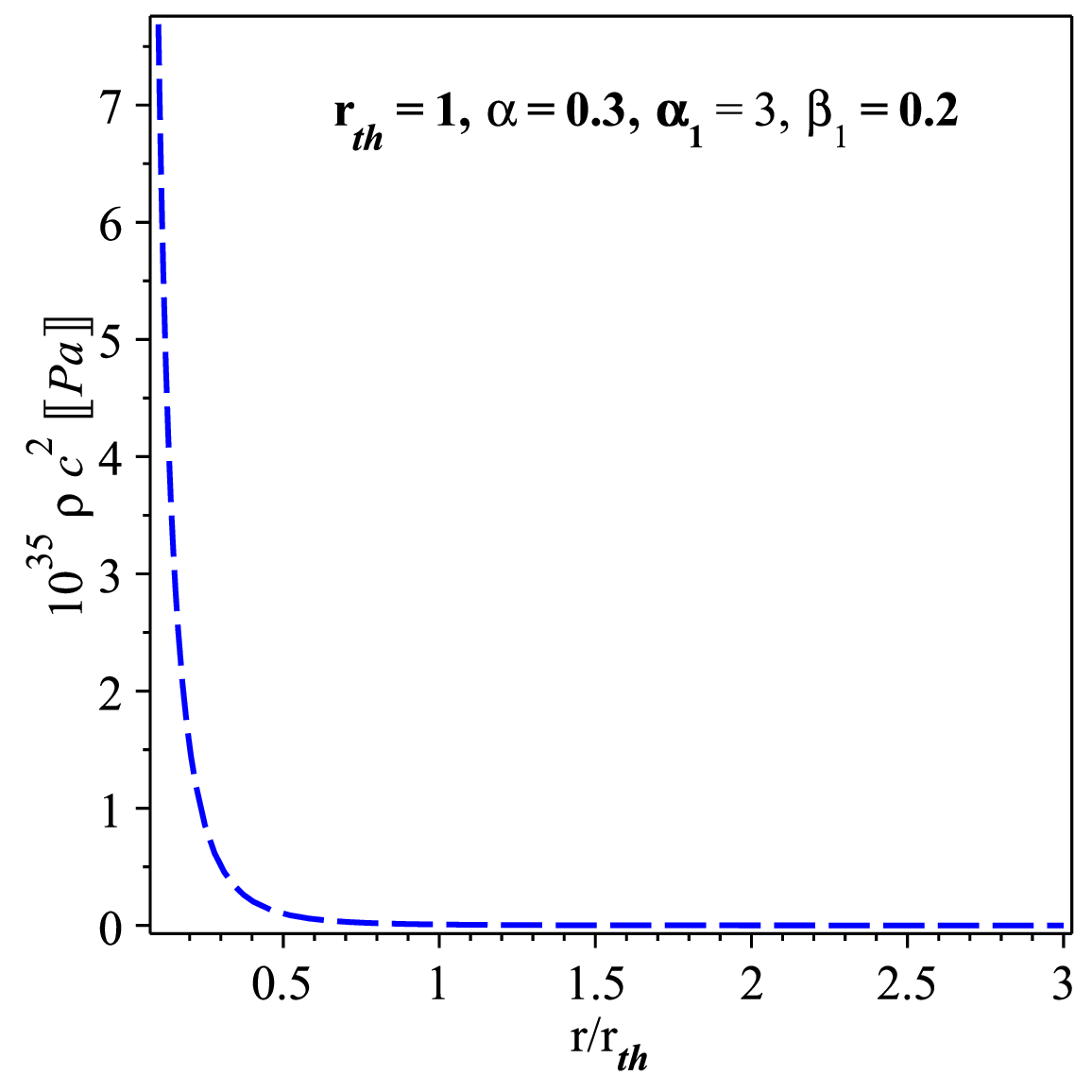}}\\
\subfigure[NEC for radial pressure]{\includegraphics[width=.25\textwidth]{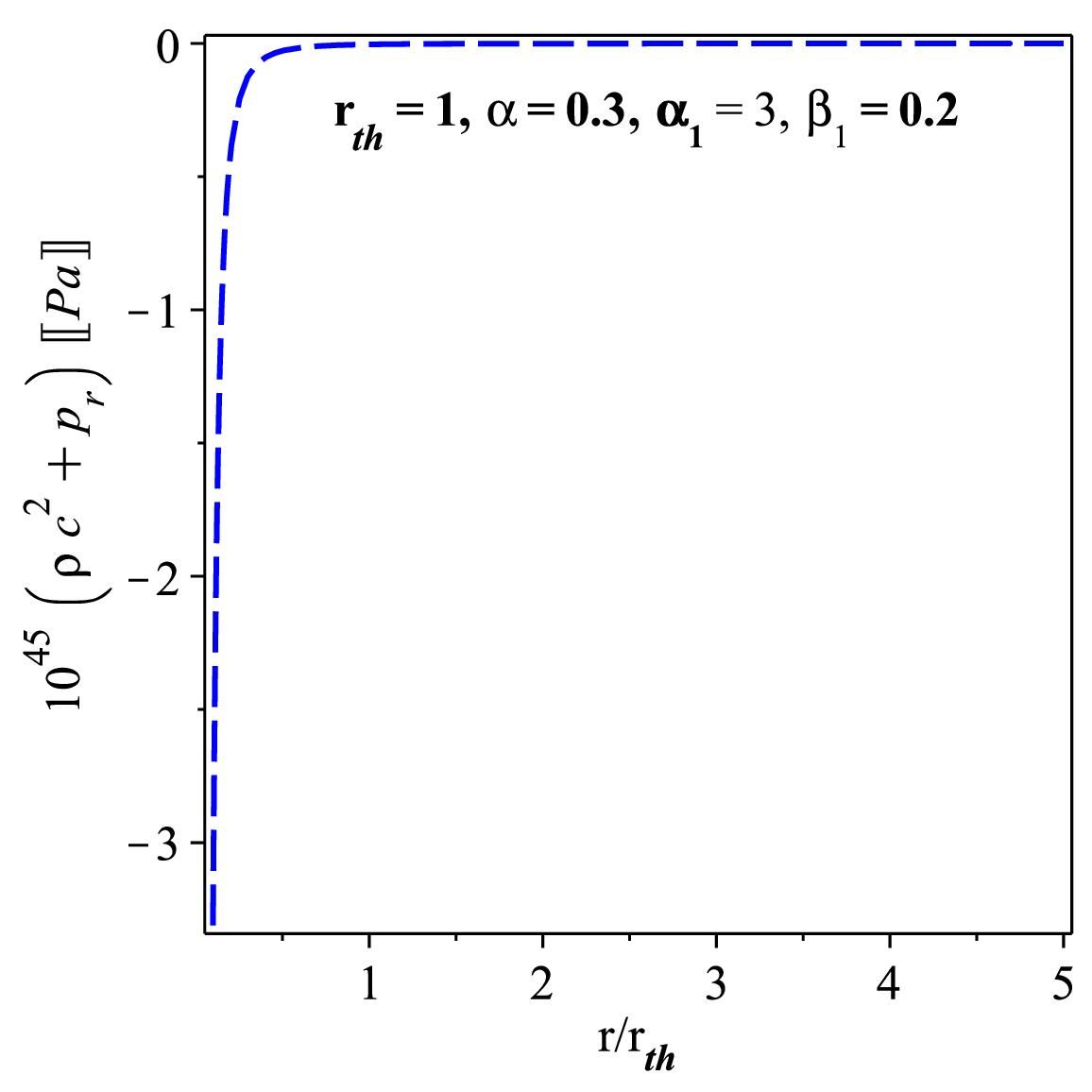}}\hspace{1cm}
\subfigure[NEC for tangential pressure]{\includegraphics[width=.25\textwidth]{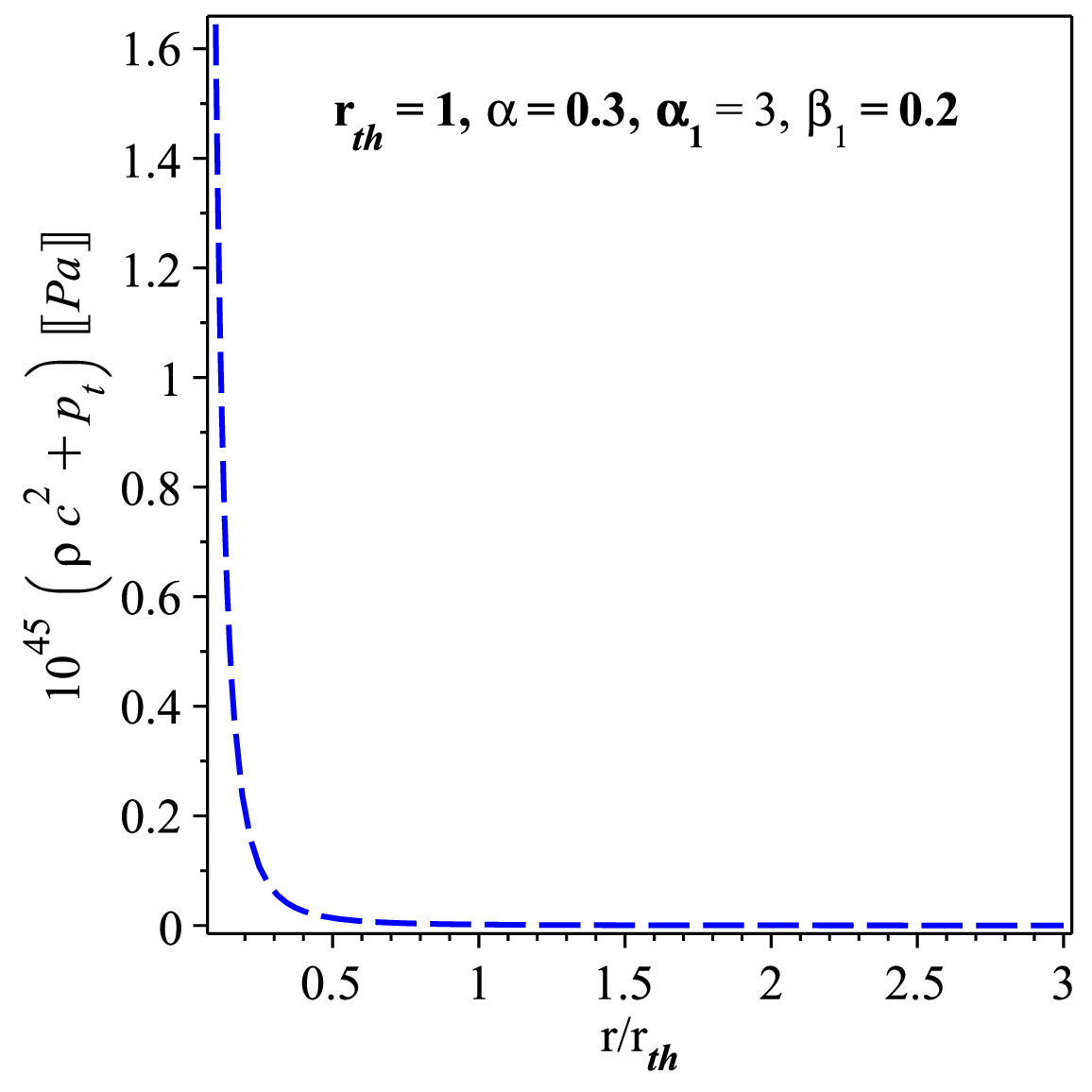}}
\caption{Density from Eq.~\eqref{Mad}: variation of the shape function and energy conditions with $r$ for chosen parameters.}
\label{Fig:Model2_beta}
\end{figure}

Figure~\ref{Fig:Model2_beta} illustrates the shape function~\eqref{212} and its compliance with the necessary geometric constraints, along with the corresponding energy condition profiles for selected model parameters.

\subsection{Third model: Holographic dark energy density constrained by the BekensteinHawking entropy principle}
Based on the holographic principle and the entropyarea connection, the BekensteinHawking HDE model links dark energy density to the BekensteinHawking entropy, which scales with the horizon's surface area. In the context of wormhole physics, traversable wormholes demand negative energy density. The BekensteinHawking HDE can provide this density, keeping the throat from collapsing. The exotic matterlike features of this model are thus vital to maintaining traversable wormholes. Here, the null energy condition, required to be violated for stability, is naturally broken. Moreover, when holographic modifications are incorporated into the dark energy equation of state, the potential for wormhole formation and persistence is broadened.

\begin{figure}
\centering
\subfigure[~{ Asymptotic behavior of $s(r)/r$}]{\includegraphics[width=.25\textwidth]{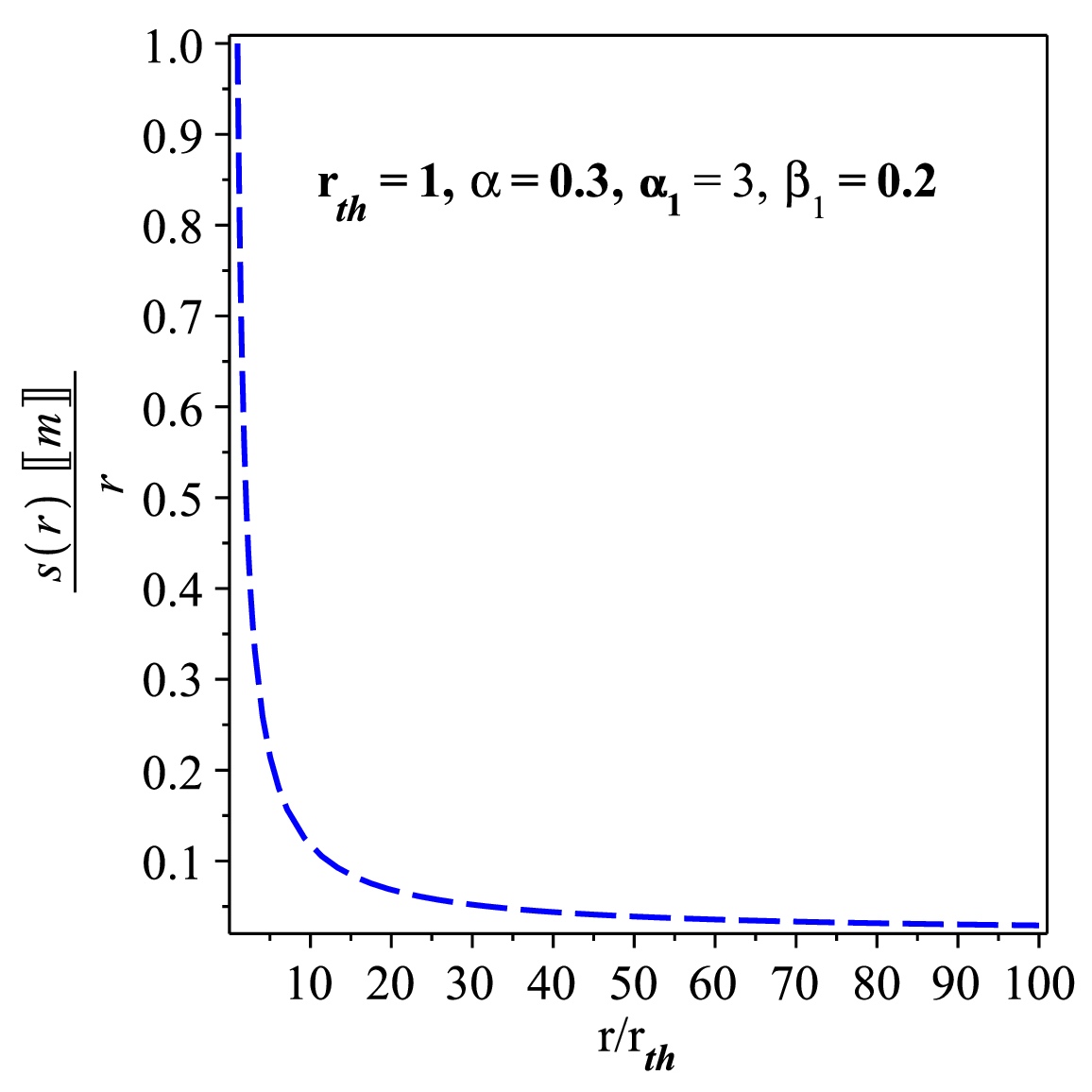}}\hspace{1cm}
\subfigure[Flaring-out at the throat]{\includegraphics[width=.25\textwidth]{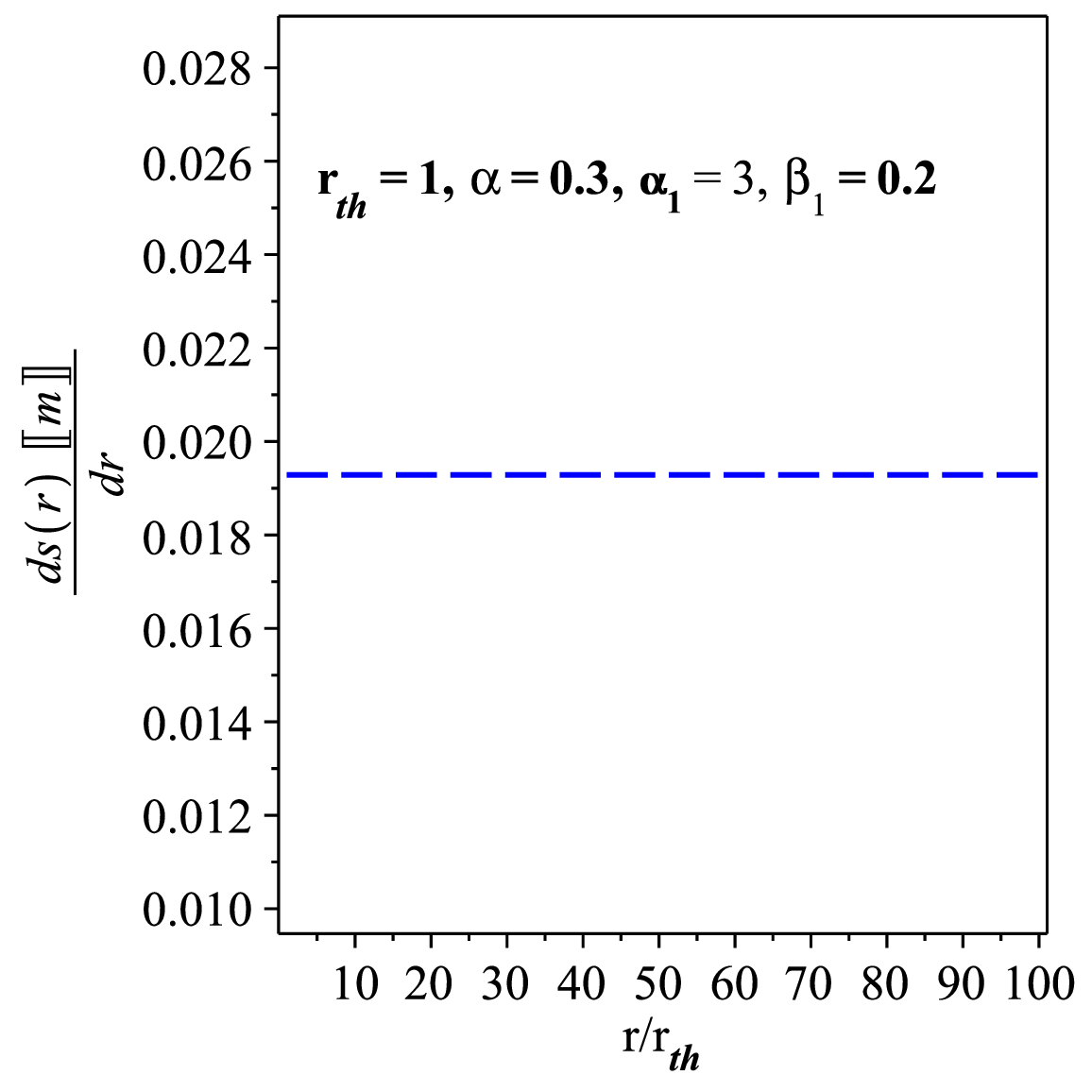}}
\subfigure[Energy density]{\includegraphics[width=.25\textwidth]{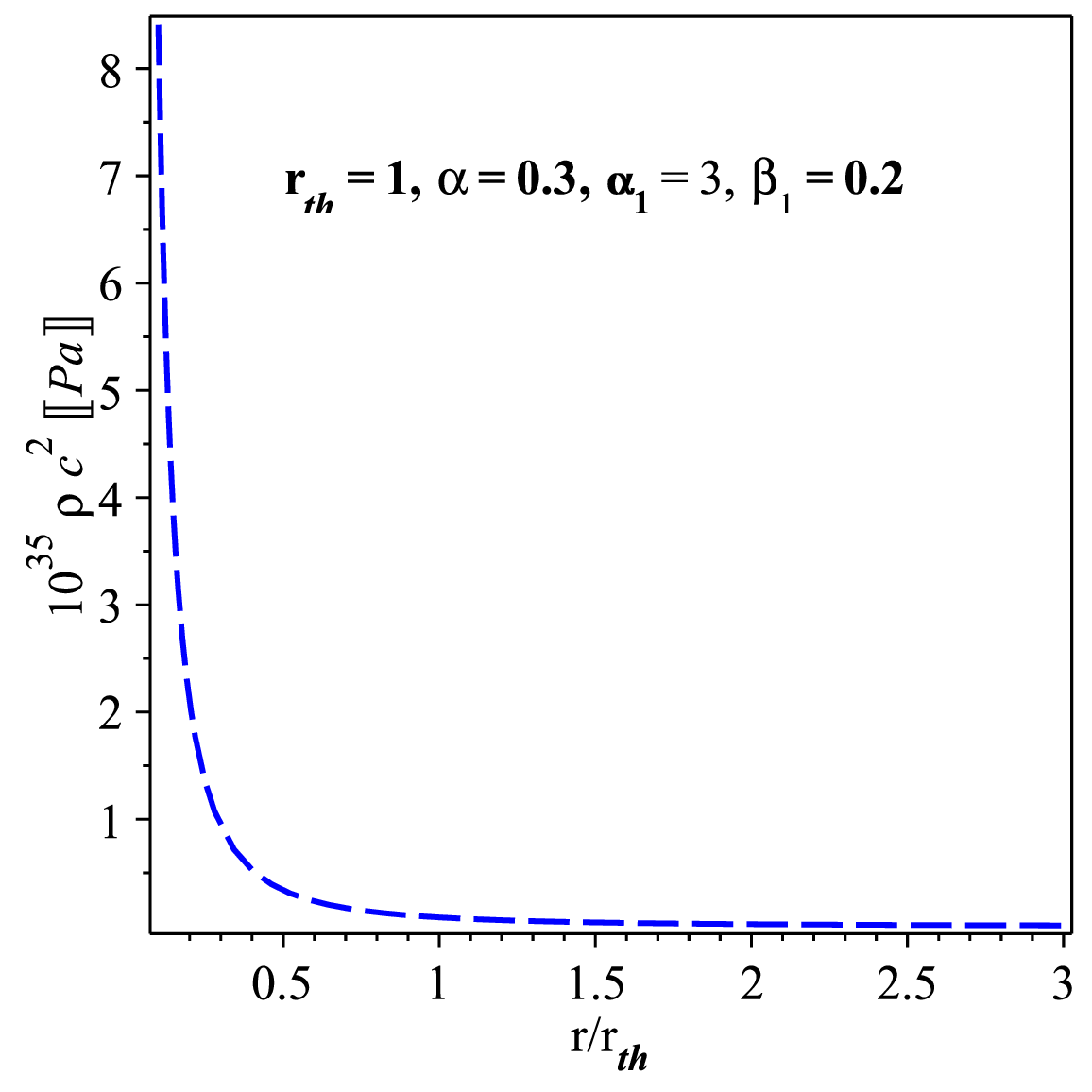}}\\
\subfigure[NEC for radial pressure]{\includegraphics[width=.25\textwidth]{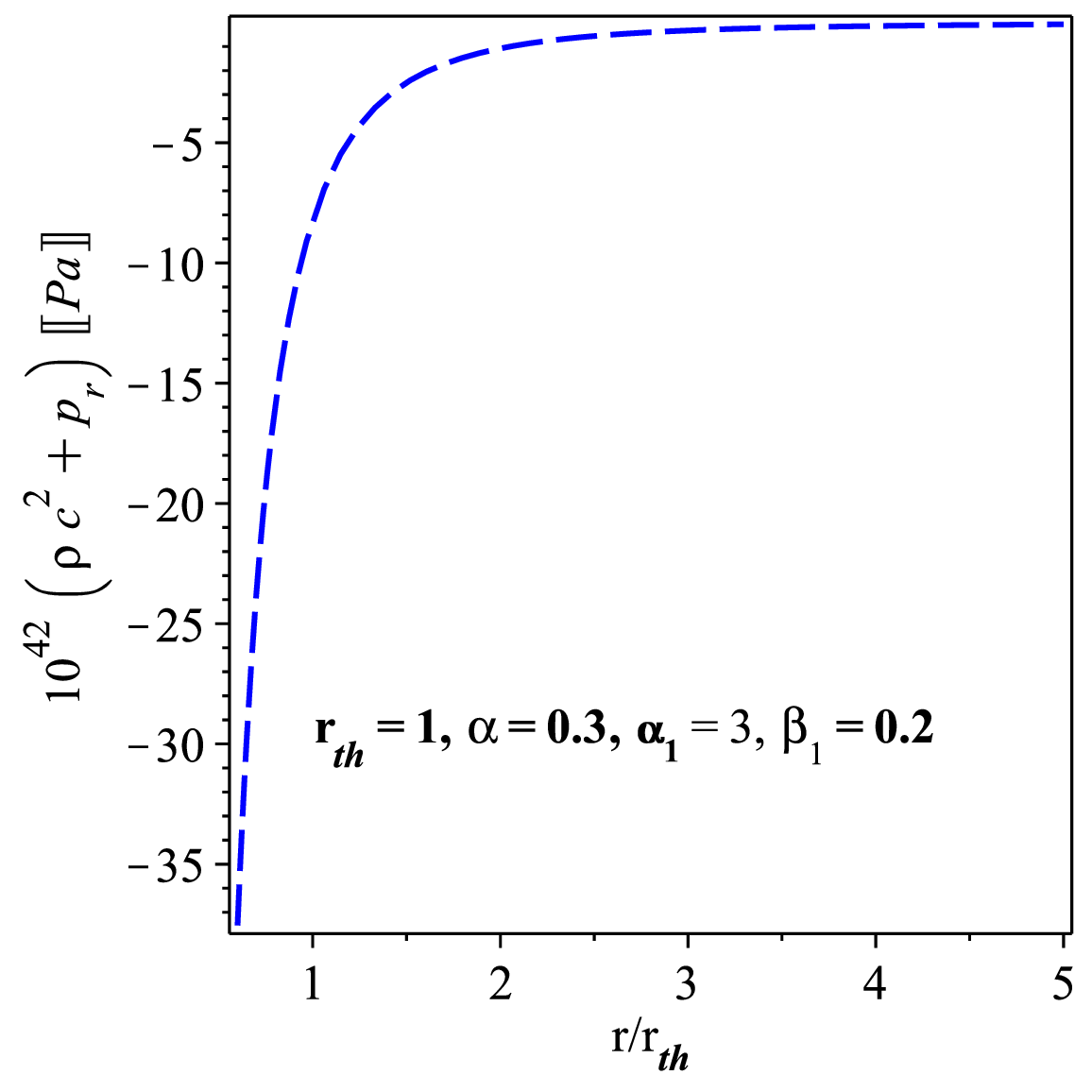}}\hspace{1cm}
\subfigure[NEC for tangential pressure]{\includegraphics[width=.25\textwidth]{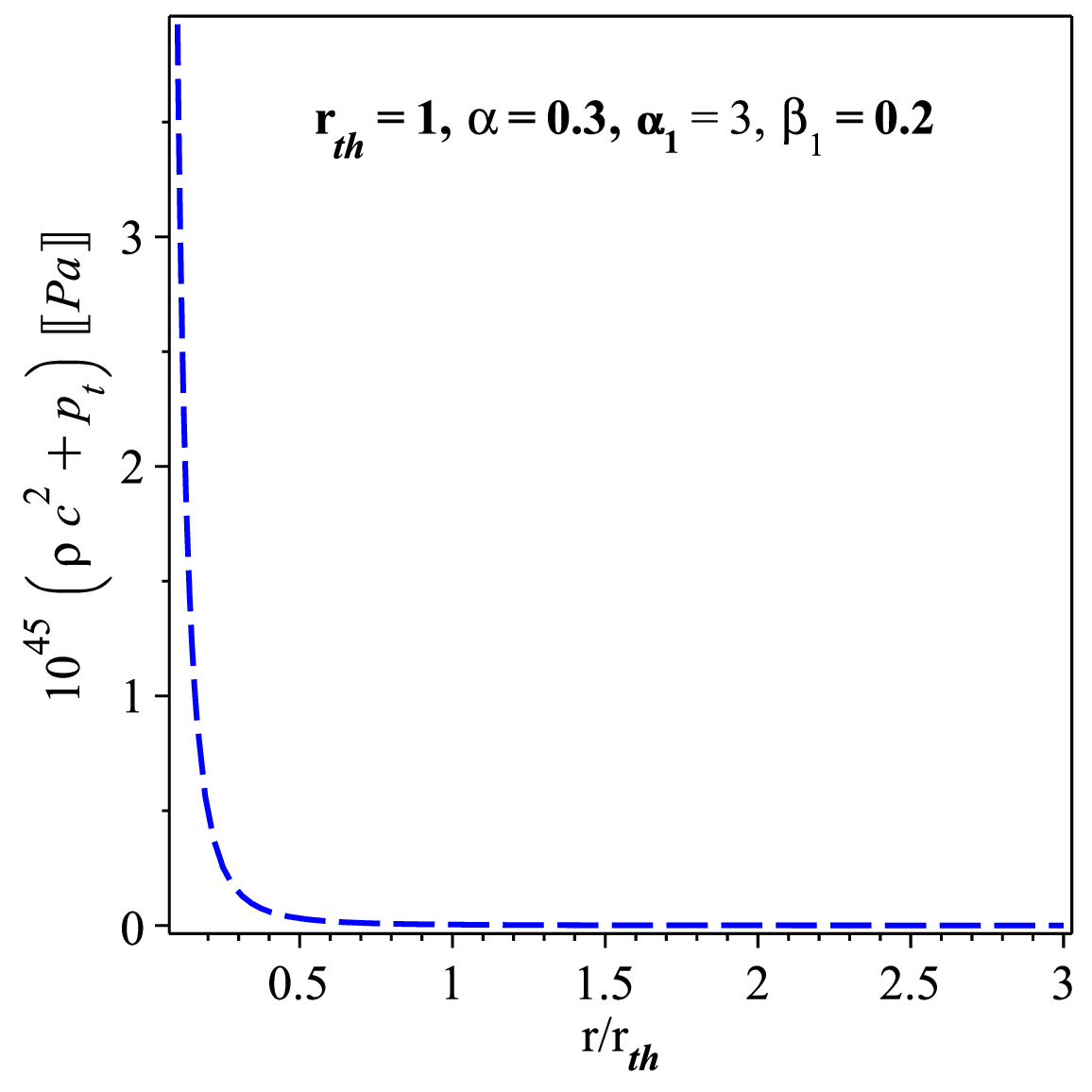}}
\caption{The density given by Eq.\eqref{BH}: Variation of the shape function and energy conditions for the non-asymptotically flat BekensteinHawking model, illustrating a local wormhole interior.}
\label{Fig:Model3_beta}
\end{figure}

According to the Bekenstein-Hawking entropy-area relation, the conventional HDE model emerges, where the energy density plays the role of exotic matter, satisfying the throat's flaringout condition in wormhole geometries. Because the energy density is tied directly to the system's size, it reduces dependence on hypothetical exotic fluids and instead supplies a physically inspired, holographically coherent basis for the development of traversable wormholes.

\begin{align}\label{BH}
\rho_{BH}=\frac{\alpha}{4 c^2\kappa r^2}, \quad \mbox{where $\alpha$ has a dimension of inverse $length^2$}.
\end{align}
Unlike the R\'enyi  and Moradpour profiles, the BekensteinHawking density leads generically to non-asymptotically flat wormhole geometries.
By equating Eqs. \eqref{sfe}  and \eqref{BH}, we get the shape function to have the form:
\begin{equation}\label{36}
s(r) = {\frac {24\,\alpha \left( r-r_{th} \right) {\beta_1}^{2}+ \left(
 \left( 18\,r-18\,r_{th} \right) \alpha+32\,r_{th} \right) \beta_1+
 \left( 3\,r-3\,r_{th} \right) \alpha+12r_{th}}{32\,\beta_1+12}}.
\end{equation}
{
 Equation (\ref{36}),  asymptotically behaves as
\begin{align}\label{ainf}
s(r)=A_{\infty}r+\beta+\mathcal{O}\!\left(\frac{1}{r}\right), \qquad
A_{\infty}=\lim_{r\to\infty}\frac{s(r)}{r}
=\frac{3\alpha(2\beta_1+1)(4\beta_1+1)}{4(8\beta_1+3)}.
\end{align}
Thus, asymptotic flatness ($s(r)/r\to 0$) requires $A_{\infty}=0$, which occurs only for
\[
\alpha=0, \qquad \beta_1=-\frac{1}{2}, \qquad \beta_1=-\frac{1}{4}.
\]
However, the values $\beta_1=-1/2$ and $\beta_1=-1/4$ make the field equations singular
(see denominators in Eq.~(\ref{sfe})), and must be excluded. Therefore, a non-trivial BH-supported wormhole
($\alpha\neq 0$) is always \emph{non-asymptotically flat}, with $s(r)/r\to A_{\infty}\neq 0$.
Our BH solution should thus be interpreted as a local wormhole interior that would need
to be matched to an exterior vacuum at a finite radius.
}
Substituting Eq.~\eqref{36} into  Eqs.~\eqref{sfe} yields the energy density and pressures:
\begin{align}\label{35}
   & p_r = -\frac {16\,\alpha r{\beta_1}^{2}-24\,\alpha r_{th}\,{\beta_1}^ {2}+14\,\alpha r\beta_1+32\,r_{th}\,\beta_1-18\,Ar_{th}\,\beta_1+3\,\alpha r+12\,r_{th}-3\,\alpha r_{th}}{4\kappa\,{r}^{3} \left( 1+2\,\beta_1 \right)  \left( 8\,\beta_1+3 \right) },\nonumber\\
    &p_t(r) =\frac {16\,\alpha r{\beta_1}^{2}-24\,\alpha r_{th}\,{\beta_1}^{
2}+8\,\alpha r\beta_1-18\,\alpha r_{th}\,\beta_1+32\,r_{th}\,\beta_1+12\,r_{th}-3\,\alpha r_{th}}{8\kappa\,{r}^{3} \left( 1+2\,\beta_1 \right)
 \left( 8\,\beta_1+3 \right) }
\end{align}
where the density of this model, $\rho(r)$, is given by Eq.\eqref{BH}.
Figure~\ref{Fig:Model3_beta} illustrates the shape function~\eqref{212} and its compliance with the necessary geometric constraints, along with the corresponding energy condition profiles for selected model parameters.

\subsection*{Summary of Results}

In this section, wormhole solutions have been explored for three forms of HDE: R\'enyi, Moradpour, and BekensteinHawking.
Figures~\ref{Fig:Model1_beta}, \ref{Fig:Model2_beta}, and \ref{Fig:Model3_beta} present the associated shape functions and their compliance with the geometric conditions for traversable wormholes.

The upper panels demonstrate that the shape function fulfills {asymptotic flatness} ($s(r)/r \to 0$ as $r \to \infty$) along with the {flaring-out condition} ($s'(r_0) < 1$ at the throat $r_0=1$).

The lower panels reveal the null energy condition (NEC) behavior. In all models, $\rho + p_r$ is negative near the throat, pointing to a violation of the radial NEC, while $\rho + p_t$ stays non-negative, hence satisfying the tangential NEC. The radial NEC violation arises naturally within the $f(\mathcal{R},\mathbb{T})$ scenario and is a characteristic aspect of traversable wormhole spacetimes.
\section{Gravitational Lensing Effects Produced by Wormholes}

This section focuses on the analysis of photon propagation in the neighborhood of a WH  throat, a phenomenon known as {gravitational lensing}. Lensing effects have been investigated for a wide range of WH space-times, including those with electric charge~\cite{Godani2021},
massless wormhole solutions~\cite{Nandi2006}, Janis--Newman--Winnicour geometries~\cite{Dey2008}, and the Ellis class~\cite{Tsukamoto2016}.
Beyond wormholes, gravitational lensing has served as a valuable method for examining the causal and geometric features of black holes
and naked singularities~\cite{Virbhadra2000,Virbhadra2002,Bozza2010,Bozza2002}.

{
For a constant redshift function, it is convenient to use the proper radial coordinate $\ell$, for which the metric is regular at the throat. The null geodesic equation yields an effective potential $V_{\textrm eff}=L^2/r(\ell)^2$, whose extremum determines circular photon orbits. Since the throat corresponds to a minimum of the areal radius, $r'(0)=0$, the condition $dV_{\textrm eff}/d\ell=0$ is automatically satisfied there. Moreover, the flaring--out condition $s'(r_{th})<1$ implies $r''(0)>0$, and hence $d^2V_{\textrm eff}/d\ell^2|_{\ell=0}<0$, showing that the throat acts as an \emph{unstable} photon sphere. The apparent nonvanishing of $dV_{\textrm eff}/dr$ in Schwarzschild-like coordinates is therefore a coordinate artifact \cite{Claudel:2000yi}.
}

Our analysis follows a conventional method frequently applied in the literature~\cite{Godani2021,Shaikh2019}. A comprehensive derivation of the relevant relations can be found in~\cite{Weinberg1972}. We begin with the Lagrangian for a null trajectory $\gamma(s) = \left(t(s), r(s), \theta(s), \phi(s)\right)$ in the spacetime metric given by Eq.~(\ref{1}), where $\dot{\phantom{x}} \equiv \partial \gamma / \partial s$:
\begin{equation}
\mathcal{L} = -\frac{e^{-2\zeta} \dot{t}^2}{2} + \frac{\dot{r}^2}{2\left(1 - \frac{s}{r}\right)} + \frac{r^2 \dot{\theta}^2}{2} + \frac{r^2 \sin^2 \theta \, \dot{\phi}^2}{2} = 0.
\label{eq:lagrangian}
\end{equation}
{ For null geodesics, the conserved quantities associated with the metric~(\ref{1}) are
\begin{equation}
E=\Upsilon(r)\,\dot t,\quad
L=r^{2}\dot\varphi,\quad
u=\frac{L}{E}, \quad \mbox{and the corresponding effective potential is} \quad
V_{\textrm eff}(r)=\Upsilon(r)\frac{L^{2}}{r^{2}} .
\end{equation}
The radial equation therefore becomes
\begin{equation}
\dot r^{2}
=
\Bigl(1-\frac{s(r)}{r}\Bigr)
\left(
\frac{E^{2}}{\Upsilon(r)}-\frac{L^{2}}{r^{2}}
\right).
\label{eq:radial}
\end{equation}

Because the redshift function is constant, $\Upsilon(r)=e^{2\zeta_{0}}=const.$, the usual condition
$V_{\textrm eff}'(r)=0$ does not yield a photon sphere in $r$--coordinates.
Introducing instead the proper radial coordinate,
\begin{equation}
\ell(r)=\pm\int_{r_{\textrm th}}^{r}
\frac{dr'}{\sqrt{1-s(r')/r'}},
\end{equation}
one finds that the wormhole throat ($\ell=0$) corresponds to an unstable circular null
orbit for all three HDE models.
Hence the critical impact parameter is
\begin{equation}
u_{0}
  =\sqrt{\frac{C(r_{\textrm th})}{\Upsilon(r_{\textrm th})}}
  =r_{\textrm th} e^{-\zeta_{\textrm th}} .
\end{equation}

Following Bozza's strong-deflection method~\cite{Bozza:2002zj,Bozza:2002af},
the deflection angle for $u\to u_{0}^{+}$ takes the universal form
\begin{equation}
\alpha(u)
=
-\bar a\,
\ln\!\left(\frac{u}{u_{0}}-1\right)
+\bar b, \quad \mbox{with} \quad
\bar a=\frac{1}{\sqrt{\kappa}},\qquad
\kappa=\tfrac12\!\left[\ln\!\left(\frac{C}{\Upsilon}\right)\right]''_{\ell=0},
\end{equation}
and $\bar b$ obtained from the regular part of the integral.
Since $\Upsilon={\textrm const.}$, the expansion
\begin{equation}
r(\ell)=r_{\textrm th}+\tfrac12\,r_{2}\,\ell^{2}+\cdots \quad \mbox{gives} \quad
\bar a=\sqrt{\frac{r_{\textrm th}}{r_{2}}},
\end{equation}
where $r_{2}$ depends on the second derivative of the shape function  evaluated at the throat.

The observable strong-field lensing quantities then read
\begin{equation}
\theta_{\infty}=\frac{u_{0}}{D_{\textrm OL}},\qquad
s\simeq\theta_{\infty}\exp\!\left(\frac{\bar b-2\pi}{\bar a}\right),\qquad
r_{\textrm mag}=\frac{5\pi}{\bar a\ln 10},
\end{equation}
where $D_{\textrm OL}$ is the angular-diameter distance between the observer \(O\) and the lens \(L\), i.e. the wormhole throat and $\theta_{\infty}$ is the asymptotic angular position of the relativistic images seen by the observer.
These expressions apply to all three HDE wormhole models; their numerical values differ only
through the model-dependent second derivative of $s(r)$ at the throat. This is a special case of the general characterization of photon surfaces in static spherically symmetric spacetimes \cite{Claudel:2000yi}.}
\newpage
{ \begin{table}[t]
\centering
\caption{ANEC volume-integral quantifier for radial null geodesics \cite{Visser:2003yf}.
We compute
$\mathcal{I}_{\textrm ANEC}=\int_{r_{\textrm th}}^{r_{\max}}(\rho+p_r)\,4\pi r^2\,dr$
for each holographic profile using the same parameter set as in the figures:
$r_{\textrm th}=1$, $\alpha=0.3$, $\alpha_1=3$, and $\beta_1=0.2$. The GR benchmark is obtained by taking $\beta_1\to 0$.
A less negative $\mathcal{I}_{\textrm ANEC}$ indicates a smaller amount of exotic matter.}
\label{tab:anec_viq}
\begin{tabular}{lccccc}
\hline
Model & $r_{\max}$ &
$\mathcal{I}_{\textrm ANEC}$ $(\beta_1=0.2)$ &
$\mathcal{I}_{\textrm ANEC}^{\textrm GR}$ $(\beta_1=0)$ &
$\Delta\mathcal{I}$ &
Reduction (\%) \\
\hline
R\'enyi HDE & 50 & $-37.59$ & $-51.27$ & $+13.68$ & $26.68$ \\
Moradpour HDE & 50 & $-35.18$ & $-49.22$ & $+14.04$ & $28.52$ \\
Bekenstein--Hawking HDE & 50 & $-30.79$ & $-45.47$ & $+14.69$ & $32.30$ \\
\hline
\end{tabular}

\vspace{1mm}
\footnotesize{
$\Delta\mathcal{I}=\mathcal{I}_{\textrm ANEC}-\mathcal{I}_{\textrm ANEC}^{\textrm GR}$.
Reduction (\%) $=100\left(1-\frac{|\mathcal{I}_{\textrm ANEC}|}{|\mathcal{I}_{\textrm ANEC}^{\textrm GR}|}\right)$.
}
\end{table}
 We choose a finite cutoff $r_{\max}=50$ because the ANEC volume integral is evaluated numerically and, for the asymptotically flat R\'enyi and Moradpour profiles, the integrand $(\rho+p_r)\,4\pi r^2$ decays rapidly at large $r$ (the densities fall off at least as $\rho\sim r^{-4}$, up to a $\ln r$ factor in the R\'enyi case). Consequently, $\mathcal{I}_{\textrm ANEC}$ effectively converges well before $r=50$, and extending the upper limit further produces only a negligible change in the quoted values. Using the same fixed cutoff for all models also enables a consistent like-for-like comparison of the exotic-matter content. For the Bekenstein--Hawking case, which is generically non-asymptotically flat in our setup, the solution is interpreted as a local wormhole interior; hence a finite $r_{\max}$ is in any case required.}

\section{Tolman--Oppenheimer--Volkoff constraint}

In this section, we investigate the stability properties of the WH solutions obtained.
An established approach for analyzing stability in self-gravitating configurations is provided by the hydrostatic equilibrium condition,
formulated through the Tolman--Oppenheimer--Volkoff (TOV) equation~\cite{Oppenheimer:1939ne,Gorini:2008zj,Kuhfittig:2020fue}.
Compliance with the TOV equation ensures that the wormhole maintains a state of dynamic equilibrium.

From the field equations~\eqref{GR+T effects EOM}, one observes that the matter sector is generally not conserved,
that is, $\nabla_\mu \mathbb{T}^{\mu}{}_{\nu} \neq 0$, owing to the explicit dependence on $\mathbb{T}$.
In contrast, the effective stress-energy tensor fulfills $\nabla_\mu \mathbb{T}^{(eff)}_{\mu \nu} = 0$,
a condition guaranteed by the Bianchi identity on the geometric side of the field equations. Hence, the Tolman--Oppenheimer--Volkoff equation is modified by an additional force term,
which originates from the non-minimal interaction between matter and geometry.

For the theory under consideration, the generalized TOV equation becomes
\begin{eqnarray}
\label{eq:TOV}
p'_r = {-\frac{3(1+2\xi)(\rho c^2 + p_r)}{3+7\beta_1} \,\zeta'
+ \beta_1 \,\frac{3\rho' c^2 - 2 p'_t}{3+7\beta_1}
+ \frac{6(1+2\xi)(p_\theta - p_r)}{(3+7\beta_1)r}}.
\end{eqnarray}

We decompose Eq.~\eqref{eq:TOV} into its contributing forces, namely:
\begin{eqnarray}
\label{eq:TOV_forces}
F_H &=& -p'_r, \quad
F_g = -\frac{3(1+2\beta_1)(\rho c^2 + p_r)}{3+7\beta_1} \,\zeta', \quad
F_c = \beta_1 \,\frac{3\rho' c^2 - 2 p'_t}{3+7\beta_1}, \quad
F_a = \frac{6(1+2\beta_1)(p_t - p_r)}{(3+7\beta_1)r}.
\end{eqnarray}

When the coupling constant is set to $\beta_1 = 0$, the coupling term $F_c$ vanishes, and Eq.~\eqref{eq:TOV} reduces to the familiar GR form:
\begin{equation*}
p'_r = -(\rho c^2 + p_r)\,\zeta' + \frac{2}{r}(p_t - p_r).
\end{equation*}

\begin{figure}
\centering
\subfigure[ TOV equation of the case $p_r = \omega\,\rho$]{\includegraphics[width=.25\textwidth]{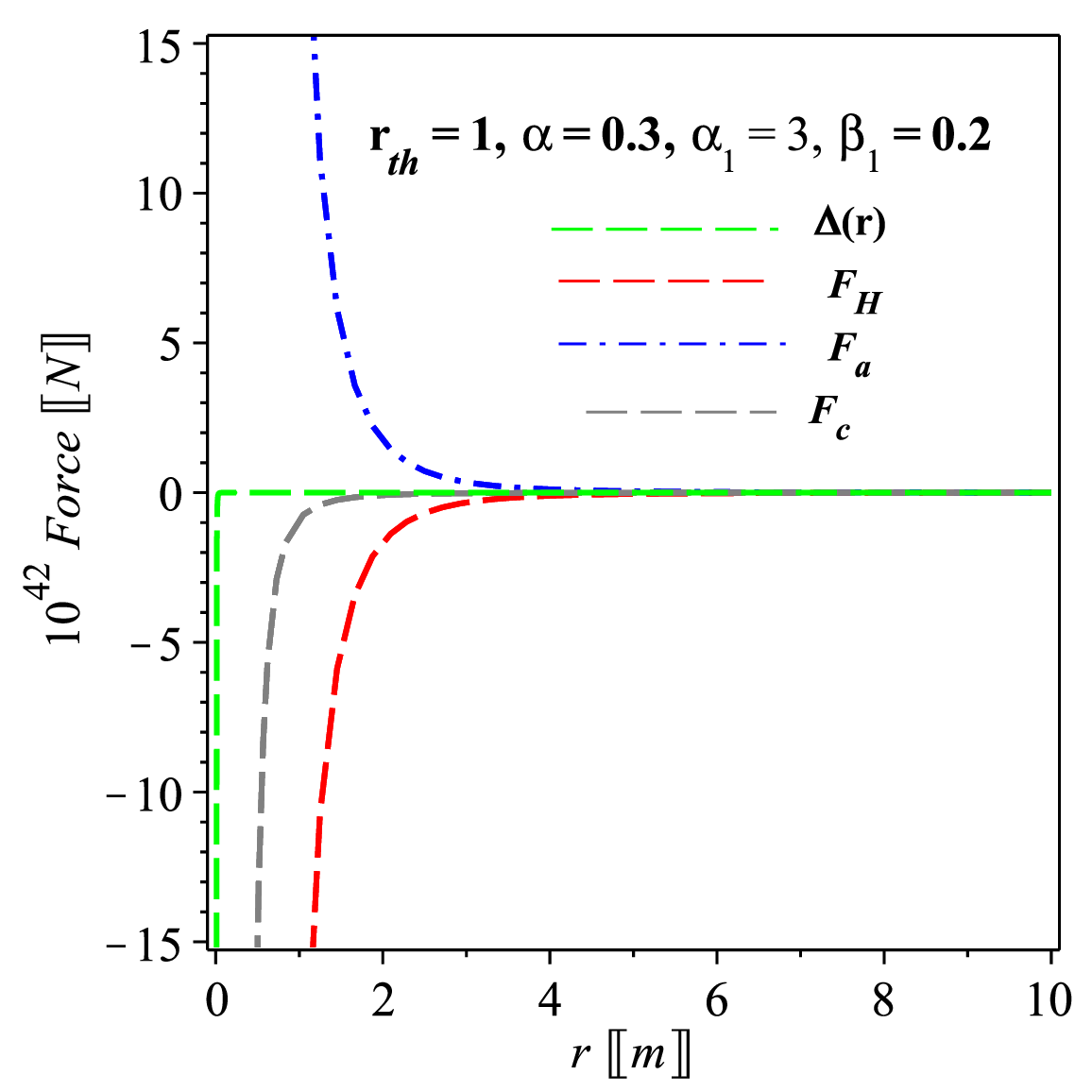}}\hspace{1cm}
\subfigure[TOV equation of the case $p_t = np_r$]{\includegraphics[width=.25\textwidth]{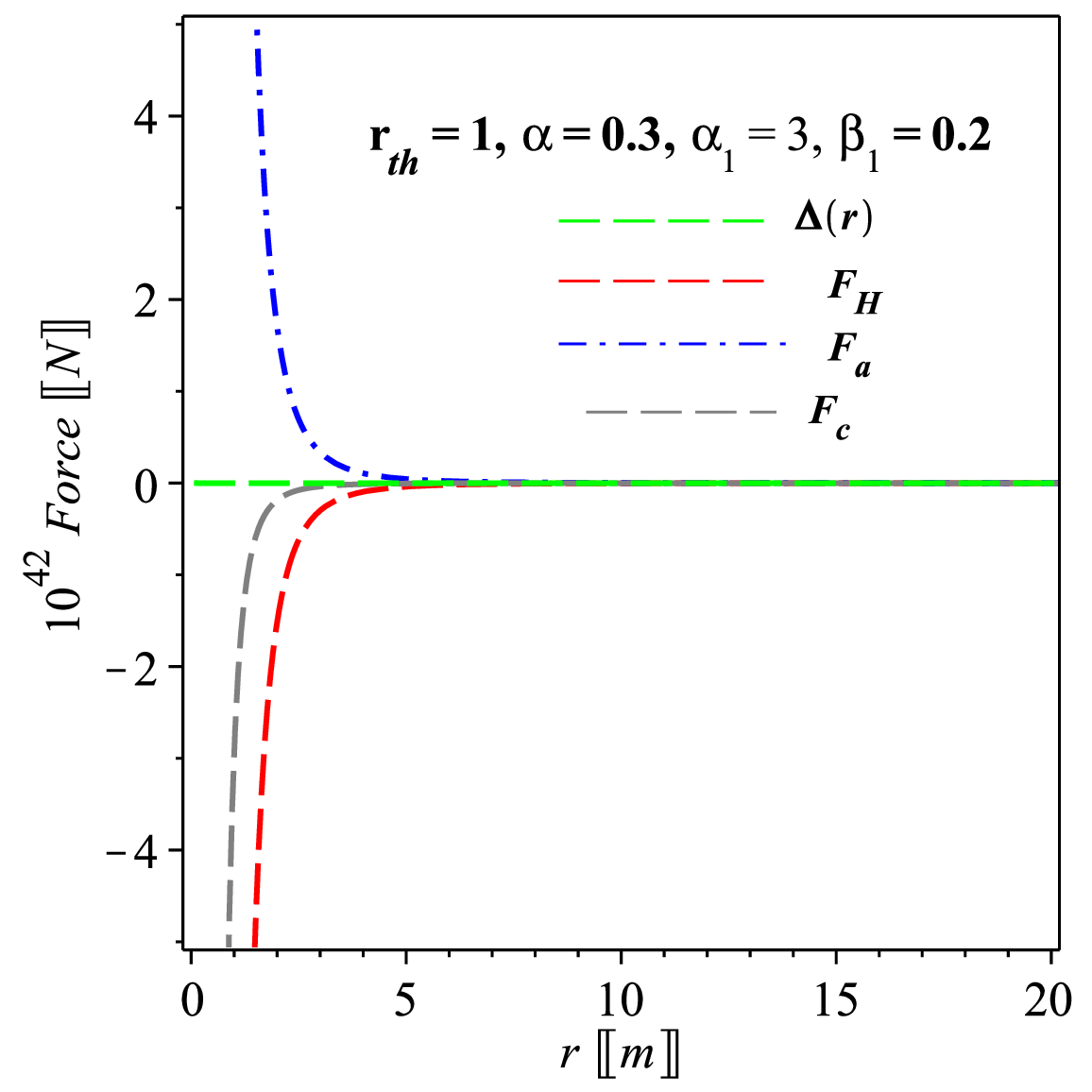}}
\subfigure[TOV equation of the case $p_r=\frac{A}{\rho_{\mathrm{ch}}^{\,\eta}}$ ]{\includegraphics[width=.25\textwidth]{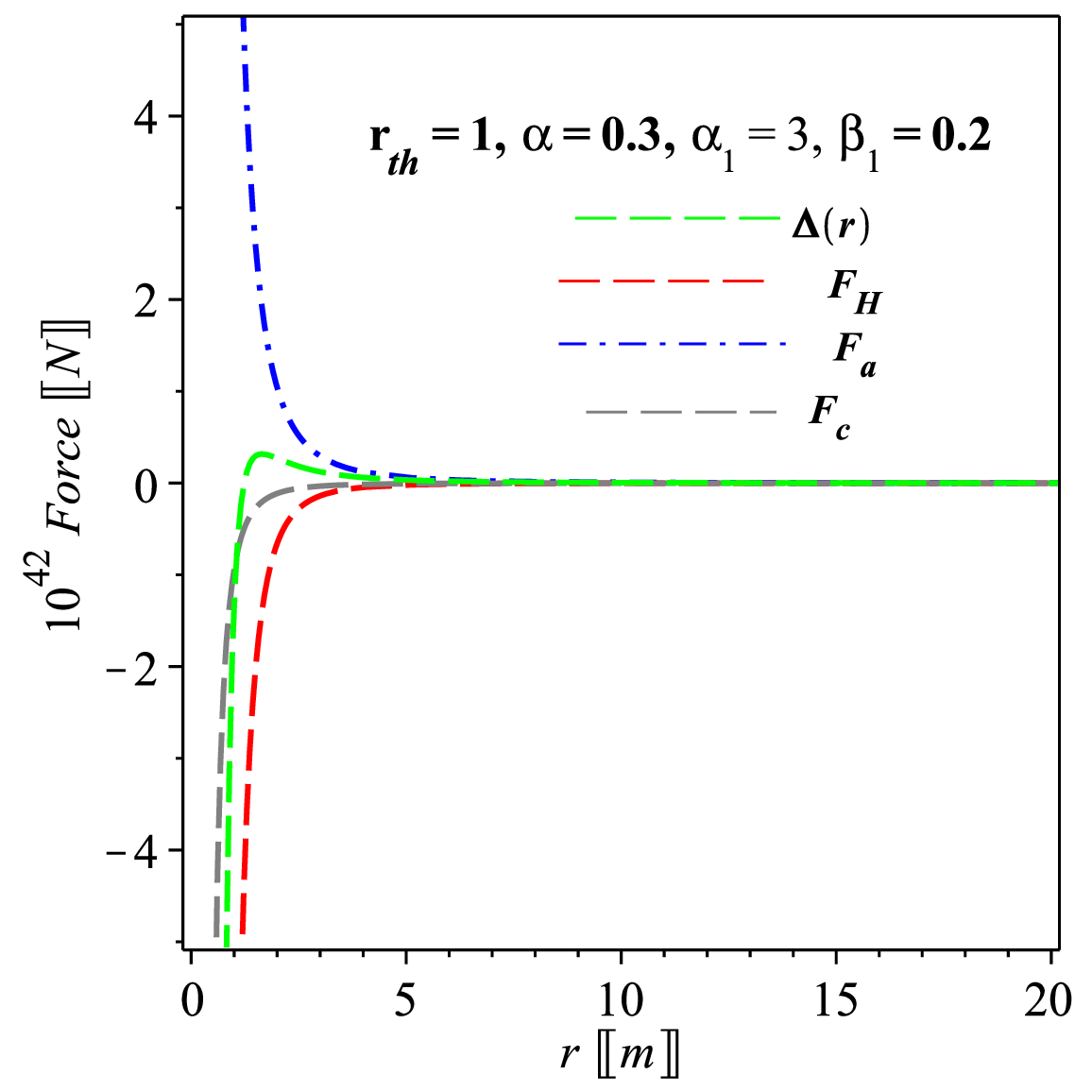}}
\caption{Variation of the generalized Tolman--Oppenheimer--Volkoff (TOV) equation as a function of the radial coordinate $r$ for the first holographic model in $f(R,T)=R+\beta T$ gravity with $r_{\textrm th}=1$, $\alpha=0.3$, $\alpha_1=3$, and $\beta_1=0.2$. The hydrostatic force $F_H$ (red dashed), anisotropic force $F_a$ (blue dash-dotted), and matter--geometry coupling force $F_c$ (gray dashed) are shown together with their sum $\Delta(r)=F_H+F_c+F_a$ (green dashed). Panels (a)--(c) correspond to different equations of state: (a) $p_r=\omega\rho$, (b) $p_t=n p_r$, and (c) $p_r=A/\rho^{\alpha_{\textrm ch}}$. Although the individual force components are nonzero and dominant near the throat, their exact cancellation yields $\Delta(r)\approx 0$ throughout the spacetime, confirming that the wormhole configurations satisfy the generalized TOV equilibrium condition.}

\label{Fig:6}
\end{figure}

For stability, the sum of all forces must vanish:
\[
F_H + F_g + F_c + F_a = 0.
\]
{ With the choice $\zeta(r) = \zeta_0$ (constant redshift), the gravitational term $F_g$ disappears as clear from Eq.~\eqref{eq:TOV_forces}, simplifying the equilibrium condition to:}
\begin{equation}
\label{eq:TOV_forces2}
\Delta(r)\equiv F_H + F_c + F_a = 0.
\end{equation}
{  For each model and for the same parameter sets used in Figs.~1--4 (e.g.\ $r_{\mathrm{th}}=1$, $\alpha=0.3$, $\alpha_1=3$, $\beta_1=0.2$), we evaluate the numerical residual $\Delta(r) $ of Eq.~\eqref{eq:TOV_forces2}.
In all three cases, $\Delta(r)$ vanishes throughout the domain $r\ge r_{\mathrm{th}}$ up to numerical precision, confirming that the wormhole configurations are in exact hydrostatic equilibrium. The residual $\Delta(r)$ is shown explicitly in Fig.~\ref{Fig:6} for each holographic profile.}

It is worth noting that the anisotropic force coefficient in Eq.~\eqref{eq:TOV_forces},
\[
\frac{1+2\beta_1}{3+7\beta_1} = \frac{2}{7} + \frac{1}{7(3+7\beta_1)} \longrightarrow \frac{2}{7} \quad \text{for} \quad |\beta_1| \to \infty,
\]
is always positive for $\beta_1 < -\tfrac12$, as required by the NEC. In such cases, the anisotropic term is repulsive if $p_r < p_t$ and attractive if $p_r > p_t$.

For the theory under consideration with constant $\zeta(r)$, the forces can be expressed in terms of the shape function $s(r)$ as:
\begin{align}
F_H &= -\frac{3s}{(1+2\beta_1)\kappa r^4}
+ \frac{(3+20\beta_1) s'}{3(1+2\beta_1)(1+4\beta_1)\kappa r^3}
- \frac{4\beta_1 s''}{3(1+2\beta_1)(1+4\beta_1)\kappa r^2}, \label{eq:hforce} \\
F_c &= \frac{9\beta_1 s}{(1+2\beta_1)(3+7\beta_1)\kappa r^4}
- \frac{\beta_1(27+68\beta_1) s'}{3(1+2\beta_1)(1+4\beta_1)(3+7\beta_1)\kappa r^3}
+ \frac{4\beta_1 s''}{3(1+2\beta_1)(1+4\beta_1)\kappa r^2}, \label{eq:cforce} \\
F_a &= \frac{9s}{(3+7\beta_1)\kappa r^4}
- \frac{3 s'}{(3+7\beta_1)\kappa r^3}. \label{eq:aforce}
\end{align}
These forms allow Eq.~\eqref{eq:TOV_forces2} to be satisfied directly.\\
\centerline{\bf HDE of R\'enyi }
\begin{align}
F_H &= \frac{144 \left(4 \beta_1+1 \right)}{{\alpha_1} \left( 1+\pi\,\alpha_1{r}^{2} \right){r}^5 \left( 8\,\beta_1+3 \right) {r_{th}} \left( 2\,\beta_1+1 \right) \left( 4\,\beta_1+1 \right) {\kappa}}  \left[ \frac{r}2 \left( \beta_1+\frac{1}4 \right)  \left( \frac{1}2+\beta_1 \right)  \left( 1+\pi\,\alpha_1{r}^{2} \right) \alpha\ln  \left( 1+\pi\,\alpha_1{r_{th}}^{2} \right)\right.\nonumber\\
  &\left. +r_{th}\, \left( -{ \frac {8}{9}}\, \left( \beta_1+\frac{3}{16} \right)  \left( \frac{1}2+\beta_1 \right)\left( 1+\pi\,\alpha_1{r}^{2} \right) \alpha\ln  \left( 1+\pi\,\alpha_1{r}^{2} \right) + \left(\frac{\alpha_1}9 \left[ r \left\{ \frac{3}2\,r_{th}\,r \left( \frac{3}8+\beta_1 \right) \alpha_1+ \left( \frac{1}2+\beta_1 \right) \alpha\beta_1 \right\} \pi \right.\right.\right.\right.\nonumber\\
  &\left.\left.\left.\left.+\frac{3}2\,r_{th}\, \left( \frac{3}8+\beta_1 \right)  \right]  +\sqrt {\pi\,\alpha_1} \left( \beta_1+\frac{1}4 \right)  \left( \frac{1}2+\beta_1 \right)  \left( 1+\pi\,\alpha_1{r}^{2} \right) \alpha\{\arctan\Theta  -\arctan \Theta_0\}  \right) r \right) \right],
   \tag{53} \\
F_c &= \frac{144\beta_1 \left( 4\beta_1+1 \right)}{{\alpha_1} \left( 1+\pi \alpha_1{r}^{2} \right){r}^{5} \left( 8\beta_1+3 \right){r_{th}} \left( 2\beta_1+1 \right)\left( 4\beta_1+1 \right){\kappa} \left( 7\beta_1+3 \right)}
 \left[ \frac{r\alpha}2 \left( \beta_1+\frac{1}4 \right)  \left( \frac{1}2+\beta_1 \right)  \left( 1+\pi\alpha_1{r}^{2} \right) \ln  \left( 1+\pi\alpha_1{r_{th}}^{2} \right)\right.\nonumber\\
  &\left.+r_{th} \left\{ -{\frac {20}{9}} \left( 1+\pi\alpha_1{r }^{2} \right)  \left( \frac{3}8+\beta_1 \right)  \left( \frac{1}2+\beta_1 \right) \alpha\ln  \left( 1+\pi\alpha_1{r}^{2} \right)+ \left\{ \sqrt {\pi\alpha_1} \left( \beta_1+\frac{1}4 \right)  \left( \frac{1}2+\beta_1 \right)  \left( 1+ \pi\alpha_1{r}^{2} \right) \alpha\left\{\arctan \Theta\right.\right.\right.\right.\nonumber\\
  &\left.\left.\left.\left. -\arctan\Theta_0\right\} +{\frac {7}{9}}\alpha_1 \left[ \left( \frac{3r_{th}r}{14} \left( \frac{3}8+\beta_1 \right) \alpha_1+ \left( \frac{1}2+ \beta_1 \right) \alpha \left( \beta_1+\frac{3}7 \right)  \right) r\pi+\frac{3r_{th}}{14} \left( \frac{3}8+\beta_1 \right)  \right]  \right\} r \right\} \right] .\nonumber\\
 F_a &=  \frac{432\left( 4\beta_1+1\right)}{{\alpha_1}{r}^5 \left( 8\,\beta_1+3 \right) { r_{th}} \left( 4\,\beta_1+1 \right){\kappa}\left( 7\,\beta_1+3 \right)}  \left[\frac{r}2 \left( \beta_1+\frac{1}4 \right)  \left( \frac{1}2+\beta_1 \right)\alpha\ln  \left( 1+\pi\,\alpha_1{r_{th}}^ {2} \right)+r_{th}\, \left( -\frac{2}3\, \left( \beta_1+\frac{1}4 \right) \left( \frac{1}2+\beta_1 \right) \right.\right.\nonumber\\
  &\left.\left. \alpha\ln  \left( 1+\pi\,\alpha_1{r}^{2} \right) +r \left\{ \sqrt {\pi\,\alpha_1} \left( \beta_1+\frac{1}4 \right)  \left( \frac{1}2+ \beta_1 \right)\{\arctan\Theta  -\arctan \Theta_0\} +\frac{1}6\,r_{th}\,\alpha_1 \left( \frac{3}8+\beta_1 \right)  \right\} \right)  \right].
\end{align}
\centerline{\bf HDE of Moradpour}
\begin{align}
F_H &=  {\frac {18 }{\sqrt {\pi\,\alpha_1}\left( 1+\pi\,\alpha_1{r}^{2} \right) ^{2} \left( 8\,\beta_1+3 \right) \left( 2\,\beta_1+1 \right){r}^{4}{\kappa}}} \left[  \left(\frac{1}2+\beta_1 \right)  \left( \beta_1+\frac{1}4 \right) \alpha \left( 1+\pi\,\alpha_1{r}^{2} \right) ^{2}\{\arctan\Theta  -\arctan \Theta_0\}\right.\nonumber\\
  &\left.-{\frac {7}{9}}\,\sqrt {\pi\,\alpha_1} \left\{ r \left( \pi\,\alpha_1{r}^{2}+\frac{5}7 \right)\alpha{\beta_1}^{2}+ \left\{  \left( { \frac {13}{28}}\,r+{\frac {17}{28}}\,{r}^{3}\pi\,\alpha_1 \right) \alpha-{\frac { 12}{7}}\,r_{th}\, \left( 1+\pi\,\alpha_1{r}^{2} \right) ^{2} \right\} \beta_1+{\frac {3}{56}}\, \left( 1+\pi\,\alpha_1{r}^{2} \right)  \right.\right.\nonumber\\
  &\left.\left.\left( r\alpha-12 \,r_{th}\,\pi\,\alpha_1{r}^{2}-12\,r_{th} \right)  \right\}  \right] , \tag{54} \\
F_c &=\frac{18 \beta_1}{\left( 1+\pi\,\alpha_1{r }^{2} \right) ^{2} \left( 8\,\beta_1+3 \right)  \left( 2\, \beta_1+1 \right){r}^{4}{ \kappa}\sqrt {\pi\,\alpha_1} \left( 7\,\beta_1+3 \right)}  \left[  \left( \frac{1}2+\beta_1 \right)  \left( \beta_1+\frac{1}4 \right) \alpha \left( 1+\pi\,\alpha_1{r}^{2} \right) ^{2}\{\arctan\Theta  -\arctan \Theta_0\}\right.\nonumber\\
 &\left.-{\frac {31}{9}}\,\sqrt {\pi\,\alpha_1} \left( \left( \pi\,\alpha_1{r}^{2}+{\frac {17}{31}} \right) \alpha r{\beta_1}^{2}+ \left(  \left( {\frac {113}{124}}\,{r}^{3}\pi\,\alpha_1+{\frac {61}{124}}\,r \right) \alpha-{\frac {12}{31}}\,r_{th}\, \left( 1+\pi\,\alpha_1{r}^{2} \right) ^{2} \right) \beta_1+ \left( {\frac {51}{248}}\,{r}^{3}\pi\, \alpha_1+{\frac {27}{248}}\,r \right) \alpha\right.\right.\nonumber\\
 &\left.\left.-{\frac {9}{62}}\,r_{th}\, \left( 1+ \pi\,\alpha_1{r}^{2} \right) ^{2} \right)  \right], \tag{55} \\
F_a &=  \frac {54 \left( 4\beta_1+1 \right)}{\sqrt {\pi\,\alpha_1}\left( 8\,\beta_1+3 \right) \left( 1+\pi\,\alpha_1{r}^{2} \right) \left( 4\,\beta_1+1 \right) {r}^{4}{\kappa} \left( 7\,\beta_1+3 \right) }  \left[ \left( \frac{1}2+\beta_1 \right)  \left( \beta_1+\frac{1}4 \right) \alpha \left( 1+\pi\,\alpha_1{r}^{2} \right) \{\arctan\Theta  -\arctan \Theta_0\} \right.\nonumber\\
 &\left.-\frac{1}3\, \left( \alpha r{\bet\alpha_1}^{2}+ \left( \frac{3}4\,r\alpha-4\,r_{th}\,\pi\,\alpha_1{r}^{2}-4\,r_{th} \right) \beta_1+\frac{1}8\,r\alpha-\frac{3}2\,r_{th}\, \left( 1+\pi\,\alpha_1{r}^{2} \right)  \right) \sqrt {\pi\,\alpha_1} \right]. \tag{55}
\end{align}
\centerline{ HDE of Bekenstein-Hawking}
\begin{align}
F_H &=\frac{8}{\left( 8\,\beta_1+3 \right) \left( 2\,\beta_1+1 \right)   {r}^{4}{\kappa}}  \left[  \left( r-\frac{9}4\,r_{th} \right) \alpha{\beta_1}^{2}+ \left(  \left( {\frac {7}{8}}\,r-{\frac {27} {16}}\,r_{th} \right) \alpha+3\,r_{th} \right) \beta_1+ \left( -{ \frac {9}{32}} r_{th}+\frac{3r}{16} \right) \alpha+{\frac {9}{8}}\,r_{th} \right], \tag{57}\\
F_c&=-\frac{16 \beta_1}{ \left( 8\beta_1+3 \right) \left( 2\beta_1+1 \right) {r}^{4}{\kappa} \left( 7\beta_1+3 \right) } \left[ \alpha \left( r+{\frac {9r_{th}}{8}} \right) {\beta_1}^{2}+ \left(  \left( {\frac {17r}{ 16}}+{\frac {27r_{th}}{32}} \right) \alpha-\frac{3r_{th}}2 \right) \beta_1+ \left( {\frac {9r_{th}}{64}}+{\frac {9}{32}}r \right) \alpha-{\frac {9r_{th}}{16}} \right],\tag{58}\\
F_a&=\frac{36}{\left( 8\,\beta_1+3 \right) {r}^{4}{\kappa} \left( 7\, \beta_1+3 \right)} \left[\alpha \left( r-\frac{3r_{th}}2 \right) {\beta_1}^{2}+ \left(  \left( -{\frac {9r_{th}}{8}}\,+\frac{3r}4  \right) \alpha+2r_{th} \right) \beta_1+ \left( -\frac{3r_{th}}{16}+\frac{r}8 \right)\alpha+\frac{3r_{th}}4 \right].
\end{align}

Figure~\ref{Fig:6} presents the hydrostatic and anisotropic forces for the three examined HDE models.
Although these forces act in opposite directions, their overall trends remain consistent, indicating a form of dynamic balance that sustains the wormhole's stability. The close resemblance in their profiles further emphasizes the resilience of these configurations.
\section{Conclusion and Discussion}

In this work, we have explored traversable wormhole solutions within the framework of $f(R,T)$ gravity, supported by three distinct HDE models: R\'enyi, Moradpour, and Bekenstein-Hawking. The constructed shape functions were shown to satisfy the essential conditions for traversability, including asymptotic flatness and the flaring-out criterion at the throat. { We also examined the asymptotic behaviour of the Bekenstein--Hawking--motivated
density profile. Using Eq.~(\ref{36}), we found that the corresponding shape
function behaves as $s(r)\sim A_{\infty} r$ at large radii, with
$A_{\infty}$ given by Eq.~\eqref{ainf}. Since $A_{\infty}=0$ occurs
only for $\alpha=0$ or for the singular values $\beta_1=-1/2,-1/4$ of the
field equations, a non-trivial Bekenstein--Hawking--supported wormhole within
the present $f(\mathcal{R}, \mathbb{T})$ model is generically non-asymptotically flat, and our
BH analysis should be interpreted as describing a physically relevant local
wormhole geometry that could be matched to an external vacuum at a finite
radius.
}

Our analysis demonstrates that while the radial null energy condition (NEC) is violated near the throat, the tangential NEC remains satisfied across all cases, highlighting the role of modified gravity in reducing the reliance on exotic matter. Furthermore, the TolmanOppenheimerVolkoff (TOV) equilibrium analysis revealed that the hydrostatic and anisotropic forces balance each other, ensuring dynamical stability of the wormhole configurations.

The embedding diagrams, energy condition profiles, and gravitational lensing behavior consistently confirm the robustness of the obtained solutions. Among the considered HDE models, all exhibit comparable qualitative behavior, with slight quantitative differences reflecting the entropy correction schemes. These results suggest that $f(R,T)$ gravity, in conjunction with HDE, provides a viable and physically consistent framework for sustaining traversable wormholes without invoking large amounts of exotic matter.
\subsection*{Acknowledgments}
 This work was supported and funded by the Deanship of Scientific Research at Imam Mohammad Ibn Saud Islamic University (IMSIU) (grant number IMSIU-DDRSP2602).


\begin{thebibliography}{109}%
\makeatletter
\providecommand \@ifxundefined [1]{%
 \@ifx{#1\undefined}
}%
\providecommand \@ifnum [1]{%
 \ifnum #1\expandafter \@firstoftwo
 \else \expandafter \@secondoftwo
 \fi
}%
\providecommand \@ifx [1]{%
 \ifx #1\expandafter \@firstoftwo
 \else \expandafter \@secondoftwo
 \fi
}%
\providecommand \natexlab [1]{#1}%
\providecommand \enquote  [1]{``#1''}%
\providecommand \bibnamefont  [1]{#1}%
\providecommand \bibfnamefont [1]{#1}%
\providecommand \citenamefont [1]{#1}%
\providecommand \href@noop [0]{\@secondoftwo}%
\providecommand \href [0]{\begingroup \@sanitize@url \@href}%
\providecommand \@href[1]{\@@startlink{#1}\@@href}%
\providecommand \@@href[1]{\endgroup#1\@@endlink}%
\providecommand \@sanitize@url [0]{\catcode `\\12\catcode `\$12\catcode
  `\&12\catcode `\#12\catcode `\^12\catcode `\_12\catcode `\%12\relax}%
\providecommand \@@startlink[1]{}%
\providecommand \@@endlink[0]{}%
\providecommand \url  [0]{\begingroup\@sanitize@url \@url }%
\providecommand \@url [1]{\endgroup\@href {#1}{\urlprefix }}%
\providecommand \urlprefix  [0]{URL }%
\providecommand \Eprint [0]{\href }%
\providecommand \doibase [0]{http://dx.doi.org/}%
\providecommand \selectlanguage [0]{\@gobble}%
\providecommand \bibinfo  [0]{\@secondoftwo}%
\providecommand \bibfield  [0]{\@secondoftwo}%
\providecommand \translation [1]{[#1]}%
\providecommand \BibitemOpen [0]{}%
\providecommand \bibitemStop [0]{}%
\providecommand \bibitemNoStop [0]{.\EOS\space}%
\providecommand \EOS [0]{\spacefactor3000\relax}%
\providecommand \BibitemShut  [1]{\csname bibitem#1\endcsname}%
\let\auto@bib@innerbib\@empty
\bibitem [{\citenamefont {Einstein}\ and\ \citenamefont
  {Rosen}(1935)}]{Einstein:1935tc}%
  \BibitemOpen
  \bibfield  {author} {\bibinfo {author} {\bibfnamefont {A.}~\bibnamefont
  {Einstein}}\ and\ \bibinfo {author} {\bibfnamefont {N.}~\bibnamefont
  {Rosen}},\ }\href {\doibase 10.1103/PhysRev.48.73} {\bibfield  {journal}
  {\bibinfo  {journal} {Phys. Rev.}\ }\textbf {\bibinfo {volume} {48}},\
  \bibinfo {pages} {73} (\bibinfo {year} {1935})}\BibitemShut {NoStop}%
\bibitem [{\citenamefont {Misner}\ and\ \citenamefont
  {Wheeler}(1957)}]{Misner:1957mt}%
  \BibitemOpen
  \bibfield  {author} {\bibinfo {author} {\bibfnamefont {C.~W.}\ \bibnamefont
  {Misner}}\ and\ \bibinfo {author} {\bibfnamefont {J.~A.}\ \bibnamefont
  {Wheeler}},\ }\href {\doibase 10.1016/0003-4916(57)90049-0} {\bibfield
  {journal} {\bibinfo  {journal} {Annals Phys.}\ }\textbf {\bibinfo {volume}
  {2}},\ \bibinfo {pages} {525} (\bibinfo {year} {1957})}\BibitemShut {NoStop}%
\bibitem [{\citenamefont {Morris}\ and\ \citenamefont
  {Thorne}(1988{\natexlab{a}})}]{Morris:1988cz}%
  \BibitemOpen
  \bibfield  {author} {\bibinfo {author} {\bibfnamefont {M.~S.}\ \bibnamefont
  {Morris}}\ and\ \bibinfo {author} {\bibfnamefont {K.~S.}\ \bibnamefont
  {Thorne}},\ }\href {\doibase 10.1119/1.15620} {\bibfield  {journal} {\bibinfo
   {journal} {Am. J. Phys.}\ }\textbf {\bibinfo {volume} {56}},\ \bibinfo
  {pages} {395} (\bibinfo {year} {1988}{\natexlab{a}})}\BibitemShut {NoStop}%
\bibitem [{\citenamefont {Abbott}\ \emph
  {et~al.}(2016{\natexlab{a}})\citenamefont {Abbott} \emph
  {et~al.}}]{LIGOScientific:2016aoc}%
  \BibitemOpen
  \bibfield  {author} {\bibinfo {author} {\bibfnamefont {B.~P.}\ \bibnamefont
  {Abbott}} \emph {et~al.} (\bibinfo {collaboration} {LIGO Scientific,
  Virgo}),\ }\href {\doibase 10.1103/PhysRevLett.116.061102} {\bibfield
  {journal} {\bibinfo  {journal} {Phys. Rev. Lett.}\ }\textbf {\bibinfo
  {volume} {116}},\ \bibinfo {pages} {061102} (\bibinfo {year}
  {2016}{\natexlab{a}})},\ \Eprint {http://arxiv.org/abs/1602.03837}
  {arXiv:1602.03837 [gr-qc]} \BibitemShut {NoStop}%
\bibitem [{\citenamefont {Khodadi}(2022)}]{Khodadi:2022dff}%
  \BibitemOpen
  \bibfield  {author} {\bibinfo {author} {\bibfnamefont {M.}~\bibnamefont
  {Khodadi}},\ }\href {\doibase 10.1103/PhysRevD.105.023025} {\bibfield
  {journal} {\bibinfo  {journal} {Phys. Rev. D}\ }\textbf {\bibinfo {volume}
  {105}},\ \bibinfo {pages} {023025} (\bibinfo {year} {2022})},\ \Eprint
  {http://arxiv.org/abs/2201.02765} {arXiv:2201.02765 [gr-qc]} \BibitemShut
  {NoStop}%
\bibitem [{\citenamefont {Abbott}\ \emph
  {et~al.}(2016{\natexlab{b}})\citenamefont {Abbott} \emph
  {et~al.}}]{LIGOScientific:2016lio}%
  \BibitemOpen
  \bibfield  {author} {\bibinfo {author} {\bibfnamefont {B.~P.}\ \bibnamefont
  {Abbott}} \emph {et~al.} (\bibinfo {collaboration} {LIGO Scientific,
  Virgo}),\ }\href {\doibase 10.1103/PhysRevLett.116.221101} {\bibfield
  {journal} {\bibinfo  {journal} {Phys. Rev. Lett.}\ }\textbf {\bibinfo
  {volume} {116}},\ \bibinfo {pages} {221101} (\bibinfo {year}
  {2016}{\natexlab{b}})},\ \bibinfo {note} {[Erratum: Phys.Rev.Lett. 121,
  129902 (2018)]},\ \Eprint {http://arxiv.org/abs/1602.03841} {arXiv:1602.03841
  [gr-qc]} \BibitemShut {NoStop}%
\bibitem [{\citenamefont {Errehymy}\ \emph {et~al.}(2023)\citenamefont
  {Errehymy}, \citenamefont {Maurya}, \citenamefont {Mustafa}, \citenamefont
  {Hansraj}, \citenamefont {Alrebdi},\ and\ \citenamefont
  {Abdel-Aty}}]{Errehymy:2023xpc}%
  \BibitemOpen
  \bibfield  {author} {\bibinfo {author} {\bibfnamefont {A.}~\bibnamefont
  {Errehymy}}, \bibinfo {author} {\bibfnamefont {S.~K.}\ \bibnamefont
  {Maurya}}, \bibinfo {author} {\bibfnamefont {G.}~\bibnamefont {Mustafa}},
  \bibinfo {author} {\bibfnamefont {S.}~\bibnamefont {Hansraj}}, \bibinfo
  {author} {\bibfnamefont {H.~I.}\ \bibnamefont {Alrebdi}}, \ and\ \bibinfo
  {author} {\bibfnamefont {A.-H.}\ \bibnamefont {Abdel-Aty}},\ }\href {\doibase
  10.1002/prop.202300052} {\bibfield  {journal} {\bibinfo  {journal} {Fortsch.
  Phys.}\ }\textbf {\bibinfo {volume} {71}},\ \bibinfo {pages} {2300052}
  (\bibinfo {year} {2023})}\BibitemShut {NoStop}%
\bibitem [{\citenamefont {De~Felice}\ and\ \citenamefont
  {Tsujikawa}(2010)}]{DeFelice:2010aj}%
  \BibitemOpen
  \bibfield  {author} {\bibinfo {author} {\bibfnamefont {A.}~\bibnamefont
  {De~Felice}}\ and\ \bibinfo {author} {\bibfnamefont {S.}~\bibnamefont
  {Tsujikawa}},\ }\href {\doibase 10.12942/lrr-2010-3} {\bibfield  {journal}
  {\bibinfo  {journal} {Living Rev. Rel.}\ }\textbf {\bibinfo {volume} {13}},\
  \bibinfo {pages} {3} (\bibinfo {year} {2010})},\ \Eprint
  {http://arxiv.org/abs/1002.4928} {arXiv:1002.4928 [gr-qc]} \BibitemShut
  {NoStop}%
\bibitem [{\citenamefont {Nashed}\ \emph {et~al.}(2021)\citenamefont {Nashed},
  \citenamefont {Odintsov},\ and\ \citenamefont {Oikonomou}}]{Nashed:2021gkp}%
  \BibitemOpen
  \bibfield  {author} {\bibinfo {author} {\bibfnamefont {G.~G.~L.}\
  \bibnamefont {Nashed}}, \bibinfo {author} {\bibfnamefont {S.~D.}\
  \bibnamefont {Odintsov}}, \ and\ \bibinfo {author} {\bibfnamefont {V.~K.}\
  \bibnamefont {Oikonomou}},\ }\href {\doibase 10.1140/epjc/s10052-021-09321-3}
  {\bibfield  {journal} {\bibinfo  {journal} {Eur. Phys. J. C}\ }\textbf
  {\bibinfo {volume} {81}},\ \bibinfo {pages} {528} (\bibinfo {year} {2021})},\
  \Eprint {http://arxiv.org/abs/2106.13607} {arXiv:2106.13607 [gr-qc]}
  \BibitemShut {NoStop}%
\bibitem [{\citenamefont {Starobinsky}(1979)}]{starobinsky1979relict}%
  \BibitemOpen
  \bibfield  {author} {\bibinfo {author} {\bibfnamefont {A.~A.}\ \bibnamefont
  {Starobinsky}},\ }\href@noop {} {\bibfield  {journal} {\bibinfo  {journal}
  {JETP lett}\ }\textbf {\bibinfo {volume} {30}},\ \bibinfo {pages} {131}
  (\bibinfo {year} {1979})}\BibitemShut {NoStop}%
\bibitem [{\citenamefont {Harko}\ \emph {et~al.}(2011)\citenamefont {Harko},
  \citenamefont {Lobo}, \citenamefont {Nojiri},\ and\ \citenamefont
  {Odintsov}}]{Harko:2011kv}%
  \BibitemOpen
  \bibfield  {author} {\bibinfo {author} {\bibfnamefont {T.}~\bibnamefont
  {Harko}}, \bibinfo {author} {\bibfnamefont {F.~S.~N.}\ \bibnamefont {Lobo}},
  \bibinfo {author} {\bibfnamefont {S.}~\bibnamefont {Nojiri}}, \ and\ \bibinfo
  {author} {\bibfnamefont {S.~D.}\ \bibnamefont {Odintsov}},\ }\href {\doibase
  10.1103/PhysRevD.84.024020} {\bibfield  {journal} {\bibinfo  {journal} {Phys.
  Rev. D}\ }\textbf {\bibinfo {volume} {84}},\ \bibinfo {pages} {024020}
  (\bibinfo {year} {2011})},\ \Eprint {http://arxiv.org/abs/1104.2669}
  {arXiv:1104.2669 [gr-qc]} \BibitemShut {NoStop}%
\bibitem [{\citenamefont {Nashed}(2023{\natexlab{a}})}]{Nashed:2023pxd}%
  \BibitemOpen
  \bibfield  {author} {\bibinfo {author} {\bibfnamefont {G.~G.~L.}\
  \bibnamefont {Nashed}},\ }\href {\doibase 10.3847/1538-4357/acd182}
  {\bibfield  {journal} {\bibinfo  {journal} {Astrophys. J.}\ }\textbf
  {\bibinfo {volume} {950}},\ \bibinfo {pages} {129} (\bibinfo {year}
  {2023}{\natexlab{a}})},\ \Eprint {http://arxiv.org/abs/2306.10273}
  {arXiv:2306.10273 [gr-qc]} \BibitemShut {NoStop}%
\bibitem [{\citenamefont {Errehymy}\ \emph
  {et~al.}(2024{\natexlab{a}})\citenamefont {Errehymy}, \citenamefont {Maurya},
  \citenamefont {V{\^\i}lcu}, \citenamefont {Khan},\ and\ \citenamefont
  {Daoud}}]{Errehymy:2024spg}%
  \BibitemOpen
  \bibfield  {author} {\bibinfo {author} {\bibfnamefont {A.}~\bibnamefont
  {Errehymy}}, \bibinfo {author} {\bibfnamefont {S.~K.}\ \bibnamefont
  {Maurya}}, \bibinfo {author} {\bibfnamefont {G.-E.}\ \bibnamefont
  {V{\^\i}lcu}}, \bibinfo {author} {\bibfnamefont {M.~A.}\ \bibnamefont
  {Khan}}, \ and\ \bibinfo {author} {\bibfnamefont {M.}~\bibnamefont {Daoud}},\
  }\href {\doibase 10.1016/j.astropartphys.2024.102972} {\bibfield  {journal}
  {\bibinfo  {journal} {Astropart. Phys.}\ }\textbf {\bibinfo {volume} {160}},\
  \bibinfo {pages} {102972} (\bibinfo {year} {2024}{\natexlab{a}})}\BibitemShut
  {NoStop}%
\bibitem [{\citenamefont {Rahaman}\ \emph {et~al.}(2015)\citenamefont
  {Rahaman}, \citenamefont {Ray}, \citenamefont {Khadekar}, \citenamefont
  {Kuhfittig},\ and\ \citenamefont {Karar}}]{Rahaman:2013ywa}%
  \BibitemOpen
  \bibfield  {author} {\bibinfo {author} {\bibfnamefont {F.}~\bibnamefont
  {Rahaman}}, \bibinfo {author} {\bibfnamefont {S.}~\bibnamefont {Ray}},
  \bibinfo {author} {\bibfnamefont {G.~S.}\ \bibnamefont {Khadekar}}, \bibinfo
  {author} {\bibfnamefont {P.~K.~F.}\ \bibnamefont {Kuhfittig}}, \ and\
  \bibinfo {author} {\bibfnamefont {I.}~\bibnamefont {Karar}},\ }\href
  {\doibase 10.1007/s10773-014-2262-y} {\bibfield  {journal} {\bibinfo
  {journal} {Int. J. Theor. Phys.}\ }\textbf {\bibinfo {volume} {54}},\
  \bibinfo {pages} {699} (\bibinfo {year} {2015})},\ \Eprint
  {http://arxiv.org/abs/1305.4539} {arXiv:1305.4539 [physics.gen-ph]}
  \BibitemShut {NoStop}%
\bibitem [{\citenamefont {Rahaman}\ \emph {et~al.}(2012)\citenamefont
  {Rahaman}, \citenamefont {Islam}, \citenamefont {Kuhfittig},\ and\
  \citenamefont {Ray}}]{Rahaman:2012pg}%
  \BibitemOpen
  \bibfield  {author} {\bibinfo {author} {\bibfnamefont {F.}~\bibnamefont
  {Rahaman}}, \bibinfo {author} {\bibfnamefont {S.}~\bibnamefont {Islam}},
  \bibinfo {author} {\bibfnamefont {P.~K.~F.}\ \bibnamefont {Kuhfittig}}, \
  and\ \bibinfo {author} {\bibfnamefont {S.}~\bibnamefont {Ray}},\ }\href
  {\doibase 10.1103/PhysRevD.86.106010} {\bibfield  {journal} {\bibinfo
  {journal} {Phys. Rev. D}\ }\textbf {\bibinfo {volume} {86}},\ \bibinfo
  {pages} {106010} (\bibinfo {year} {2012})},\ \Eprint
  {http://arxiv.org/abs/1209.2917} {arXiv:1209.2917 [gr-qc]} \BibitemShut
  {NoStop}%
\bibitem [{\citenamefont {Mustafa}\ \emph
  {et~al.}(2024{\natexlab{a}})\citenamefont {Mustafa}, \citenamefont {Javed},
  \citenamefont {Maurya},\ and\ \citenamefont {Errehymy}}]{Mustafa:2024ark}%
  \BibitemOpen
  \bibfield  {author} {\bibinfo {author} {\bibfnamefont {G.}~\bibnamefont
  {Mustafa}}, \bibinfo {author} {\bibfnamefont {F.}~\bibnamefont {Javed}},
  \bibinfo {author} {\bibfnamefont {S.~K.}\ \bibnamefont {Maurya}}, \ and\
  \bibinfo {author} {\bibfnamefont {A.}~\bibnamefont {Errehymy}},\ }\href
  {\doibase 10.1002/andp.202400155} {\bibfield  {journal} {\bibinfo  {journal}
  {Annalen Phys.}\ }\textbf {\bibinfo {volume} {536}},\ \bibinfo {pages}
  {2400155} (\bibinfo {year} {2024}{\natexlab{a}})},\ \Eprint
  {http://arxiv.org/abs/2404.04649} {arXiv:2404.04649 [gr-qc]} \BibitemShut
  {NoStop}%
\bibitem [{\citenamefont {Ashraf}\ \emph {et~al.}(2024)\citenamefont {Ashraf},
  \citenamefont {Mumtaz}, \citenamefont {Javed},\ and\ \citenamefont
  {Zhang}}]{Ashraf:2023bfg}%
  \BibitemOpen
  \bibfield  {author} {\bibinfo {author} {\bibfnamefont {A.}~\bibnamefont
  {Ashraf}}, \bibinfo {author} {\bibfnamefont {S.}~\bibnamefont {Mumtaz}},
  \bibinfo {author} {\bibfnamefont {F.}~\bibnamefont {Javed}}, \ and\ \bibinfo
  {author} {\bibfnamefont {Z.}~\bibnamefont {Zhang}},\ }\href {\doibase
  10.1142/S0219887824502086} {\bibfield  {journal} {\bibinfo  {journal} {Int.
  J. Geom. Meth. Mod. Phys.}\ }\textbf {\bibinfo {volume} {21}},\ \bibinfo
  {pages} {2450208} (\bibinfo {year} {2024})},\ \Eprint
  {http://arxiv.org/abs/2304.06256} {arXiv:2304.06256 [gr-qc]} \BibitemShut
  {NoStop}%
\bibitem [{\citenamefont {Mustafa}\ \emph {et~al.}(2023)\citenamefont
  {Mustafa}, \citenamefont {Maurya},\ and\ \citenamefont
  {Ray}}]{Mustafa:2023kqt}%
  \BibitemOpen
  \bibfield  {author} {\bibinfo {author} {\bibfnamefont {G.}~\bibnamefont
  {Mustafa}}, \bibinfo {author} {\bibfnamefont {S.~K.}\ \bibnamefont {Maurya}},
  \ and\ \bibinfo {author} {\bibfnamefont {S.}~\bibnamefont {Ray}},\ }\href
  {\doibase 10.1002/prop.202200129} {\bibfield  {journal} {\bibinfo  {journal}
  {Fortsch. Phys.}\ }\textbf {\bibinfo {volume} {71}},\ \bibinfo {pages}
  {2200129} (\bibinfo {year} {2023})}\BibitemShut {NoStop}%
\bibitem [{\citenamefont {Kiroriwal}\ \emph {et~al.}(2024)\citenamefont
  {Kiroriwal}, \citenamefont {Kumar}, \citenamefont {Maurya}, \citenamefont
  {Chaudhary},\ and\ \citenamefont {Aziz}}]{Kiroriwal:2024ymu}%
  \BibitemOpen
  \bibfield  {author} {\bibinfo {author} {\bibfnamefont {S.}~\bibnamefont
  {Kiroriwal}}, \bibinfo {author} {\bibfnamefont {J.}~\bibnamefont {Kumar}},
  \bibinfo {author} {\bibfnamefont {S.~K.}\ \bibnamefont {Maurya}}, \bibinfo
  {author} {\bibfnamefont {S.}~\bibnamefont {Chaudhary}}, \ and\ \bibinfo
  {author} {\bibfnamefont {A.}~\bibnamefont {Aziz}},\ }\href {\doibase
  10.1002/prop.202300197} {\bibfield  {journal} {\bibinfo  {journal} {Fortsch.
  Phys.}\ }\textbf {\bibinfo {volume} {72}},\ \bibinfo {pages} {2300197}
  (\bibinfo {year} {2024})}\BibitemShut {NoStop}%
\bibitem [{\citenamefont {Kiroriwal}\ \emph {et~al.}(2023)\citenamefont
  {Kiroriwal}, \citenamefont {Kumar}, \citenamefont {Maurya},\ and\
  \citenamefont {Chaudhary}}]{Kiroriwal:2023nul}%
  \BibitemOpen
  \bibfield  {author} {\bibinfo {author} {\bibfnamefont {S.}~\bibnamefont
  {Kiroriwal}}, \bibinfo {author} {\bibfnamefont {J.}~\bibnamefont {Kumar}},
  \bibinfo {author} {\bibfnamefont {S.~K.}\ \bibnamefont {Maurya}}, \ and\
  \bibinfo {author} {\bibfnamefont {S.}~\bibnamefont {Chaudhary}},\ }\href
  {\doibase 10.1088/1402-4896/ad0820} {\bibfield  {journal} {\bibinfo
  {journal} {Phys. Scripta}\ }\textbf {\bibinfo {volume} {98}},\ \bibinfo
  {pages} {125305} (\bibinfo {year} {2023})}\BibitemShut {NoStop}%
\bibitem [{\citenamefont {Agrawal}\ \emph {et~al.}(2023)\citenamefont
  {Agrawal}, \citenamefont {Mishra},\ and\ \citenamefont
  {Moraes}}]{Agrawal:2022atn}%
  \BibitemOpen
  \bibfield  {author} {\bibinfo {author} {\bibfnamefont {A.~S.}\ \bibnamefont
  {Agrawal}}, \bibinfo {author} {\bibfnamefont {B.}~\bibnamefont {Mishra}}, \
  and\ \bibinfo {author} {\bibfnamefont {P.~H. R.~S.}\ \bibnamefont {Moraes}},\
  }\href {\doibase 10.1140/epjp/s13360-023-03872-y} {\bibfield  {journal}
  {\bibinfo  {journal} {Eur. Phys. J. Plus}\ }\textbf {\bibinfo {volume}
  {138}},\ \bibinfo {pages} {275} (\bibinfo {year} {2023})},\ \Eprint
  {http://arxiv.org/abs/2201.08392} {arXiv:2201.08392 [gr-qc]} \BibitemShut
  {NoStop}%
\bibitem [{\citenamefont {Errehymy}\ \emph
  {et~al.}(2024{\natexlab{b}})\citenamefont {Errehymy}, \citenamefont
  {Banerjee}, \citenamefont {Hansraj}, \citenamefont {Donmez}, \citenamefont
  {Nisar},\ and\ \citenamefont {Abdel-Aty}}]{Errehymy:2024lhl}%
  \BibitemOpen
  \bibfield  {author} {\bibinfo {author} {\bibfnamefont {A.}~\bibnamefont
  {Errehymy}}, \bibinfo {author} {\bibfnamefont {A.}~\bibnamefont {Banerjee}},
  \bibinfo {author} {\bibfnamefont {S.}~\bibnamefont {Hansraj}}, \bibinfo
  {author} {\bibfnamefont {O.}~\bibnamefont {Donmez}}, \bibinfo {author}
  {\bibfnamefont {K.~S.}\ \bibnamefont {Nisar}}, \ and\ \bibinfo {author}
  {\bibfnamefont {A.-H.}\ \bibnamefont {Abdel-Aty}},\ }\href {\doibase
  10.1140/epjc/s10052-024-12929-w} {\bibfield  {journal} {\bibinfo  {journal}
  {Eur. Phys. J. C}\ }\textbf {\bibinfo {volume} {84}},\ \bibinfo {pages} {573}
  (\bibinfo {year} {2024}{\natexlab{b}})}\BibitemShut {NoStop}%
\bibitem [{\citenamefont {Battista}\ \emph {et~al.}(2024)\citenamefont
  {Battista}, \citenamefont {Capozziello},\ and\ \citenamefont
  {Errehymy}}]{Battista:2024gud}%
  \BibitemOpen
  \bibfield  {author} {\bibinfo {author} {\bibfnamefont {E.}~\bibnamefont
  {Battista}}, \bibinfo {author} {\bibfnamefont {S.}~\bibnamefont
  {Capozziello}}, \ and\ \bibinfo {author} {\bibfnamefont {A.}~\bibnamefont
  {Errehymy}},\ }\href {\doibase 10.1140/epjc/s10052-024-13656-y} {\bibfield
  {journal} {\bibinfo  {journal} {Eur. Phys. J. C}\ }\textbf {\bibinfo {volume}
  {84}},\ \bibinfo {pages} {1314} (\bibinfo {year} {2024})},\ \Eprint
  {http://arxiv.org/abs/2409.09750} {arXiv:2409.09750 [gr-qc]} \BibitemShut
  {NoStop}%
\bibitem [{\citenamefont {Mehdizadeh}\ \emph {et~al.}(2015)\citenamefont
  {Mehdizadeh}, \citenamefont {Kord~Zangeneh},\ and\ \citenamefont
  {Lobo}}]{Mehdizadeh:2015jra}%
  \BibitemOpen
  \bibfield  {author} {\bibinfo {author} {\bibfnamefont {M.~R.}\ \bibnamefont
  {Mehdizadeh}}, \bibinfo {author} {\bibfnamefont {M.}~\bibnamefont
  {Kord~Zangeneh}}, \ and\ \bibinfo {author} {\bibfnamefont {F.~S.~N.}\
  \bibnamefont {Lobo}},\ }\href {\doibase 10.1103/PhysRevD.91.084004}
  {\bibfield  {journal} {\bibinfo  {journal} {Phys. Rev. D}\ }\textbf {\bibinfo
  {volume} {91}},\ \bibinfo {pages} {084004} (\bibinfo {year} {2015})},\
  \Eprint {http://arxiv.org/abs/1501.04773} {arXiv:1501.04773 [gr-qc]}
  \BibitemShut {NoStop}%
\bibitem [{\citenamefont {Nashed}(2021)}]{Nashed:2021pkc}%
  \BibitemOpen
  \bibfield  {author} {\bibinfo {author} {\bibfnamefont {G.~G.~L.}\
  \bibnamefont {Nashed}},\ }\href {\doibase 10.3847/1538-4357/ac19bb}
  {\bibfield  {journal} {\bibinfo  {journal} {Astrophys. J.}\ }\textbf
  {\bibinfo {volume} {919}},\ \bibinfo {pages} {113} (\bibinfo {year}
  {2021})},\ \Eprint {http://arxiv.org/abs/2108.04060} {arXiv:2108.04060
  [gr-qc]} \BibitemShut {NoStop}%
\bibitem [{\citenamefont {Boehmer}\ \emph {et~al.}(2012)\citenamefont
  {Boehmer}, \citenamefont {Harko},\ and\ \citenamefont
  {Lobo}}]{Boehmer:2012uyw}%
  \BibitemOpen
  \bibfield  {author} {\bibinfo {author} {\bibfnamefont {C.~G.}\ \bibnamefont
  {Boehmer}}, \bibinfo {author} {\bibfnamefont {T.}~\bibnamefont {Harko}}, \
  and\ \bibinfo {author} {\bibfnamefont {F.~S.~N.}\ \bibnamefont {Lobo}},\
  }\href {\doibase 10.1103/PhysRevD.85.044033} {\bibfield  {journal} {\bibinfo
  {journal} {Phys. Rev. D}\ }\textbf {\bibinfo {volume} {85}},\ \bibinfo
  {pages} {044033} (\bibinfo {year} {2012})},\ \Eprint
  {http://arxiv.org/abs/1110.5756} {arXiv:1110.5756 [gr-qc]} \BibitemShut
  {NoStop}%
\bibitem [{\citenamefont {Nashed}\ and\ \citenamefont
  {Saridakis}(2022)}]{Nashed:2021pah}%
  \BibitemOpen
  \bibfield  {author} {\bibinfo {author} {\bibfnamefont {G.~G.~L.}\
  \bibnamefont {Nashed}}\ and\ \bibinfo {author} {\bibfnamefont {E.~N.}\
  \bibnamefont {Saridakis}},\ }\href {\doibase 10.1088/1475-7516/2022/05/017}
  {\bibfield  {journal} {\bibinfo  {journal} {JCAP}\ }\textbf {\bibinfo
  {volume} {05}},\ \bibinfo {pages} {017} (\bibinfo {year} {2022})},\ \Eprint
  {http://arxiv.org/abs/2111.06359} {arXiv:2111.06359 [gr-qc]} \BibitemShut
  {NoStop}%
\bibitem [{\citenamefont {Mustafa}\ \emph {et~al.}(2020)\citenamefont
  {Mustafa}, \citenamefont {Waheed}, \citenamefont {Zubair},\ and\
  \citenamefont {Xia}}]{Mustafa:2020kng}%
  \BibitemOpen
  \bibfield  {author} {\bibinfo {author} {\bibfnamefont {G.}~\bibnamefont
  {Mustafa}}, \bibinfo {author} {\bibfnamefont {S.}~\bibnamefont {Waheed}},
  \bibinfo {author} {\bibfnamefont {M.}~\bibnamefont {Zubair}}, \ and\ \bibinfo
  {author} {\bibfnamefont {T.-C.}\ \bibnamefont {Xia}},\ }\href {\doibase
  10.1016/j.cjph.2020.02.008} {\bibfield  {journal} {\bibinfo  {journal} {Chin.
  J. Phys.}\ }\textbf {\bibinfo {volume} {65}},\ \bibinfo {pages} {163}
  (\bibinfo {year} {2020})}\BibitemShut {NoStop}%
\bibitem [{\citenamefont {Shirafuji}\ \emph {et~al.}(1996)\citenamefont
  {Shirafuji}, \citenamefont {Nashed},\ and\ \citenamefont
  {Kobayashi}}]{Shirafuji:1996im}%
  \BibitemOpen
  \bibfield  {author} {\bibinfo {author} {\bibfnamefont {T.}~\bibnamefont
  {Shirafuji}}, \bibinfo {author} {\bibfnamefont {G.~G.~L.}\ \bibnamefont
  {Nashed}}, \ and\ \bibinfo {author} {\bibfnamefont {Y.}~\bibnamefont
  {Kobayashi}},\ }\href {\doibase 10.1143/PTP.96.933} {\bibfield  {journal}
  {\bibinfo  {journal} {Prog. Theor. Phys.}\ }\textbf {\bibinfo {volume}
  {96}},\ \bibinfo {pages} {933} (\bibinfo {year} {1996})},\ \Eprint
  {http://arxiv.org/abs/gr-qc/9609060} {arXiv:gr-qc/9609060} \BibitemShut
  {NoStop}%
\bibitem [{\citenamefont {Chaudhary}\ \emph {et~al.}(2023)\citenamefont
  {Chaudhary}, \citenamefont {Maurya}, \citenamefont {Kumar}, \citenamefont
  {Kiroriwal},\ and\ \citenamefont {Aziz}}]{Chaudhary:2023mfx}%
  \BibitemOpen
  \bibfield  {author} {\bibinfo {author} {\bibfnamefont {S.}~\bibnamefont
  {Chaudhary}}, \bibinfo {author} {\bibfnamefont {S.~K.}\ \bibnamefont
  {Maurya}}, \bibinfo {author} {\bibfnamefont {J.}~\bibnamefont {Kumar}},
  \bibinfo {author} {\bibfnamefont {S.}~\bibnamefont {Kiroriwal}}, \ and\
  \bibinfo {author} {\bibfnamefont {A.}~\bibnamefont {Aziz}},\ }\href {\doibase
  10.1016/j.cjph.2023.10.027} {\bibfield  {journal} {\bibinfo  {journal} {Chin.
  J. Phys.}\ }\textbf {\bibinfo {volume} {86}},\ \bibinfo {pages} {578}
  (\bibinfo {year} {2023})}\BibitemShut {NoStop}%
\bibitem [{\citenamefont {De~Falco}\ \emph {et~al.}(2021)\citenamefont
  {De~Falco}, \citenamefont {Battista}, \citenamefont {Capozziello},\ and\
  \citenamefont {De~Laurentis}}]{DeFalco:2021ksd}%
  \BibitemOpen
  \bibfield  {author} {\bibinfo {author} {\bibfnamefont {V.}~\bibnamefont
  {De~Falco}}, \bibinfo {author} {\bibfnamefont {E.}~\bibnamefont {Battista}},
  \bibinfo {author} {\bibfnamefont {S.}~\bibnamefont {Capozziello}}, \ and\
  \bibinfo {author} {\bibfnamefont {M.}~\bibnamefont {De~Laurentis}},\ }\href
  {\doibase 10.1140/epjc/s10052-021-08958-4} {\bibfield  {journal} {\bibinfo
  {journal} {Eur. Phys. J. C}\ }\textbf {\bibinfo {volume} {81}},\ \bibinfo
  {pages} {157} (\bibinfo {year} {2021})},\ \Eprint
  {http://arxiv.org/abs/2102.01123} {arXiv:2102.01123 [gr-qc]} \BibitemShut
  {NoStop}%
\bibitem [{\citenamefont {Malik}\ \emph {et~al.}(2023)\citenamefont {Malik},
  \citenamefont {Ashraf}, \citenamefont {Mofarreh}, \citenamefont {Ali},\ and\
  \citenamefont {Shoaib}}]{Malik:2023rov}%
  \BibitemOpen
  \bibfield  {author} {\bibinfo {author} {\bibfnamefont {A.}~\bibnamefont
  {Malik}}, \bibinfo {author} {\bibfnamefont {A.}~\bibnamefont {Ashraf}},
  \bibinfo {author} {\bibfnamefont {F.}~\bibnamefont {Mofarreh}}, \bibinfo
  {author} {\bibfnamefont {A.}~\bibnamefont {Ali}}, \ and\ \bibinfo {author}
  {\bibfnamefont {M.}~\bibnamefont {Shoaib}},\ }\href {\doibase
  10.1142/S0219887823501451} {\bibfield  {journal} {\bibinfo  {journal} {Int.
  J. Geom. Meth. Mod. Phys.}\ }\textbf {\bibinfo {volume} {20}},\ \bibinfo
  {pages} {2350145} (\bibinfo {year} {2023})}\BibitemShut {NoStop}%
\bibitem [{\citenamefont {Errehymy}(2024)}]{Errehymy:2024yey}%
  \BibitemOpen
  \bibfield  {author} {\bibinfo {author} {\bibfnamefont {A.}~\bibnamefont
  {Errehymy}},\ }\href {\doibase 10.1016/j.dark.2024.101438} {\bibfield
  {journal} {\bibinfo  {journal} {Phys. Dark Univ.}\ }\textbf {\bibinfo
  {volume} {44}},\ \bibinfo {pages} {101438} (\bibinfo {year}
  {2024})}\BibitemShut {NoStop}%
\bibitem [{\citenamefont {Mishra}\ \emph {et~al.}(2021)\citenamefont {Mishra},
  \citenamefont {Agrawal}, \citenamefont {Tripathy},\ and\ \citenamefont
  {Ray}}]{Mishra:2021ato}%
  \BibitemOpen
  \bibfield  {author} {\bibinfo {author} {\bibfnamefont {B.}~\bibnamefont
  {Mishra}}, \bibinfo {author} {\bibfnamefont {A.~S.}\ \bibnamefont {Agrawal}},
  \bibinfo {author} {\bibfnamefont {S.~K.}\ \bibnamefont {Tripathy}}, \ and\
  \bibinfo {author} {\bibfnamefont {S.}~\bibnamefont {Ray}},\ }\href {\doibase
  10.1142/S0218271821500619} {\bibfield  {journal} {\bibinfo  {journal} {Int.
  J. Mod. Phys. D}\ }\textbf {\bibinfo {volume} {30}},\ \bibinfo {pages}
  {2150061} (\bibinfo {year} {2021})},\ \Eprint
  {http://arxiv.org/abs/2104.05440} {arXiv:2104.05440 [gr-qc]} \BibitemShut
  {NoStop}%
\bibitem [{\citenamefont {Hussain}\ and\ \citenamefont
  {Mustafa}(2022)}]{Hussain:2022lxb}%
  \BibitemOpen
  \bibfield  {author} {\bibinfo {author} {\bibfnamefont {I.}~\bibnamefont
  {Hussain}}\ and\ \bibinfo {author} {\bibfnamefont {G.}~\bibnamefont
  {Mustafa}},\ }\href {\doibase 10.1142/S0219887822500748} {\bibfield
  {journal} {\bibinfo  {journal} {Int. J. Geom. Meth. Mod. Phys.}\ }\textbf
  {\bibinfo {volume} {19}},\ \bibinfo {pages} {2250074} (\bibinfo {year}
  {2022})}\BibitemShut {NoStop}%
\bibitem [{\citenamefont {Eiroa}\ and\ \citenamefont
  {Figueroa~Aguirre}(2012)}]{Eiroa:2012nv}%
  \BibitemOpen
  \bibfield  {author} {\bibinfo {author} {\bibfnamefont {E.~F.}\ \bibnamefont
  {Eiroa}}\ and\ \bibinfo {author} {\bibfnamefont {G.}~\bibnamefont
  {Figueroa~Aguirre}},\ }\href {\doibase 10.1140/epjc/s10052-012-2240-6}
  {\bibfield  {journal} {\bibinfo  {journal} {Eur. Phys. J. C}\ }\textbf
  {\bibinfo {volume} {72}},\ \bibinfo {pages} {2240} (\bibinfo {year}
  {2012})},\ \Eprint {http://arxiv.org/abs/1205.2685} {arXiv:1205.2685 [gr-qc]}
  \BibitemShut {NoStop}%
\bibitem [{\citenamefont {Harada}\ \emph {et~al.}(2009)\citenamefont {Harada},
  \citenamefont {Nakao},\ and\ \citenamefont {Nolan}}]{Harada:2008rx}%
  \BibitemOpen
  \bibfield  {author} {\bibinfo {author} {\bibfnamefont {T.}~\bibnamefont
  {Harada}}, \bibinfo {author} {\bibfnamefont {K.-i.}\ \bibnamefont {Nakao}}, \
  and\ \bibinfo {author} {\bibfnamefont {B.~C.}\ \bibnamefont {Nolan}},\ }\href
  {\doibase 10.1103/PhysRevD.80.024025} {\bibfield  {journal} {\bibinfo
  {journal} {Phys. Rev. D}\ }\textbf {\bibinfo {volume} {80}},\ \bibinfo
  {pages} {024025} (\bibinfo {year} {2009})},\ \bibinfo {note} {[Erratum:
  Phys.Rev.D 80, 109903 (2009)]},\ \Eprint {http://arxiv.org/abs/0812.3462}
  {arXiv:0812.3462 [gr-qc]} \BibitemShut {NoStop}%
\bibitem [{\citenamefont {Bronnikov}\ and\ \citenamefont
  {Galiakhmetov}(2015)}]{Bronnikov:2015pha}%
  \BibitemOpen
  \bibfield  {author} {\bibinfo {author} {\bibfnamefont {K.~A.}\ \bibnamefont
  {Bronnikov}}\ and\ \bibinfo {author} {\bibfnamefont {A.~M.}\ \bibnamefont
  {Galiakhmetov}},\ }\href {\doibase 10.1134/S0202289315040027} {\bibfield
  {journal} {\bibinfo  {journal} {Grav. Cosmol.}\ }\textbf {\bibinfo {volume}
  {21}},\ \bibinfo {pages} {283} (\bibinfo {year} {2015})},\ \Eprint
  {http://arxiv.org/abs/1508.01114} {arXiv:1508.01114 [gr-qc]} \BibitemShut
  {NoStop}%
\bibitem [{\citenamefont {Nashed}(2010)}]{Nashed:2009hn}%
  \BibitemOpen
  \bibfield  {author} {\bibinfo {author} {\bibfnamefont {G.~G.~L.}\
  \bibnamefont {Nashed}},\ }\href {\doibase 10.1088/1674-1056/19/2/020401}
  {\bibfield  {journal} {\bibinfo  {journal} {Chin. Phys. B}\ }\textbf
  {\bibinfo {volume} {19}},\ \bibinfo {pages} {020401} (\bibinfo {year}
  {2010})},\ \Eprint {http://arxiv.org/abs/0910.5124} {arXiv:0910.5124 [gr-qc]}
  \BibitemShut {NoStop}%
\bibitem [{\citenamefont {Bronnikov}\ and\ \citenamefont
  {Galiakhmetov}(2016)}]{Bronnikov:2016xvj}%
  \BibitemOpen
  \bibfield  {author} {\bibinfo {author} {\bibfnamefont {K.~A.}\ \bibnamefont
  {Bronnikov}}\ and\ \bibinfo {author} {\bibfnamefont {A.~M.}\ \bibnamefont
  {Galiakhmetov}},\ }\href {\doibase 10.1103/PhysRevD.94.124006} {\bibfield
  {journal} {\bibinfo  {journal} {Phys. Rev. D}\ }\textbf {\bibinfo {volume}
  {94}},\ \bibinfo {pages} {124006} (\bibinfo {year} {2016})},\ \Eprint
  {http://arxiv.org/abs/1607.07791} {arXiv:1607.07791 [gr-qc]} \BibitemShut
  {NoStop}%
\bibitem [{\citenamefont {El~Hanafy}\ and\ \citenamefont
  {Nashed}(2016)}]{ElHanafy:2014efn}%
  \BibitemOpen
  \bibfield  {author} {\bibinfo {author} {\bibfnamefont {W.}~\bibnamefont
  {El~Hanafy}}\ and\ \bibinfo {author} {\bibfnamefont {G.~L.}\ \bibnamefont
  {Nashed}},\ }\href {\doibase 10.1007/s10509-016-2786-0} {\bibfield  {journal}
  {\bibinfo  {journal} {Astrophys. Space Sci.}\ }\textbf {\bibinfo {volume}
  {361}},\ \bibinfo {pages} {197} (\bibinfo {year} {2016})},\ \Eprint
  {http://arxiv.org/abs/1410.2467} {arXiv:1410.2467 [hep-th]} \BibitemShut
  {NoStop}%
\bibitem [{\citenamefont {Nashed}(2006)}]{Nashed:2005kn}%
  \BibitemOpen
  \bibfield  {author} {\bibinfo {author} {\bibfnamefont {G.~G.~L.}\
  \bibnamefont {Nashed}},\ }\href {\doibase 10.1142/S0217751X06031478}
  {\bibfield  {journal} {\bibinfo  {journal} {Int. J. Mod. Phys. A}\ }\textbf
  {\bibinfo {volume} {21}},\ \bibinfo {pages} {3181} (\bibinfo {year}
  {2006})},\ \Eprint {http://arxiv.org/abs/gr-qc/0501002} {arXiv:gr-qc/0501002}
  \BibitemShut {NoStop}%
\bibitem [{\citenamefont {Mehdizadeh}\ and\ \citenamefont
  {Ziaie}(2017)}]{Mehdizadeh:2017tcf}%
  \BibitemOpen
  \bibfield  {author} {\bibinfo {author} {\bibfnamefont {M.~R.}\ \bibnamefont
  {Mehdizadeh}}\ and\ \bibinfo {author} {\bibfnamefont {A.~H.}\ \bibnamefont
  {Ziaie}},\ }\href {\doibase 10.1103/PhysRevD.95.064049} {\bibfield  {journal}
  {\bibinfo  {journal} {Phys. Rev. D}\ }\textbf {\bibinfo {volume} {95}},\
  \bibinfo {pages} {064049} (\bibinfo {year} {2017})},\ \Eprint
  {http://arxiv.org/abs/1704.06923} {arXiv:1704.06923 [gr-qc]} \BibitemShut
  {NoStop}%
\bibitem [{\citenamefont {La~Camera}(2003)}]{LaCamera:2003zd}%
  \BibitemOpen
  \bibfield  {author} {\bibinfo {author} {\bibfnamefont {M.}~\bibnamefont
  {La~Camera}},\ }\href {\doibase 10.1016/j.physletb.2003.08.042} {\bibfield
  {journal} {\bibinfo  {journal} {Phys. Lett. B}\ }\textbf {\bibinfo {volume}
  {573}},\ \bibinfo {pages} {27} (\bibinfo {year} {2003})},\ \Eprint
  {http://arxiv.org/abs/gr-qc/0306017} {arXiv:gr-qc/0306017} \BibitemShut
  {NoStop}%
\bibitem [{\citenamefont {Parsaei}\ and\ \citenamefont
  {Riazi}(2020)}]{Parsaei:2020hke}%
  \BibitemOpen
  \bibfield  {author} {\bibinfo {author} {\bibfnamefont {F.}~\bibnamefont
  {Parsaei}}\ and\ \bibinfo {author} {\bibfnamefont {N.}~\bibnamefont
  {Riazi}},\ }\href {\doibase 10.1103/PhysRevD.102.044003} {\bibfield
  {journal} {\bibinfo  {journal} {Phys. Rev. D}\ }\textbf {\bibinfo {volume}
  {102}},\ \bibinfo {pages} {044003} (\bibinfo {year} {2020})},\ \Eprint
  {http://arxiv.org/abs/2004.01750} {arXiv:2004.01750 [gr-qc]} \BibitemShut
  {NoStop}%
\bibitem [{\citenamefont {Kar}\ \emph {et~al.}(2015)\citenamefont {Kar},
  \citenamefont {Lahiri},\ and\ \citenamefont {SenGupta}}]{Kar:2015lma}%
  \BibitemOpen
  \bibfield  {author} {\bibinfo {author} {\bibfnamefont {S.}~\bibnamefont
  {Kar}}, \bibinfo {author} {\bibfnamefont {S.}~\bibnamefont {Lahiri}}, \ and\
  \bibinfo {author} {\bibfnamefont {S.}~\bibnamefont {SenGupta}},\ }\href
  {\doibase 10.1016/j.physletb.2015.09.039} {\bibfield  {journal} {\bibinfo
  {journal} {Phys. Lett. B}\ }\textbf {\bibinfo {volume} {750}},\ \bibinfo
  {pages} {319} (\bibinfo {year} {2015})},\ \Eprint
  {http://arxiv.org/abs/1505.06831} {arXiv:1505.06831 [gr-qc]} \BibitemShut
  {NoStop}%
\bibitem [{\citenamefont {Javed}\ \emph {et~al.}(2024)\citenamefont {Javed},
  \citenamefont {Waseem}, \citenamefont {Mustafa},\ and\ \citenamefont
  {G{\"u}dekli}}]{Javed:2023jqk}%
  \BibitemOpen
  \bibfield  {author} {\bibinfo {author} {\bibfnamefont {F.}~\bibnamefont
  {Javed}}, \bibinfo {author} {\bibfnamefont {A.}~\bibnamefont {Waseem}},
  \bibinfo {author} {\bibfnamefont {G.}~\bibnamefont {Mustafa}}, \ and\
  \bibinfo {author} {\bibfnamefont {E.}~\bibnamefont {G{\"u}dekli}},\ }\href
  {\doibase 10.1142/S0219887824500610} {\bibfield  {journal} {\bibinfo
  {journal} {Int. J. Geom. Meth. Mod. Phys.}\ }\textbf {\bibinfo {volume}
  {21}},\ \bibinfo {pages} {2450061} (\bibinfo {year} {2024})}\BibitemShut
  {NoStop}%
\bibitem [{\citenamefont {Eiroa}\ and\ \citenamefont
  {Simeone}(2007)}]{Eiroa:2007qz}%
  \BibitemOpen
  \bibfield  {author} {\bibinfo {author} {\bibfnamefont {E.~F.}\ \bibnamefont
  {Eiroa}}\ and\ \bibinfo {author} {\bibfnamefont {C.}~\bibnamefont
  {Simeone}},\ }\href {\doibase 10.1103/PhysRevD.76.024021} {\bibfield
  {journal} {\bibinfo  {journal} {Phys. Rev. D}\ }\textbf {\bibinfo {volume}
  {76}},\ \bibinfo {pages} {024021} (\bibinfo {year} {2007})},\ \Eprint
  {http://arxiv.org/abs/0704.1136} {arXiv:0704.1136 [gr-qc]} \BibitemShut
  {NoStop}%
\bibitem [{\citenamefont {Sharif}\ and\ \citenamefont
  {Yousaf}(2014)}]{Sharif:2014xwa}%
  \BibitemOpen
  \bibfield  {author} {\bibinfo {author} {\bibfnamefont {M.}~\bibnamefont
  {Sharif}}\ and\ \bibinfo {author} {\bibfnamefont {Z.}~\bibnamefont
  {Yousaf}},\ }\href {\doibase 10.1007/s10509-014-1836-8} {\bibfield  {journal}
  {\bibinfo  {journal} {Astrophys. Space Sci.}\ }\textbf {\bibinfo {volume}
  {351}},\ \bibinfo {pages} {351} (\bibinfo {year} {2014})},\ \Eprint
  {http://arxiv.org/abs/1410.4757} {arXiv:1410.4757 [physics.gen-ph]}
  \BibitemShut {NoStop}%
\bibitem [{\citenamefont {Nashed}(2015)}]{Nashed:2015pga}%
  \BibitemOpen
  \bibfield  {author} {\bibinfo {author} {\bibfnamefont {G.~G.~L.}\
  \bibnamefont {Nashed}},\ }\href {\doibase 10.1140/epjp/i2015-15124-3}
  {\bibfield  {journal} {\bibinfo  {journal} {Eur. Phys. J. Plus}\ }\textbf
  {\bibinfo {volume} {130}},\ \bibinfo {pages} {124} (\bibinfo {year}
  {2015})}\BibitemShut {NoStop}%
\bibitem [{\citenamefont {Lobo}(2006)}]{Lobo:2005vc}%
  \BibitemOpen
  \bibfield  {author} {\bibinfo {author} {\bibfnamefont {F.~S.~N.}\
  \bibnamefont {Lobo}},\ }\href {\doibase 10.1103/PhysRevD.73.064028}
  {\bibfield  {journal} {\bibinfo  {journal} {Phys. Rev. D}\ }\textbf {\bibinfo
  {volume} {73}},\ \bibinfo {pages} {064028} (\bibinfo {year} {2006})},\
  \Eprint {http://arxiv.org/abs/gr-qc/0511003} {arXiv:gr-qc/0511003}
  \BibitemShut {NoStop}%
\bibitem [{\citenamefont {Esmakhanova}\ \emph {et~al.}(2012)\citenamefont
  {Esmakhanova}, \citenamefont {Myrzakulov}, \citenamefont {Nugmanova},\ and\
  \citenamefont {Myrzakulov}}]{Esmakhanova:2011ar}%
  \BibitemOpen
  \bibfield  {author} {\bibinfo {author} {\bibfnamefont {K.}~\bibnamefont
  {Esmakhanova}}, \bibinfo {author} {\bibfnamefont {Y.}~\bibnamefont
  {Myrzakulov}}, \bibinfo {author} {\bibfnamefont {G.}~\bibnamefont
  {Nugmanova}}, \ and\ \bibinfo {author} {\bibfnamefont {R.}~\bibnamefont
  {Myrzakulov}},\ }\href {\doibase 10.1007/s10773-011-0996-3} {\bibfield
  {journal} {\bibinfo  {journal} {Int. J. Theor. Phys.}\ }\textbf {\bibinfo
  {volume} {51}},\ \bibinfo {pages} {1204} (\bibinfo {year} {2012})},\ \Eprint
  {http://arxiv.org/abs/1102.4456} {arXiv:1102.4456 [physics.gen-ph]}
  \BibitemShut {NoStop}%
\bibitem [{\citenamefont {Javed}\ \emph {et~al.}(2023)\citenamefont {Javed},
  \citenamefont {Waseem},\ and\ \citenamefont {Almutairi}}]{Javed:2023coi}%
  \BibitemOpen
  \bibfield  {author} {\bibinfo {author} {\bibfnamefont {F.}~\bibnamefont
  {Javed}}, \bibinfo {author} {\bibfnamefont {A.}~\bibnamefont {Waseem}}, \
  and\ \bibinfo {author} {\bibfnamefont {B.}~\bibnamefont {Almutairi}},\ }\href
  {\doibase 10.1140/epjc/s10052-023-11990-1} {\bibfield  {journal} {\bibinfo
  {journal} {Eur. Phys. J. C}\ }\textbf {\bibinfo {volume} {83}},\ \bibinfo
  {pages} {811} (\bibinfo {year} {2023})}\BibitemShut {NoStop}%
\bibitem [{\citenamefont {Mustafa}\ \emph
  {et~al.}(2024{\natexlab{b}})\citenamefont {Mustafa}, \citenamefont {Javed},
  \citenamefont {Maurya},\ and\ \citenamefont {Ray}}]{Mustafa:2022obk}%
  \BibitemOpen
  \bibfield  {author} {\bibinfo {author} {\bibfnamefont {G.}~\bibnamefont
  {Mustafa}}, \bibinfo {author} {\bibfnamefont {F.}~\bibnamefont {Javed}},
  \bibinfo {author} {\bibfnamefont {S.~K.}\ \bibnamefont {Maurya}}, \ and\
  \bibinfo {author} {\bibfnamefont {S.}~\bibnamefont {Ray}},\ }\href {\doibase
  10.1016/j.cjph.2023.12.035} {\bibfield  {journal} {\bibinfo  {journal} {Chin.
  J. Phys.}\ }\textbf {\bibinfo {volume} {88}},\ \bibinfo {pages} {32}
  (\bibinfo {year} {2024}{\natexlab{b}})},\ \Eprint
  {http://arxiv.org/abs/2211.10778} {arXiv:2211.10778 [gr-qc]} \BibitemShut
  {NoStop}%
\bibitem [{\citenamefont {Sadiq}\ \emph {et~al.}(2024)\citenamefont {Sadiq},
  \citenamefont {Atif}, \citenamefont {Javed},\ and\ \citenamefont
  {Saleem}}]{Sadiq:2024srd}%
  \BibitemOpen
  \bibfield  {author} {\bibinfo {author} {\bibfnamefont {S.}~\bibnamefont
  {Sadiq}}, \bibinfo {author} {\bibfnamefont {A.}~\bibnamefont {Atif}},
  \bibinfo {author} {\bibfnamefont {F.}~\bibnamefont {Javed}}, \ and\ \bibinfo
  {author} {\bibfnamefont {R.}~\bibnamefont {Saleem}},\ }\href {\doibase
  10.1016/j.cjph.2024.06.004} {\bibfield  {journal} {\bibinfo  {journal} {Chin.
  J. Phys.}\ }\textbf {\bibinfo {volume} {90}},\ \bibinfo {pages} {594}
  (\bibinfo {year} {2024})}\BibitemShut {NoStop}%
\bibitem [{\citenamefont {Fatima}\ \emph {et~al.}(2024)\citenamefont {Fatima},
  \citenamefont {Javed}, \citenamefont {Waseem}, \citenamefont {Mustafa},\ and\
  \citenamefont {Tchier}}]{Fatima:2024vvt}%
  \BibitemOpen
  \bibfield  {author} {\bibinfo {author} {\bibfnamefont {G.}~\bibnamefont
  {Fatima}}, \bibinfo {author} {\bibfnamefont {F.}~\bibnamefont {Javed}},
  \bibinfo {author} {\bibfnamefont {A.}~\bibnamefont {Waseem}}, \bibinfo
  {author} {\bibfnamefont {G.}~\bibnamefont {Mustafa}}, \ and\ \bibinfo
  {author} {\bibfnamefont {F.}~\bibnamefont {Tchier}},\ }\href {\doibase
  10.1142/S0219887824501986} {\bibfield  {journal} {\bibinfo  {journal} {Int.
  J. Geom. Meth. Mod. Phys.}\ }\textbf {\bibinfo {volume} {21}},\ \bibinfo
  {pages} {2450198} (\bibinfo {year} {2024})}\BibitemShut {NoStop}%
\bibitem [{\citenamefont {Fatima}\ \emph {et~al.}(2025)\citenamefont {Fatima},
  \citenamefont {Javed}, \citenamefont {Mustafa},\ and\ \citenamefont
  {Tchier}}]{Fatima:2024dqh}%
  \BibitemOpen
  \bibfield  {author} {\bibinfo {author} {\bibfnamefont {G.}~\bibnamefont
  {Fatima}}, \bibinfo {author} {\bibfnamefont {F.}~\bibnamefont {Javed}},
  \bibinfo {author} {\bibfnamefont {G.}~\bibnamefont {Mustafa}}, \ and\
  \bibinfo {author} {\bibfnamefont {F.}~\bibnamefont {Tchier}},\ }\href
  {\doibase 10.1142/S0219887824502797} {\bibfield  {journal} {\bibinfo
  {journal} {Int. J. Geom. Meth. Mod. Phys.}\ }\textbf {\bibinfo {volume}
  {22}},\ \bibinfo {pages} {2450279} (\bibinfo {year} {2025})}\BibitemShut
  {NoStop}%
\bibitem [{\citenamefont {Furey}\ and\ \citenamefont
  {DeBenedictis}(2005)}]{Furey:2004rq}%
  \BibitemOpen
  \bibfield  {author} {\bibinfo {author} {\bibfnamefont {N.}~\bibnamefont
  {Furey}}\ and\ \bibinfo {author} {\bibfnamefont {A.}~\bibnamefont
  {DeBenedictis}},\ }\href {\doibase 10.1088/0264-9381/22/2/005} {\bibfield
  {journal} {\bibinfo  {journal} {Class. Quant. Grav.}\ }\textbf {\bibinfo
  {volume} {22}},\ \bibinfo {pages} {313} (\bibinfo {year} {2005})},\ \Eprint
  {http://arxiv.org/abs/gr-qc/0410088} {arXiv:gr-qc/0410088} \BibitemShut
  {NoStop}%
\bibitem [{\citenamefont {Godani}\ and\ \citenamefont
  {Samanta}(2018)}]{Godani:2018blx}%
  \BibitemOpen
  \bibfield  {author} {\bibinfo {author} {\bibfnamefont {N.}~\bibnamefont
  {Godani}}\ and\ \bibinfo {author} {\bibfnamefont {G.~C.}\ \bibnamefont
  {Samanta}},\ }\href {\doibase 10.1142/S0218271819500391} {\bibfield
  {journal} {\bibinfo  {journal} {Int. J. Mod. Phys. D}\ }\textbf {\bibinfo
  {volume} {28}},\ \bibinfo {pages} {1950039} (\bibinfo {year} {2018})},\
  \Eprint {http://arxiv.org/abs/1809.00341} {arXiv:1809.00341 [gr-qc]}
  \BibitemShut {NoStop}%
\bibitem [{\citenamefont {Xu}\ \emph {et~al.}(2016)\citenamefont {Xu},
  \citenamefont {Harko},\ and\ \citenamefont {Liang}}]{Xu:2016rdf}%
  \BibitemOpen
  \bibfield  {author} {\bibinfo {author} {\bibfnamefont {M.-X.}\ \bibnamefont
  {Xu}}, \bibinfo {author} {\bibfnamefont {T.}~\bibnamefont {Harko}}, \ and\
  \bibinfo {author} {\bibfnamefont {S.-D.}\ \bibnamefont {Liang}},\ }\href
  {\doibase 10.1140/epjc/s10052-016-4303-6} {\bibfield  {journal} {\bibinfo
  {journal} {Eur. Phys. J. C}\ }\textbf {\bibinfo {volume} {76}},\ \bibinfo
  {pages} {449} (\bibinfo {year} {2016})},\ \Eprint
  {http://arxiv.org/abs/1608.00113} {arXiv:1608.00113 [gr-qc]} \BibitemShut
  {NoStop}%
\bibitem [{\citenamefont {Nashed}(2023{\natexlab{b}})}]{Nashed:2023uvk}%
  \BibitemOpen
  \bibfield  {author} {\bibinfo {author} {\bibfnamefont {G.~G.~L.}\
  \bibnamefont {Nashed}},\ }\href {\doibase 10.1140/epjc/s10052-023-11882-4}
  {\bibfield  {journal} {\bibinfo  {journal} {Eur. Phys. J. C}\ }\textbf
  {\bibinfo {volume} {83}},\ \bibinfo {pages} {698} (\bibinfo {year}
  {2023}{\natexlab{b}})},\ \Eprint {http://arxiv.org/abs/2308.08565}
  {arXiv:2308.08565 [gr-qc]} \BibitemShut {NoStop}%
\bibitem [{\citenamefont {Jeakel}\ \emph {et~al.}(2024)\citenamefont {Jeakel},
  \citenamefont {Pinheiro~da Silva},\ and\ \citenamefont
  {Velten}}]{Jeakel:2023hss}%
  \BibitemOpen
  \bibfield  {author} {\bibinfo {author} {\bibfnamefont {A.~P.}\ \bibnamefont
  {Jeakel}}, \bibinfo {author} {\bibfnamefont {J.}~\bibnamefont {Pinheiro~da
  Silva}}, \ and\ \bibinfo {author} {\bibfnamefont {H.}~\bibnamefont
  {Velten}},\ }\href {\doibase 10.1016/j.dark.2023.101401} {\bibfield
  {journal} {\bibinfo  {journal} {Phys. Dark Univ.}\ }\textbf {\bibinfo
  {volume} {43}},\ \bibinfo {pages} {101401} (\bibinfo {year} {2024})},\
  \Eprint {http://arxiv.org/abs/2303.15208} {arXiv:2303.15208 [astro-ph.CO]}
  \BibitemShut {NoStop}%
\bibitem [{\citenamefont {Khatri}\ \emph {et~al.}(2024)\citenamefont {Khatri},
  \citenamefont {Chhakchhuak},\ and\ \citenamefont
  {Lalchhuangliana}}]{Khatri:2024fef}%
  \BibitemOpen
  \bibfield  {author} {\bibinfo {author} {\bibfnamefont {M.}~\bibnamefont
  {Khatri}}, \bibinfo {author} {\bibfnamefont {Z.}~\bibnamefont {Chhakchhuak}},
  \ and\ \bibinfo {author} {\bibfnamefont {A.}~\bibnamefont
  {Lalchhuangliana}},\ }\href {\doibase 10.1016/j.aop.2024.169788} {\bibfield
  {journal} {\bibinfo  {journal} {Annals Phys.}\ }\textbf {\bibinfo {volume}
  {470}},\ \bibinfo {pages} {169788} (\bibinfo {year} {2024})}\BibitemShut
  {NoStop}%
\bibitem [{\citenamefont {Rajabi}\ and\ \citenamefont
  {Nozari}(2017)}]{Rajabi:2017alf}%
  \BibitemOpen
  \bibfield  {author} {\bibinfo {author} {\bibfnamefont {F.}~\bibnamefont
  {Rajabi}}\ and\ \bibinfo {author} {\bibfnamefont {K.}~\bibnamefont
  {Nozari}},\ }\href {\doibase 10.1103/PhysRevD.96.084061} {\bibfield
  {journal} {\bibinfo  {journal} {Phys. Rev. D}\ }\textbf {\bibinfo {volume}
  {96}},\ \bibinfo {pages} {084061} (\bibinfo {year} {2017})},\ \Eprint
  {http://arxiv.org/abs/1710.01910} {arXiv:1710.01910 [gr-qc]} \BibitemShut
  {NoStop}%
\bibitem [{\citenamefont {Zwicky}(1937)}]{Zwicky:1937zza}%
  \BibitemOpen
  \bibfield  {author} {\bibinfo {author} {\bibfnamefont {F.}~\bibnamefont
  {Zwicky}},\ }\href {\doibase 10.1086/143864} {\bibfield  {journal} {\bibinfo
  {journal} {Astrophys. J.}\ }\textbf {\bibinfo {volume} {86}},\ \bibinfo
  {pages} {217} (\bibinfo {year} {1937})}\BibitemShut {NoStop}%
\bibitem [{\citenamefont {Trujillo-Gomez}\ \emph {et~al.}(2011)\citenamefont
  {Trujillo-Gomez}, \citenamefont {Klypin}, \citenamefont {Primack},\ and\
  \citenamefont {Romanowsky}}]{Trujillo-Gomez:2010jbn}%
  \BibitemOpen
  \bibfield  {author} {\bibinfo {author} {\bibfnamefont {S.}~\bibnamefont
  {Trujillo-Gomez}}, \bibinfo {author} {\bibfnamefont {A.}~\bibnamefont
  {Klypin}}, \bibinfo {author} {\bibfnamefont {J.}~\bibnamefont {Primack}}, \
  and\ \bibinfo {author} {\bibfnamefont {A.~J.}\ \bibnamefont {Romanowsky}},\
  }\href {\doibase 10.1088/0004-637X/742/1/16} {\bibfield  {journal} {\bibinfo
  {journal} {Astrophys. J.}\ }\textbf {\bibinfo {volume} {742}},\ \bibinfo
  {pages} {16} (\bibinfo {year} {2011})},\ \Eprint
  {http://arxiv.org/abs/1005.1289} {arXiv:1005.1289 [astro-ph.CO]} \BibitemShut
  {NoStop}%
\bibitem [{\citenamefont {Bekenstein}(1973)}]{Bekenstein:1973ur}%
  \BibitemOpen
  \bibfield  {author} {\bibinfo {author} {\bibfnamefont {J.~D.}\ \bibnamefont
  {Bekenstein}},\ }\href {\doibase 10.1103/PhysRevD.7.2333} {\bibfield
  {journal} {\bibinfo  {journal} {Phys. Rev. D}\ }\textbf {\bibinfo {volume}
  {7}},\ \bibinfo {pages} {2333} (\bibinfo {year} {1973})}\BibitemShut
  {NoStop}%
\bibitem [{\citenamefont {Hawking}(1975)}]{Hawking:1975vcx}%
  \BibitemOpen
  \bibfield  {author} {\bibinfo {author} {\bibfnamefont {S.~W.}\ \bibnamefont
  {Hawking}},\ }\href {\doibase 10.1007/BF02345020} {\bibfield  {journal}
  {\bibinfo  {journal} {Commun. Math. Phys.}\ }\textbf {\bibinfo {volume}
  {43}},\ \bibinfo {pages} {199} (\bibinfo {year} {1975})},\ \bibinfo {note}
  {[Erratum: Commun.Math.Phys. 46, 206 (1976)]}\BibitemShut {NoStop}%
\bibitem [{\citenamefont {Nashed}(2002)}]{Nashed:2001im}%
  \BibitemOpen
  \bibfield  {author} {\bibinfo {author} {\bibfnamefont {G.~G.~L.}\
  \bibnamefont {Nashed}},\ }\href@noop {} {\bibfield  {journal} {\bibinfo
  {journal} {Nuovo Cim. B}\ }\textbf {\bibinfo {volume} {117}},\ \bibinfo
  {pages} {521} (\bibinfo {year} {2002})},\ \Eprint
  {http://arxiv.org/abs/gr-qc/0109017} {arXiv:gr-qc/0109017} \BibitemShut
  {NoStop}%
\bibitem [{\citenamefont {Cohen}\ \emph {et~al.}(1999)\citenamefont {Cohen},
  \citenamefont {Kaplan},\ and\ \citenamefont {Nelson}}]{Cohen:1998zx}%
  \BibitemOpen
  \bibfield  {author} {\bibinfo {author} {\bibfnamefont {A.~G.}\ \bibnamefont
  {Cohen}}, \bibinfo {author} {\bibfnamefont {D.~B.}\ \bibnamefont {Kaplan}}, \
  and\ \bibinfo {author} {\bibfnamefont {A.~E.}\ \bibnamefont {Nelson}},\
  }\href {\doibase 10.1103/PhysRevLett.82.4971} {\bibfield  {journal} {\bibinfo
   {journal} {Phys. Rev. Lett.}\ }\textbf {\bibinfo {volume} {82}},\ \bibinfo
  {pages} {4971} (\bibinfo {year} {1999})},\ \Eprint
  {http://arxiv.org/abs/hep-th/9803132} {arXiv:hep-th/9803132} \BibitemShut
  {NoStop}%
\bibitem [{\citenamefont {Chen}\ \emph {et~al.}(2023)\citenamefont {Chen},
  \citenamefont {Mustafa}, \citenamefont {Caliskan},\ and\ \citenamefont
  {G{\"u}dekli}}]{Chen:2023azy}%
  \BibitemOpen
  \bibfield  {author} {\bibinfo {author} {\bibfnamefont {R.-Y.}\ \bibnamefont
  {Chen}}, \bibinfo {author} {\bibfnamefont {G.}~\bibnamefont {Mustafa}},
  \bibinfo {author} {\bibfnamefont {A.}~\bibnamefont {Caliskan}}, \ and\
  \bibinfo {author} {\bibfnamefont {E.}~\bibnamefont {G{\"u}dekli}},\ }\href
  {\doibase 10.1142/S021988782350130X} {\bibfield  {journal} {\bibinfo
  {journal} {Int. J. Geom. Meth. Mod. Phys.}\ }\textbf {\bibinfo {volume}
  {20}},\ \bibinfo {pages} {2350130} (\bibinfo {year} {2023})}\BibitemShut
  {NoStop}%
\bibitem [{\citenamefont {Moradpour}\ \emph {et~al.}(2017)\citenamefont
  {Moradpour}, \citenamefont {Heydarzade}, \citenamefont {Darabi},\ and\
  \citenamefont {Salako}}]{Moradpour:2017shy}%
  \BibitemOpen
  \bibfield  {author} {\bibinfo {author} {\bibfnamefont {H.}~\bibnamefont
  {Moradpour}}, \bibinfo {author} {\bibfnamefont {Y.}~\bibnamefont
  {Heydarzade}}, \bibinfo {author} {\bibfnamefont {F.}~\bibnamefont {Darabi}},
  \ and\ \bibinfo {author} {\bibfnamefont {I.~G.}\ \bibnamefont {Salako}},\
  }\href {\doibase 10.1140/epjc/s10052-017-4811-z} {\bibfield  {journal}
  {\bibinfo  {journal} {Eur. Phys. J. C}\ }\textbf {\bibinfo {volume} {77}},\
  \bibinfo {pages} {259} (\bibinfo {year} {2017})},\ \Eprint
  {http://arxiv.org/abs/1704.02458} {arXiv:1704.02458 [gr-qc]} \BibitemShut
  {NoStop}%
\bibitem [{\citenamefont {Tsallis}(1988)}]{Tsallis:1987eu}%
  \BibitemOpen
  \bibfield  {author} {\bibinfo {author} {\bibfnamefont {C.}~\bibnamefont
  {Tsallis}},\ }\href {\doibase 10.1007/BF01016429} {\bibfield  {journal}
  {\bibinfo  {journal} {J. Statist. Phys.}\ }\textbf {\bibinfo {volume} {52}},\
  \bibinfo {pages} {479} (\bibinfo {year} {1988})}\BibitemShut {NoStop}%
\bibitem [{\citenamefont {Chaudhary}\ \emph {et~al.}(2025)\citenamefont
  {Chaudhary}, \citenamefont {Maurya}, \citenamefont {Kumar}, \citenamefont
  {Errehymy}, \citenamefont {Alessa},\ and\ \citenamefont
  {Abdel-Aty}}]{Chaudhary:2025wsd}%
  \BibitemOpen
  \bibfield  {author} {\bibinfo {author} {\bibfnamefont {S.}~\bibnamefont
  {Chaudhary}}, \bibinfo {author} {\bibfnamefont {S.~K.}\ \bibnamefont
  {Maurya}}, \bibinfo {author} {\bibfnamefont {J.}~\bibnamefont {Kumar}},
  \bibinfo {author} {\bibfnamefont {A.}~\bibnamefont {Errehymy}}, \bibinfo
  {author} {\bibfnamefont {N.}~\bibnamefont {Alessa}}, \ and\ \bibinfo {author}
  {\bibfnamefont {A.~H.}\ \bibnamefont {Abdel-Aty}},\ }\href {\doibase
  10.1140/epjc/s10052-025-14177-y} {\bibfield  {journal} {\bibinfo  {journal}
  {Eur. Phys. J. C}\ }\textbf {\bibinfo {volume} {85}},\ \bibinfo {pages} {478}
  (\bibinfo {year} {2025})},\ \bibinfo {note} {[Erratum: Eur.Phys.J.C 85, 612
  (2025)]}\BibitemShut {NoStop}%
\bibitem [{\citenamefont {Paul}\ \emph {et~al.}(2025)\citenamefont {Paul},
  \citenamefont {Maurya},\ and\ \citenamefont {Kumar}}]{Paul:2025vem}%
  \BibitemOpen
  \bibfield  {author} {\bibinfo {author} {\bibfnamefont {S.}~\bibnamefont
  {Paul}}, \bibinfo {author} {\bibfnamefont {S.~K.}\ \bibnamefont {Maurya}}, \
  and\ \bibinfo {author} {\bibfnamefont {J.}~\bibnamefont {Kumar}},\ }\href
  {\doibase 10.1016/j.nuclphysb.2025.116886} {\bibfield  {journal} {\bibinfo
  {journal} {Nucl. Phys. B}\ }\textbf {\bibinfo {volume} {1014}},\ \bibinfo
  {pages} {116886} (\bibinfo {year} {2025})}\BibitemShut {NoStop}%
\bibitem [{\citenamefont {Wu}\ \emph {et~al.}(2018)\citenamefont {Wu},
  \citenamefont {Li}, \citenamefont {Harko},\ and\ \citenamefont
  {Liang}}]{Wu:2018idg}%
  \BibitemOpen
  \bibfield  {author} {\bibinfo {author} {\bibfnamefont {J.}~\bibnamefont
  {Wu}}, \bibinfo {author} {\bibfnamefont {G.}~\bibnamefont {Li}}, \bibinfo
  {author} {\bibfnamefont {T.}~\bibnamefont {Harko}}, \ and\ \bibinfo {author}
  {\bibfnamefont {S.-D.}\ \bibnamefont {Liang}},\ }\href {\doibase
  10.1140/epjc/s10052-018-5923-9} {\bibfield  {journal} {\bibinfo  {journal}
  {Eur. Phys. J. C}\ }\textbf {\bibinfo {volume} {78}},\ \bibinfo {pages} {430}
  (\bibinfo {year} {2018})},\ \Eprint {http://arxiv.org/abs/1805.07419}
  {arXiv:1805.07419 [gr-qc]} \BibitemShut {NoStop}%
\bibitem [{\citenamefont {Barrientos~O.}\ and\ \citenamefont
  {Rubilar}(2014)}]{BarrientosO:2014mys}%
  \BibitemOpen
  \bibfield  {author} {\bibinfo {author} {\bibfnamefont {J.}~\bibnamefont
  {Barrientos~O.}}\ and\ \bibinfo {author} {\bibfnamefont {G.~F.}\ \bibnamefont
  {Rubilar}},\ }\href {\doibase 10.1103/PhysRevD.90.028501} {\bibfield
  {journal} {\bibinfo  {journal} {Phys. Rev. D}\ }\textbf {\bibinfo {volume}
  {90}},\ \bibinfo {pages} {028501} (\bibinfo {year} {2014})}\BibitemShut
  {NoStop}%
\bibitem [{\citenamefont {Pretel}\ \emph
  {et~al.}(2021{\natexlab{a}})\citenamefont {Pretel}, \citenamefont
  {Jor{\'a}s}, \citenamefont {Reis},\ and\ \citenamefont
  {Arba{\~n}il}}]{Pretel:2021kgl}%
  \BibitemOpen
  \bibfield  {author} {\bibinfo {author} {\bibfnamefont {J.~M.~Z.}\
  \bibnamefont {Pretel}}, \bibinfo {author} {\bibfnamefont {S.~E.}\
  \bibnamefont {Jor{\'a}s}}, \bibinfo {author} {\bibfnamefont {R.~R.~R.}\
  \bibnamefont {Reis}}, \ and\ \bibinfo {author} {\bibfnamefont {J.~D.~V.}\
  \bibnamefont {Arba{\~n}il}},\ }\href {\doibase 10.1088/1475-7516/2021/08/055}
  {\bibfield  {journal} {\bibinfo  {journal} {JCAP}\ }\textbf {\bibinfo
  {volume} {08}},\ \bibinfo {pages} {055} (\bibinfo {year}
  {2021}{\natexlab{a}})},\ \Eprint {http://arxiv.org/abs/2105.07573}
  {arXiv:2105.07573 [gr-qc]} \BibitemShut {NoStop}%
\bibitem [{\citenamefont {Hansraj}\ and\ \citenamefont
  {Banerjee}(2018)}]{Hansraj:2018jzb}%
  \BibitemOpen
  \bibfield  {author} {\bibinfo {author} {\bibfnamefont {S.}~\bibnamefont
  {Hansraj}}\ and\ \bibinfo {author} {\bibfnamefont {A.}~\bibnamefont
  {Banerjee}},\ }\href {\doibase 10.1103/PhysRevD.97.104020} {\bibfield
  {journal} {\bibinfo  {journal} {Phys. Rev. D}\ }\textbf {\bibinfo {volume}
  {97}},\ \bibinfo {pages} {104020} (\bibinfo {year} {2018})}\BibitemShut
  {NoStop}%
\bibitem [{\citenamefont {Bhar}\ \emph {et~al.}(2022)\citenamefont {Bhar},
  \citenamefont {Rej},\ and\ \citenamefont {Zubair}}]{Bhar:2021uqr}%
  \BibitemOpen
  \bibfield  {author} {\bibinfo {author} {\bibfnamefont {P.}~\bibnamefont
  {Bhar}}, \bibinfo {author} {\bibfnamefont {P.}~\bibnamefont {Rej}}, \ and\
  \bibinfo {author} {\bibfnamefont {M.}~\bibnamefont {Zubair}},\ }\href
  {\doibase 10.1016/j.cjph.2021.11.013} {\bibfield  {journal} {\bibinfo
  {journal} {Chin. J. Phys.}\ }\textbf {\bibinfo {volume} {77}},\ \bibinfo
  {pages} {2201} (\bibinfo {year} {2022})},\ \Eprint
  {http://arxiv.org/abs/2112.07581} {arXiv:2112.07581 [gr-qc]} \BibitemShut
  {NoStop}%
\bibitem [{\citenamefont {Pretel}\ \emph
  {et~al.}(2021{\natexlab{b}})\citenamefont {Pretel}, \citenamefont
  {Jor{\'a}s}, \citenamefont {Reis},\ and\ \citenamefont
  {Arba{\~n}il}}]{Pretel:2020oae}%
  \BibitemOpen
  \bibfield  {author} {\bibinfo {author} {\bibfnamefont {J.~M.~Z.}\
  \bibnamefont {Pretel}}, \bibinfo {author} {\bibfnamefont {S.~E.}\
  \bibnamefont {Jor{\'a}s}}, \bibinfo {author} {\bibfnamefont {R.~R.~R.}\
  \bibnamefont {Reis}}, \ and\ \bibinfo {author} {\bibfnamefont {J.~D.~V.}\
  \bibnamefont {Arba{\~n}il}},\ }\href {\doibase 10.1088/1475-7516/2021/04/064}
  {\bibfield  {journal} {\bibinfo  {journal} {JCAP}\ }\textbf {\bibinfo
  {volume} {04}},\ \bibinfo {pages} {064} (\bibinfo {year}
  {2021}{\natexlab{b}})},\ \Eprint {http://arxiv.org/abs/2012.03342}
  {arXiv:2012.03342 [gr-qc]} \BibitemShut {NoStop}%
\bibitem [{\citenamefont {Rosa}(2021)}]{Rosa:2021teg}%
  \BibitemOpen
  \bibfield  {author} {\bibinfo {author} {\bibfnamefont {J.~L.}\ \bibnamefont
  {Rosa}},\ }\href {\doibase 10.1103/PhysRevD.103.104069} {\bibfield  {journal}
  {\bibinfo  {journal} {Phys. Rev. D}\ }\textbf {\bibinfo {volume} {103}},\
  \bibinfo {pages} {104069} (\bibinfo {year} {2021})},\ \Eprint
  {http://arxiv.org/abs/2103.11698} {arXiv:2103.11698 [gr-qc]} \BibitemShut
  {NoStop}%
\bibitem [{\citenamefont {Rosa}\ and\ \citenamefont
  {Rubiera-Garcia}(2022)}]{Rosa:2022cen}%
  \BibitemOpen
  \bibfield  {author} {\bibinfo {author} {\bibfnamefont {J.~L.}\ \bibnamefont
  {Rosa}}\ and\ \bibinfo {author} {\bibfnamefont {D.}~\bibnamefont
  {Rubiera-Garcia}},\ }\href {\doibase 10.1103/PhysRevD.106.064007} {\bibfield
  {journal} {\bibinfo  {journal} {Phys. Rev. D}\ }\textbf {\bibinfo {volume}
  {106}},\ \bibinfo {pages} {064007} (\bibinfo {year} {2022})},\ \Eprint
  {http://arxiv.org/abs/2204.12944} {arXiv:2204.12944 [gr-qc]} \BibitemShut
  {NoStop}%
\bibitem [{\citenamefont {Morris}\ and\ \citenamefont
  {Thorne}(1988{\natexlab{b}})}]{M_S_Morris}%
  \BibitemOpen
  \bibfield  {author} {\bibinfo {author} {\bibfnamefont {M.~S.}\ \bibnamefont
  {Morris}}\ and\ \bibinfo {author} {\bibfnamefont {K.~S.}\ \bibnamefont
  {Thorne}},\ }\href@noop {} {\bibfield  {journal} {\bibinfo  {journal}
  {American Journal of Physics}\ }\textbf {\bibinfo {volume} {56}},\ \bibinfo
  {pages} {395} (\bibinfo {year} {1988}{\natexlab{b}})}\BibitemShut {NoStop}%
\bibitem [{\citenamefont {et~al.}(2022)}]{Tayde_1}%
  \BibitemOpen
  \bibfield  {author} {\bibinfo {author} {\bibfnamefont {M.~T.}\ \bibnamefont
  {et~al.}},\ }\href@noop {} {\bibfield  {journal} {\bibinfo  {journal}
  {Chinese Physics C}\ }\textbf {\bibinfo {volume} {46}},\ \bibinfo {pages}
  {115101} (\bibinfo {year} {2022})}\BibitemShut {NoStop}%
\bibitem [{\citenamefont {et~al.}(2019)}]{Y_Xu}%
  \BibitemOpen
  \bibfield  {author} {\bibinfo {author} {\bibfnamefont {Y.~X.}\ \bibnamefont
  {et~al.}},\ }\href@noop {} {\bibfield  {journal} {\bibinfo  {journal}
  {European Physical Journal C}\ }\textbf {\bibinfo {volume} {79}},\ \bibinfo
  {pages} {708} (\bibinfo {year} {2019})}\BibitemShut {NoStop}%
\bibitem [{\citenamefont {Avik}\ and\ \citenamefont {Loo}(2023)}]{Avik2023CQG}%
  \BibitemOpen
  \bibfield  {author} {\bibinfo {author} {\bibfnamefont {D.}~\bibnamefont
  {Avik}}\ and\ \bibinfo {author} {\bibfnamefont {T.-H.}\ \bibnamefont {Loo}},\
  }\href {\doibase 10.1088/1361-6382/accccf} {\bibfield  {journal} {\bibinfo
  {journal} {Classical and Quantum Gravity}\ }\textbf {\bibinfo {volume}
  {40}},\ \bibinfo {pages} {115007} (\bibinfo {year} {2023})}\BibitemShut
  {NoStop}%
\bibitem [{\citenamefont {Hu}\ and\ \citenamefont {Ling}(2006)}]{Hu:2006ar}%
  \BibitemOpen
  \bibfield  {author} {\bibinfo {author} {\bibfnamefont {B.}~\bibnamefont
  {Hu}}\ and\ \bibinfo {author} {\bibfnamefont {Y.}~\bibnamefont {Ling}},\
  }\href {\doibase 10.1103/PhysRevD.73.123510} {\bibfield  {journal} {\bibinfo
  {journal} {Phys. Rev. D}\ }\textbf {\bibinfo {volume} {73}},\ \bibinfo
  {pages} {123510} (\bibinfo {year} {2006})},\ \Eprint
  {http://arxiv.org/abs/hep-th/0601093} {arXiv:hep-th/0601093} \BibitemShut
  {NoStop}%
\bibitem [{\citenamefont {Myung}(2005)}]{Myung:2004ch}%
  \BibitemOpen
  \bibfield  {author} {\bibinfo {author} {\bibfnamefont {Y.~S.}\ \bibnamefont
  {Myung}},\ }\href {\doibase 10.1016/j.physletb.2005.02.006} {\bibfield
  {journal} {\bibinfo  {journal} {Phys. Lett. B}\ }\textbf {\bibinfo {volume}
  {610}},\ \bibinfo {pages} {18} (\bibinfo {year} {2005})},\ \Eprint
  {http://arxiv.org/abs/hep-th/0412224} {arXiv:hep-th/0412224} \BibitemShut
  {NoStop}%
\bibitem [{\citenamefont {Manoharan}\ \emph {et~al.}(2023)\citenamefont
  {Manoharan}, \citenamefont {Shaji},\ and\ \citenamefont
  {Mathew}}]{Manoharan:2022qll}%
  \BibitemOpen
  \bibfield  {author} {\bibinfo {author} {\bibfnamefont {M.~T.}\ \bibnamefont
  {Manoharan}}, \bibinfo {author} {\bibfnamefont {N.}~\bibnamefont {Shaji}}, \
  and\ \bibinfo {author} {\bibfnamefont {T.~K.}\ \bibnamefont {Mathew}},\
  }\href {\doibase 10.1140/epjc/s10052-023-11202-w} {\bibfield  {journal}
  {\bibinfo  {journal} {Eur. Phys. J. C}\ }\textbf {\bibinfo {volume} {83}},\
  \bibinfo {pages} {19} (\bibinfo {year} {2023})},\ \Eprint
  {http://arxiv.org/abs/2208.08736} {arXiv:2208.08736 [gr-qc]} \BibitemShut
  {NoStop}%
\bibitem [{\citenamefont {Elizalde}\ \emph {et~al.}(2020)\citenamefont
  {Elizalde}, \citenamefont {Khurshudyan}, \citenamefont {Odintsov},\ and\
  \citenamefont {Myrzakulov}}]{Elizalde:2020mfs}%
  \BibitemOpen
  \bibfield  {author} {\bibinfo {author} {\bibfnamefont {E.}~\bibnamefont
  {Elizalde}}, \bibinfo {author} {\bibfnamefont {M.}~\bibnamefont
  {Khurshudyan}}, \bibinfo {author} {\bibfnamefont {S.~D.}\ \bibnamefont
  {Odintsov}}, \ and\ \bibinfo {author} {\bibfnamefont {R.}~\bibnamefont
  {Myrzakulov}},\ }\href {\doibase 10.1103/PhysRevD.102.123501} {\bibfield
  {journal} {\bibinfo  {journal} {Phys. Rev. D}\ }\textbf {\bibinfo {volume}
  {102}},\ \bibinfo {pages} {123501} (\bibinfo {year} {2020})},\ \Eprint
  {http://arxiv.org/abs/2006.01879} {arXiv:2006.01879 [gr-qc]} \BibitemShut
  {NoStop}%
\bibitem [{\citenamefont {Moradpour}\ \emph {et~al.}(2018)\citenamefont
  {Moradpour}, \citenamefont {Sheykhi}, \citenamefont {Corda},\ and\
  \citenamefont {Salako}}]{Moradpour:2017fmq}%
  \BibitemOpen
  \bibfield  {author} {\bibinfo {author} {\bibfnamefont {H.}~\bibnamefont
  {Moradpour}}, \bibinfo {author} {\bibfnamefont {A.}~\bibnamefont {Sheykhi}},
  \bibinfo {author} {\bibfnamefont {C.}~\bibnamefont {Corda}}, \ and\ \bibinfo
  {author} {\bibfnamefont {I.~G.}\ \bibnamefont {Salako}},\ }\href {\doibase
  10.1016/j.physletb.2018.06.040} {\bibfield  {journal} {\bibinfo  {journal}
  {Phys. Lett. B}\ }\textbf {\bibinfo {volume} {783}},\ \bibinfo {pages} {82}
  (\bibinfo {year} {2018})},\ \Eprint {http://arxiv.org/abs/1711.10336}
  {arXiv:1711.10336 [physics.gen-ph]} \BibitemShut {NoStop}%
\bibitem [{\citenamefont {Godani}\ and\ \citenamefont
  {Samanta}(2021)}]{Godani2021}%
  \BibitemOpen
  \bibfield  {author} {\bibinfo {author} {\bibfnamefont {N.}~\bibnamefont
  {Godani}}\ and\ \bibinfo {author} {\bibfnamefont {G.~C.}\ \bibnamefont
  {Samanta}},\ }\href {\doibase 10.1016/j.aop.2021.168460} {\bibfield
  {journal} {\bibinfo  {journal} {Annals of Physics}\ }\textbf {\bibinfo
  {volume} {429}},\ \bibinfo {pages} {168460} (\bibinfo {year}
  {2021})}\BibitemShut {NoStop}%
\bibitem [{\citenamefont {Nandi}\ \emph {et~al.}(2006)\citenamefont {Nandi},
  \citenamefont {Zhang},\ and\ \citenamefont {Zakharov}}]{Nandi2006}%
  \BibitemOpen
  \bibfield  {author} {\bibinfo {author} {\bibfnamefont {K.~K.}\ \bibnamefont
  {Nandi}}, \bibinfo {author} {\bibfnamefont {Y.~Z.}\ \bibnamefont {Zhang}}, \
  and\ \bibinfo {author} {\bibfnamefont {A.~V.}\ \bibnamefont {Zakharov}},\
  }\href {\doibase 10.1103/PhysRevD.74.024020} {\bibfield  {journal} {\bibinfo
  {journal} {Physical Review D}\ }\textbf {\bibinfo {volume} {74}},\ \bibinfo
  {pages} {024020} (\bibinfo {year} {2006})}\BibitemShut {NoStop}%
\bibitem [{\citenamefont {Dey}\ and\ \citenamefont {Sen}(2008)}]{Dey2008}%
  \BibitemOpen
  \bibfield  {author} {\bibinfo {author} {\bibfnamefont {T.~K.}\ \bibnamefont
  {Dey}}\ and\ \bibinfo {author} {\bibfnamefont {S.}~\bibnamefont {Sen}},\
  }\href {\doibase 10.1142/S0217732308026898} {\bibfield  {journal} {\bibinfo
  {journal} {Modern Physics Letters A}\ }\textbf {\bibinfo {volume} {23}},\
  \bibinfo {pages} {953} (\bibinfo {year} {2008})}\BibitemShut {NoStop}%
\bibitem [{\citenamefont {Tsukamoto}(2016)}]{Tsukamoto2016}%
  \BibitemOpen
  \bibfield  {author} {\bibinfo {author} {\bibfnamefont {N.}~\bibnamefont
  {Tsukamoto}},\ }\href {\doibase 10.1103/PhysRevD.94.124001} {\bibfield
  {journal} {\bibinfo  {journal} {Physical Review D}\ }\textbf {\bibinfo
  {volume} {94}},\ \bibinfo {pages} {124001} (\bibinfo {year}
  {2016})}\BibitemShut {NoStop}%
\bibitem [{\citenamefont {Virbhadra}\ and\ \citenamefont
  {Ellis}(2000)}]{Virbhadra2000}%
  \BibitemOpen
  \bibfield  {author} {\bibinfo {author} {\bibfnamefont {K.~S.}\ \bibnamefont
  {Virbhadra}}\ and\ \bibinfo {author} {\bibfnamefont {G.~F.~R.}\ \bibnamefont
  {Ellis}},\ }\href {\doibase 10.1103/PhysRevD.62.084003} {\bibfield  {journal}
  {\bibinfo  {journal} {Physical Review D}\ }\textbf {\bibinfo {volume} {62}},\
  \bibinfo {pages} {084003} (\bibinfo {year} {2000})}\BibitemShut {NoStop}%
\bibitem [{\citenamefont {Virbhadra}\ and\ \citenamefont
  {Ellis}(2002)}]{Virbhadra2002}%
  \BibitemOpen
  \bibfield  {author} {\bibinfo {author} {\bibfnamefont {K.~S.}\ \bibnamefont
  {Virbhadra}}\ and\ \bibinfo {author} {\bibfnamefont {G.~F.~R.}\ \bibnamefont
  {Ellis}},\ }\href {\doibase 10.1103/PhysRevD.65.103004} {\bibfield  {journal}
  {\bibinfo  {journal} {Physical Review D}\ }\textbf {\bibinfo {volume} {65}},\
  \bibinfo {pages} {103004} (\bibinfo {year} {2002})}\BibitemShut {NoStop}%
\bibitem [{\citenamefont {Bozza}(2010)}]{Bozza2010}%
  \BibitemOpen
  \bibfield  {author} {\bibinfo {author} {\bibfnamefont {V.}~\bibnamefont
  {Bozza}},\ }\href {\doibase 10.1007/s10714-010-0988-2} {\bibfield  {journal}
  {\bibinfo  {journal} {General Relativity and Gravitation}\ }\textbf {\bibinfo
  {volume} {42}},\ \bibinfo {pages} {2269} (\bibinfo {year}
  {2010})}\BibitemShut {NoStop}%
\bibitem [{\citenamefont {Bozza}(2002{\natexlab{a}})}]{Bozza2002}%
  \BibitemOpen
  \bibfield  {author} {\bibinfo {author} {\bibfnamefont {V.}~\bibnamefont
  {Bozza}},\ }\href {\doibase 10.1103/PhysRevD.66.103001} {\bibfield  {journal}
  {\bibinfo  {journal} {Physical Review D}\ }\textbf {\bibinfo {volume} {66}},\
  \bibinfo {pages} {103001} (\bibinfo {year} {2002}{\natexlab{a}})}\BibitemShut
  {NoStop}%
\bibitem [{\citenamefont {Claudel}\ \emph {et~al.}(2001)\citenamefont
  {Claudel}, \citenamefont {Virbhadra},\ and\ \citenamefont
  {Ellis}}]{Claudel:2000yi}%
  \BibitemOpen
  \bibfield  {author} {\bibinfo {author} {\bibfnamefont {C.-M.}\ \bibnamefont
  {Claudel}}, \bibinfo {author} {\bibfnamefont {K.~S.}\ \bibnamefont
  {Virbhadra}}, \ and\ \bibinfo {author} {\bibfnamefont {G.~F.~R.}\
  \bibnamefont {Ellis}},\ }\href {\doibase 10.1063/1.1308507} {\bibfield
  {journal} {\bibinfo  {journal} {J. Math. Phys.}\ }\textbf {\bibinfo {volume}
  {42}},\ \bibinfo {pages} {818} (\bibinfo {year} {2001})},\ \Eprint
  {http://arxiv.org/abs/gr-qc/0005050} {arXiv:gr-qc/0005050} \BibitemShut
  {NoStop}%
\bibitem [{\citenamefont {Shaikh}\ \emph {et~al.}(2019)\citenamefont {Shaikh},
  \citenamefont {Banerjee}, \citenamefont {Paul},\ and\ \citenamefont
  {Sarkar}}]{Shaikh2019}%
  \BibitemOpen
  \bibfield  {author} {\bibinfo {author} {\bibfnamefont {R.}~\bibnamefont
  {Shaikh}}, \bibinfo {author} {\bibfnamefont {P.}~\bibnamefont {Banerjee}},
  \bibinfo {author} {\bibfnamefont {S.}~\bibnamefont {Paul}}, \ and\ \bibinfo
  {author} {\bibfnamefont {T.}~\bibnamefont {Sarkar}},\ }\href {\doibase
  10.1088/1475-7516/2019/07/028} {\bibfield  {journal} {\bibinfo  {journal}
  {Journal of Cosmology and Astroparticle Physics}\ }\textbf {\bibinfo {volume}
  {07}},\ \bibinfo {pages} {028} (\bibinfo {year} {2019})}\BibitemShut
  {NoStop}%
\bibitem [{\citenamefont {Weinberg}(1972)}]{Weinberg1972}%
  \BibitemOpen
  \bibfield  {author} {\bibinfo {author} {\bibfnamefont {S.}~\bibnamefont
  {Weinberg}},\ }\href@noop {} {\emph {\bibinfo {title} {Gravitation and
  Cosmology: Principles and Applications of the General Theory of
  Relativity}}}\ (\bibinfo  {publisher} {John Wiley \& Sons},\ \bibinfo
  {address} {New York},\ \bibinfo {year} {1972})\BibitemShut {NoStop}%
\bibitem [{\citenamefont {Bozza}(2002{\natexlab{b}})}]{Bozza:2002zj}%
  \BibitemOpen
  \bibfield  {author} {\bibinfo {author} {\bibfnamefont {V.}~\bibnamefont
  {Bozza}},\ }\href {\doibase 10.1103/PhysRevD.66.103001} {\bibfield  {journal}
  {\bibinfo  {journal} {Phys. Rev. D}\ }\textbf {\bibinfo {volume} {66}},\
  \bibinfo {pages} {103001} (\bibinfo {year} {2002}{\natexlab{b}})},\ \Eprint
  {http://arxiv.org/abs/gr-qc/0208075} {arXiv:gr-qc/0208075} \BibitemShut
  {NoStop}%
\bibitem [{\citenamefont {Bozza}(2003)}]{Bozza:2002af}%
  \BibitemOpen
  \bibfield  {author} {\bibinfo {author} {\bibfnamefont {V.}~\bibnamefont
  {Bozza}},\ }\href {\doibase 10.1103/PhysRevD.67.103006} {\bibfield  {journal}
  {\bibinfo  {journal} {Phys. Rev. D}\ }\textbf {\bibinfo {volume} {67}},\
  \bibinfo {pages} {103006} (\bibinfo {year} {2003})},\ \Eprint
  {http://arxiv.org/abs/gr-qc/0210109} {arXiv:gr-qc/0210109} \BibitemShut
  {NoStop}%
\bibitem [{\citenamefont {Visser}\ \emph {et~al.}(2003)\citenamefont {Visser},
  \citenamefont {Kar},\ and\ \citenamefont {Dadhich}}]{Visser:2003yf}%
  \BibitemOpen
  \bibfield  {author} {\bibinfo {author} {\bibfnamefont {M.}~\bibnamefont
  {Visser}}, \bibinfo {author} {\bibfnamefont {S.}~\bibnamefont {Kar}}, \ and\
  \bibinfo {author} {\bibfnamefont {N.}~\bibnamefont {Dadhich}},\ }\href
  {\doibase 10.1103/PhysRevLett.90.201102} {\bibfield  {journal} {\bibinfo
  {journal} {Phys. Rev. Lett.}\ }\textbf {\bibinfo {volume} {90}},\ \bibinfo
  {pages} {201102} (\bibinfo {year} {2003})},\ \Eprint
  {http://arxiv.org/abs/gr-qc/0301003} {arXiv:gr-qc/0301003} \BibitemShut
  {NoStop}%
\bibitem [{\citenamefont {Oppenheimer}\ and\ \citenamefont
  {Volkoff}(1939)}]{Oppenheimer:1939ne}%
  \BibitemOpen
  \bibfield  {author} {\bibinfo {author} {\bibfnamefont {J.~R.}\ \bibnamefont
  {Oppenheimer}}\ and\ \bibinfo {author} {\bibfnamefont {G.~M.}\ \bibnamefont
  {Volkoff}},\ }\href {\doibase 10.1103/PhysRev.55.374} {\bibfield  {journal}
  {\bibinfo  {journal} {Phys. Rev.}\ }\textbf {\bibinfo {volume} {55}},\
  \bibinfo {pages} {374} (\bibinfo {year} {1939})}\BibitemShut {NoStop}%
\bibitem [{\citenamefont {Gorini}\ \emph {et~al.}(2008)\citenamefont {Gorini},
  \citenamefont {Moschella}, \citenamefont {Kamenshchik}, \citenamefont
  {Pasquier},\ and\ \citenamefont {Starobinsky}}]{Gorini:2008zj}%
  \BibitemOpen
  \bibfield  {author} {\bibinfo {author} {\bibfnamefont {V.}~\bibnamefont
  {Gorini}}, \bibinfo {author} {\bibfnamefont {U.}~\bibnamefont {Moschella}},
  \bibinfo {author} {\bibfnamefont {A.~Y.}\ \bibnamefont {Kamenshchik}},
  \bibinfo {author} {\bibfnamefont {V.}~\bibnamefont {Pasquier}}, \ and\
  \bibinfo {author} {\bibfnamefont {A.~A.}\ \bibnamefont {Starobinsky}},\
  }\href {\doibase 10.1103/PhysRevD.78.064064} {\bibfield  {journal} {\bibinfo
  {journal} {Phys. Rev. D}\ }\textbf {\bibinfo {volume} {78}},\ \bibinfo
  {pages} {064064} (\bibinfo {year} {2008})},\ \Eprint
  {http://arxiv.org/abs/0807.2740} {arXiv:0807.2740 [astro-ph]} \BibitemShut
  {NoStop}%
\bibitem [{\citenamefont {Kuhfittig}(2020)}]{Kuhfittig:2020fue}%
  \BibitemOpen
  \bibfield  {author} {\bibinfo {author} {\bibfnamefont {P.~K.~F.}\
  \bibnamefont {Kuhfittig}},\ }\href@noop {} {\bibfield  {journal} {\bibinfo
  {journal} {Fund. J. Mod. Phys.}\ }\textbf {\bibinfo {volume} {14}},\ \bibinfo
  {pages} {23} (\bibinfo {year} {2020})},\ \Eprint
  {http://arxiv.org/abs/2009.11179} {arXiv:2009.11179 [gr-qc]} \BibitemShut
  {NoStop}%
\end{thebibliography}
%

\end{document}